\makeatletter{}

\newif\ifms
\msfalse

\newif{\ifgrayscale}
\grayscalefalse

\ifms
  \documentclass[preprint2]{aastex}
  \usepackage[table,svgnames]{xcolor}
  \received{}
  \revised{}
  \accepted{}
  \newcommand{\cmt}[1]{\textcolor{red}{#1}}

\else

  \documentclass[revtex4,iop,numberedappendix,appendixfloats,twocolappendix]{emulateapj}
  \newcommand{\cmt}[1]{}
  \slugcomment{submitted to AJ} 

\fi

\usepackage[svgnames]{xcolor}  \usepackage[linkcolor=blue,citecolor=darkgray,colorlinks=true]{hyperref}
\usepackage{natbib}   \usepackage{amsmath}  \usepackage{graphicx}
\usepackage{enumerate}\usepackage{amssymb}  

\ifpdf
    \DeclareGraphicsExtensions{.png,.pdf,.jpg}
\else
    \DeclareGraphicsExtensions{.eps,.ps}
\fi

\def\pmstat{\ensuremath{\substack{\pm \\ \mbox{\scalebox{0.45}{stat}}}}}
\def\pmsys{\ensuremath{\substack{\pm \\ \mbox{\scalebox{0.45}{sys}}}}}

\def\fp{\ensuremath{f_{P}}}
\def\fP{\ensuremath{f_{P}}}

\def\Om{\ensuremath{\Omega_{\rm m}}}

\def\OL{\ensuremath{\Omega_{\Lambda}}}

\def\LCDM{$\Lambda$CDM}

\def\Ho{\ensuremath{H_0}}

\def\arcsec{\ensuremath{^{\prime\prime}}}

\def\Msun{\mbox{M$_{\odot}$}}
\def\Av{\mbox{$A_V$}}
\def\Rv{\mbox{$R_V$}}

\newcommand{\SNuVol}{\ensuremath{10^{-4}~\mbox{yr}^{-1}~\mbox{Mpc}^{-3}~{\mbox{h}_{70}}^{3}}}
\newcommand{\CCSN}{CC\,SN}

\newcommand{\SNIa}{SN\,Ia}

\newcommand{\SNRz}{SNR($z$)}

\def\galsnid{{\it galsnid}}

\def\HST{{\it HST}}

\newcommand{\JHU}{Department of Physics and Astronomy, The Johns Hopkins University, Baltimore, MD 21218, USA}
\newcommand{\STScI}{Space Telescope Science Institute, Baltimore, MD 21218, USA}
\newcommand{\AMNH}{Department of Astrophysics, American Museum of Natural History, New York, NY 10024, USA}
\newcommand{\Berkeley}{Department of Astronomy, University of California, Berkeley, CA 94720, USA}
\newcommand{\Riverside}{Department of Physics and Astronomy, University of California, Riverside, CA 92521, USA}
\newcommand{\Irvine}{Department Physics and Astronomy, University of California, Irvine, 92697, USA}

\newcommand{\Copenhagen}{Dark Cosmology Centre, Niels Bohr Institute, University of Copenhagen, Juliane Maries Vej 30, DK-2100 Copenhagen, Denmark.}
\newcommand{\Arizona}{Department of Astronomy, University of Arizona, Tucson, AZ 85721.}
\newcommand{\SantaCruz}{Department of Astronomy and Astrophysics, University of California, Santa Cruz, CA 92064.}
\newcommand{\NotreDame}{Department of Physics, University of Notre Dame, Notre Dame, IN 46556.}
\newcommand{\TelAviv}{Department of Astrophysics, Tel Aviv University, 69978 Tel Aviv, Israel.}
\newcommand{\Rutgers}{Department of Physics and Astronomy, Rutgers, The State University of New Jersey, Piscataway, NJ 08854.}
\newcommand{\CfA}{Harvard-Smithsonian Center for Astrophysics, Cambridge, MA 02138.}

\newcommand{\NOAO}{National Optical Astronomy Observatory, 950 North Cherry Avenue, Tucson, AZ 85719, USA}
\newcommand{\LBNL}{E.O. Lawrence Berkeley National Lab, 1 Cyclotron Rd., Berkeley, CA, 94720}
\newcommand{\GSFC}{Astrophysics Science Division, NASA Goddard Space Flight Center, Mail Code 661, Greenbelt, MD 20771, USA}
\newcommand{\JSSI}{Joint Space Science Institute, University of Maryland, College Park, Maryland 20742, USA}
\newcommand{\ESO}{European Southern Observatory, Garching bei M\"unchen, Germany}
\newcommand{\ECUmunich}{Excellence Cluster Universe, Technische Universit\"at M\"unchen, Germany}
\newcommand{\PSU}{Department of Astronomy and Astrophysics, Pennsylvania State University, University Park, PA 16802, USA}
\newcommand{\UT}{Department of Astronomy, University of Texas, Austin, TX 78712, USA}
\newcommand{\ASU}{School of Earth and Space Exploration, Arizona State University, Tempe, AZ 85287, USA}
\newcommand{\UK}{Department of Physics and Astronomy, University of Kentucky, Lexington, KY 40506, USA}

\ifms
\else

\fi

\newcommand{\insertfigwide}[2] {
\begin{figure*}
\begin{center}
\resizebox{\textwidth}{!}{\includegraphics{{#1}}}
\caption{{#2}}
\end{center}
\end{figure*}
}

\newcommand{\insertfigfloat}[2] {
\begin{figure}
\begin{center}
\resizebox{\columnwidth}{!}{\includegraphics{{#1}}}
\figcaption{{#2}}
\end{center}
\end{figure}
}

\newcommand{\insertfigquad}[5] {
\begin{figure*}
\begin{center}
\resizebox{0.48\textwidth}{!}{\includegraphics{{#1}}}
\resizebox{0.48\textwidth}{!}{\includegraphics{{#2}}}
\resizebox{0.48\textwidth}{!}{\includegraphics{{#3}}}
\resizebox{0.48\textwidth}{!}{\includegraphics{{#4}}}
\figcaption{{#5}}
\end{center}
\end{figure*}
}

\newcommand{\insertfig}[2] {
\insertfigfloat{{#1}}{{#2}}
}

\def\primo{GSD10Pri}                                                                                                                                                                                                              
                                                                   \def\wilson{UDS10Wil}

\slugcomment{AJ, in press} 

\shorttitle{CANDELS \SNIa\ Rates}
\shortauthors{Rodney et al.}

\begin{document}

\title{Type Ia Supernova Rate Measurements to Redshift 2.5 from
CANDELS:\linebreak Searching for Prompt Explosions in the Early Universe}

\makeatletter{}    \author{Steven A.~Rodney\altaffilmark{1,$\dagger$}}
    \author{Adam G.~Riess\altaffilmark{1,2}}
    \author{Louis-Gregory Strolger\altaffilmark{2}}
    \author{Tomas Dahlen\altaffilmark{2}}
    \author{Or Graur\altaffilmark{1,3,4}}

    \author{Stefano Casertano\altaffilmark{2}}
    \author{Mark E.~Dickinson\altaffilmark{5}}
    \author{Henry C.~Ferguson\altaffilmark{2}}
    \author{Peter Garnavich\altaffilmark{6}}
    \author{Brian Hayden\altaffilmark{7}}
    \author{Saurabh W.~Jha\altaffilmark{8}}
    \author{David O.~Jones\altaffilmark{1}}
    \author{Robert P.~Kirshner\altaffilmark{9}}
    \author{Anton M.~Koekemoer\altaffilmark{2}}
    \author{Curtis ~McCully\altaffilmark{8}}
    \author{Bahram Mobasher\altaffilmark{10}}
    \author{Brandon Patel\altaffilmark{8}}
    \author{Benjamin J.~Weiner\altaffilmark{11}}

        \author{S.~Bradley Cenko\altaffilmark{12,13}}
    \author{Kelsey I.~Clubb\altaffilmark{14}}
    \author{Michael Cooper\altaffilmark{15}}
    \author{Alexei V.~Filippenko\altaffilmark{14}}
    \author{Teddy F.~Frederiksen\altaffilmark{16}}
    \author{Jens Hjorth\altaffilmark{16}}
    \author{Bruno Leibundgut\altaffilmark{17,18}}
    \author{Thomas Matheson\altaffilmark{5}}
    \author{Hooshang Nayyeri\altaffilmark{10}}
    \author{Kyle Penner\altaffilmark{11}}
    \author{Jonathan Trump\altaffilmark{19,$\dagger$}}
    \author{Jeffrey M.~Silverman\altaffilmark{20}}
    \author{Vivian U\altaffilmark{10}}

    \author{K. Azalee Bostroem\altaffilmark{2}}
    \author{Peter Challis\altaffilmark{9}}
    \author{Abhijith Rajan\altaffilmark{21}}
    \author{Schuyler Wolff\altaffilmark{1}}
    
    \author{S.~M.~Faber\altaffilmark{22}}
    \author{Norman A.~Grogin\altaffilmark{2}}
    \author{Dale Kocevski\altaffilmark{23}}

    \altaffiltext{1}{\JHU}
    \altaffiltext{2}{\STScI}
    \altaffiltext{3}{\TelAviv}
    \altaffiltext{4}{\AMNH}
    \altaffiltext{5}{\NOAO}
    \altaffiltext{6}{\NotreDame}
    \altaffiltext{7}{\LBNL}
    \altaffiltext{8}{\Rutgers}
    \altaffiltext{9}{\CfA}
    \altaffiltext{10}{\Riverside}
    \altaffiltext{11}{\Arizona}
    \altaffiltext{12}{\GSFC}
    \altaffiltext{13}{\JSSI}
    \altaffiltext{14}{\Berkeley}
    \altaffiltext{15}{\Irvine}
    \altaffiltext{16}{\Copenhagen}
    \altaffiltext{17}{\ESO}
    \altaffiltext{18}{\ECUmunich}
    \altaffiltext{19}{\PSU}
    \altaffiltext{20}{\UT}
    \altaffiltext{21}{\ASU}
    \altaffiltext{22}{\SantaCruz}
    \altaffiltext{23}{\UK}
    \altaffiltext{$\dagger$}{Hubble Fellow}
    \altaffiltext{$\ddagger$}{NSF Postdoctoral Fellow}

\begin{abstract}

The Cosmic Assembly Near-infrared Deep Extragalactic Legacy Survey
(CANDELS) was a multi-cycle treasury program on the {\it Hubble Space
Telescope} (\HST) that surveyed a total area of $\sim$0.25 deg$^2$ with
$\sim$900 \HST\ orbits spread across 5 fields over 3 years.  Within
these survey images we discovered 65 supernovae (SN) of all types, out
to $z\approx2.5$.  We classify $\sim$24 of these as Type Ia SN
(\SNIa) based on host-galaxy redshifts and SN photometry (supplemented
by grism spectroscopy of 6 SN). Here we present a measurement of the
volumetric \SNIa\ rate as a function of redshift, reaching for the
first time beyond $z=2$ and putting new constraints on \SNIa\
progenitor models.  Our highest redshift bin includes detections of SN
that exploded when the universe was only $\sim$3 Gyr old and near the
peak of the cosmic star-formation history. This gives the CANDELS 
high-redshift sample unique leverage for evaluating the fraction of \SNIa\
that explode promptly after formation ($<$500 Myr). Combining the
CANDELS rates with all available \SNIa\ rate measurements in the
literature we find that this prompt \SNIa\ fraction
is \fp=0.53 \pmstat$^{0.09}_{0.10}$ \pmsys$^{0.10}_{0.26}$,
consistent with a delay time distribution that follows a simple
$t^{-1}$ power law for all times $t>40$ Myr.  However, a mild tension
is apparent between ground-based low-$z$ surveys and space-based
high-$z$ surveys.  In both CANDELS and the sister \HST\ program
CLASH, we find a low rate of \SNIa\ at $z>1$.  This
could be a hint that prompt progenitors are in fact relatively
rare, accounting for only $\sim$20\% of all \SNIa\ explosions --
though further analysis and larger samples will be needed to examine
that suggestion.
\end{abstract}

\keywords{supernovae: general; surveys; infrared: general}

\section{Introduction}

The prevailing model for a Type Ia supernova (\SNIa) progenitor system
begins with a binary system in which the primary star evolves to
become a white dwarf (WD). The WD acquires
mass from its companion star, approaches the Chandrasekhar
limit, and explodes in a thermonuclear runaway \citep[for reviews, 
see][]{Hillebrandt:2000,Livio:2001}.  The companion star that feeds
the WD and thereby sets off the thermonuclear bomb is one of the key
components of this model, but remains a topic of ongoing debate.  In
single-degenerate (SD) models, the companion is a main sequence 
or evolved giant star, transferring mass via Roche-lobe overflow,
stellar winds or other means \citep{Whelan:1973}. In double-degenerate
(DD) models the companion is another WD, merging with the primary
after a period of orbital decay driven by gravitational wave
radiation \citep{Iben:1984,Webbink:1984}.  More recent variations on
these pathways to explosion include the ``core-degenerate
scenario'' \citep{Kashi:2011} and perturbation-induced mergers in triple
systems \citep{Thompson:2011}. 

The \SNIa\ explosion rate as a function of redshift, \SNRz, can
provide an important observational test to constrain \SNIa\ progenitor
models and possibly distinguish between them.  In this paper we will
present measurements of the \SNIa\ rate as a function of redshift and
use them to place new constraints on \SNIa\ progenitor models,
particularly on the fraction of \SNIa\ progenitors that explode within
500 Myr after their formation.

Suppose we have a burst of star formation in a galaxy,
such that the star-formation rate can be approximated by a delta
function in time.  Binary population synthesis modeling gives us the initial
conditions of all the binaries (mass, orbital separation, etc.),
and a progenitor model sets the conditions necessary for explosion as
a \SNIa.  Using a stellar evolution model, one can follow the binary
systems as they evolve, measuring the {\em delay time distribution}
(DTD) between formation and explosion.  To put constraints on SNIa
progenitor models, we can translate this DTD to cosmic scales and
compare it to the observed volumetric \SNIa\ rate as a function of
look-back time, as first proposed by \citet{Madau:1998b}.

\ifgrayscale
   \insertfig{FIG/figRatesPreCC_gray}{\label{fig:RatesPreCC}
Volumetric \SNIa\ rates before completion of the CANDELS and CLASH SN
surveys.  Assorted ground-based surveys are plotted as white
circles \citep{Blanc:2004,Botticella:2008,Cappellaro:1999,Dilday:2010,Hardin:2000,Horesh:2008,Graur:2013,Li:2011a,Melinder:2012,Pain:2002,Perrett:2012,Rodney:2010b,Tonry:2003}.
Three high-redshift SN surveys are highlighted with filled markers: circles for the
Subaru Deep Field \citep[SDF,][]{Graur:2011}, downward triangles
for volumetric (not cluster) rates from the Cluster Supernova
Survey \citep[CSS,][]{Barbary:2012}, upward triangles for the
GOODS and PANS surveys \citep{Dahlen:2008}. 
  (A color version of this figure appears in the online edition.)
}
\else
   \insertfig{FIG/figRatesPreCC}{\label{fig:RatesPreCC}
Volumetric \SNIa\ rates before completion of the CANDELS and CLASH SN
surveys.  Assorted ground-based surveys are plotted as white
circles \citep{Blanc:2004,Botticella:2008,Cappellaro:1999,Dilday:2010,Hardin:2000,Horesh:2008,Graur:2013,Li:2011a,Melinder:2012,Pain:2002,Perrett:2012,Rodney:2010b,Tonry:2003}.
Three high-redshift SN surveys are highlighted: gray circles for the
Subaru Deep Field \citep[SDF,][]{Graur:2011}, blue downward triangles
for volumetric (not cluster) rates from the Cluster Supernova
Survey \citep[CSS,][]{Barbary:2012}, green upward triangles for the
GOODS and PANS surveys \citep{Dahlen:2008}. }
\fi

As shown in Figure~\ref{fig:RatesPreCC}, recent measurements 
of the \SNIa\ rate at low redshift ($z<1$) are in good agreement,
consistently finding that the \SNRz\ rises steadily to at least
$z\approx1$ \citep[e.g.,][]{Rodney:2010b,Dilday:2010,Perrett:2012}.
However, at $z>1$ the trend of the \SNRz\ curve is much less clear.
The spectral energy distribution of a \SNIa\ peaks in the rest-frame B
band with an absolute magnitude around -19.5.  At $z=1.2$ that peak
brightness becomes fainter than 25th magnitude in the observer's $z$
band -- making discovery and light curve follow-up nearly impossible
for ground based observatories.

For that reason, space-based surveys using the {\it Hubble Space
Telescope}'s Advanced Camera for Surveys (ACS) have been the
primary vehicle for tracking the \SNRz\ to $z\approx1.5$.  The GOODS+PANS
surveys were the first programs to extend rate measurements beyond
$z\approx1$ \citep{Dahlen:2004,Dahlen:2008}, and their measured rates
suggested a peak in the \SNIa\ rate at $z\approx1.2$, with a decline at
higher redshifts.  Independent examination of the same survey data
recovered the same trend \citep{Kuznetsova:2008}, although both
analyses were limited by a small sample size in the highest redshift
bin.  Subsequently, the Cluster Supernova Survey (CSS) of the
Supernova Cosmology Project used ACS to measure the
volumetric \SNIa\ rate \citep{Barbary:2012}.  These data revealed a
similar peak and decline, although with even larger uncertainty in the
high-$z$ bins.  From the ground, the Subaru Deep Field (SDF) SN survey
used the Suprime-cam imager on the {\it Subaru} telescope to reach
similar redshifts \citep{Poznanski:2007,Graur:2011}.  As can
be seen in Figure~\ref{fig:RatesPreCC}, these SDF rates formally show
no decline in the highest redshift bin, but they are consistent with
the ACS results, within the errors.

The ACS high-$z$ \SNIa\ generally have reliable classifications,
based on well-sampled multi-band light curves, spectroscopic
redshifts, and \HST\ grism spectroscopy of most \SNIa\ candidates.
However, due to the relatively small survey area, these programs have
very large statistical uncertainties (\citet{Dahlen:2008} have $\sim$3
SN in their highest redshift bin, \citet{Barbary:2012} have $\sim$1).
In contrast, the SDF survey built up a larger sample (10 \SNIa\ at
$z\approx1.5$) but their survey design introduced a potential for large
systematic biases.  The SDF epochs were spaced by $\sim$1 year,
meaning that the phase of the SN light curve at discovery was
unconstrained, and the classification of detected SN was based on only
a single epoch of photometric data in the {\it R,i$\prime$,z$\prime$}
bands.  Furthermore, redshifts for the SDF high-$z$ SN sample were
based almost exclusively on photometric redshift estimates of the SN
host galaxies, not as precise or reliable as spectroscopic redshifts
$-$ though see \citet{Frederiksen:2012b} for one spectroscopic
confirmation of a SDF host galaxy at $z=1.55$.

An apparent peak in the \SNIa\ rate at $z\approx1$ and a decline toward
$z=1.5$ has been interpreted as indicating a delay of $>1$ Gyr between
formation and explosion for
most \SNIa\ \citep{Strolger:2004,Strolger:2010}.  This would be
broadly consistent with some SD models, and inconsistent with DD
models, which typically predict a large fraction of \SNIa\ that
explode promptly after star formation (within 1 Gyr).  A clear
measurement of the shape of the \SNIa\ rate function at $z>1$ would
provide an important constraint on DTD models, and would go a long way
toward resolving the question of whether a SD or DD model could be the
dominant progenitor channel for all \SNIa\ at all redshifts.  Given
the problems with current high-$z$ SN rates, there is a clear need to
improve the measurement by expanding the sample of well-classified SN
at $z>1$.

\smallskip

In this paper we present a measurement of the \SNRz\ from a
sample of 65 SN discovered in the CANDELS SN program, extending
the \SNRz\ measurement for the first time to $z=2.5$.  This SN survey
is a joint operation of two \HST\ Multi-Cycle Treasury (MCT) programs:
the Cosmic Assembly Near-infrared Deep Extragalactic Legacy Survey
\citep[CANDELS; PIs:Faber and Ferguson;][]{Grogin:2011,Koekemoer:2011}, 
and the Cluster Lensing and Supernovae search with
Hubble \citep[CLASH; PI:Postman;][]{Postman:2012}. The SN discovery
and follow-up for both programs were allocated to the \HST\ MCT SN
program (PI:Riess). The results presented here are based on the full
five fields and $\sim$0.25 deg$^2$ of the CANDELS program, observed
from 2010 to 2013. A companion paper presents the \SNIa\ rates from
the CLASH sample \citep{Graur:2014}.  A composite analysis that
combines the CANDELS+CLASH SN sample and revisits past \HST\ surveys
will be presented in a future paper.

In Section \ref{sec:CANDELS} we describe the SN search component of
the CANDELS survey, and in Section \ref{sec:DetectionEfficiency} we
describe our detection efficiency measurements.  Our photometric SN
classifications are presented in Section \ref{sec:Classification},
properties of the SN host galaxies are described in
Section \ref{sec:HostGalaxies}, and in
Section~\ref{sec:GrismSpectroscopy} we detail new grism spectroscopy
for 4 of our SN.  The rates calculation is described in
Section \ref{sec:Rates} and we discuss the consequences for \SNIa\
progenitor models in Section \ref{sec:Discussion}. Finally, a summary
is presented in Section \ref{sec:Summary}.  In tables and figures
throughout the paper, we present the subset of 14 SN with
$z>1.5$ in the main body of the text, with the remaining 51
shown in Appendix~\ref{app:lowz}.  Throughout this work we assume a
flat \LCDM\ cosmology with \Ho=70, \Om=0.3 and \OL=0.7.

\section{The CANDELS SN Survey}\label{sec:CANDELS}

The 3-year CANDELS program was designed to probe galaxy evolution out
to $z\approx8$ with deep infrared (IR) and optical imaging of five
well-studied extragalactic fields: GOODS-S, GOODS-N, COSMOS, UDS, and
EGS.\footnote{GOODS-S/N: the Great Observatories Origins Deep Survey
South and North \citep{Giavalisco:2004}; COSMOS: the Cosmic Evolution
Survey \citep{Scoville:2007,Koekemoer:2007}; UDS: the UKIDSS Ultra
Deep Survey \citep{Lawrence:2007,Cirasuolo:2007}; EGS: the Extended
Groth Strip \citep{DavisM:2007}} As described fully
in \citet{Grogin:2011}, the CANDELS program includes both ``wide'' and
``deep'' fields.  The wide component of CANDELS comprises the COSMOS,
UDS, and EGS fields, plus one third of the GOODS-S field and one half
of the GOODS-N field -- a total survey area of 730 square
arcminutes. The CANDELS survey provides two visits to each wide field,
spaced by $\sim$50 days. The ``deep'' component of CANDELS came from
the central 67 square arcminutes of each of the GOODS-S and GOODS-N
fields. These deep regions were each visited 15 times over the course
of two years (2010-2012 for GOODS-S, 2012-2013 for GOODS-N). Only 10
of those visits are used for SN discovery (the other visits lack
template data for generating difference images), and those 10 epochs
are also spaced at a cadence of $\sim$50 days.  The CANDELS fields
analyzed in this work are described in Table~\ref{tab:Survey}.

Table~\ref{tab:Exposures} presents the exposure times and 5$\sigma$
limiting magnitudes for a typical single-epoch set of exposures.  Each
CANDELS visit includes a set of four IR exposures from the Wide
Field Camera 3 (WFC3) IR detector: two in F160W ($H$ band) and two in
F125W ($J$ band).  These are the search filters for the CANDELS SN
survey (i.e., all SN in our sample are IR detections).
Additionally, each observation set includes a broad optical band,
which helps to distinguish \SNIa\ from core-collapse supernovae
(\CCSN) and other transients (see Section \ref{sec:Classification}).
In $\sim$80\%\ of the SN search visits, this blue component is
collected within minutes of the IR exposures as a single exposure
using the WFC3 UVIS camera in the F350LP filter (a broad ``white
light'' filter that we refer to as ``W band''). In the remaining
$\sim$20\%\ of visits (in the wide fields) the W band exposure is
replaced with ACS observations in the F606W filter (broad V band), and
complemented by the ACS F814W filter (broad I band).  These ACS
observations come from coordinated parallel visits and are taken
within 3 days of the primary IR visit.

\makeatletter{}\begin{small}
\ifms 
  \begin{deluxetable}{cllclp{2in}} 
\else 
  \begin{deluxetable*}{cllclp{2in}} 
\fi
\tablecolumns{6}
\tablecaption{CANDELS SN Survey Fields\tablenotemark{a}\label{tab:Survey}}
\tablehead{ 
  \colhead{} & \colhead{R.A.} & \colhead{Decl} & \colhead{WFC3-IR} & \colhead{Searchable Area} & \colhead{SN Search}\\
  \colhead{Field}  & \colhead{(J2000)}  & \colhead{(J2000)}  & \colhead{Tiles/Epoch}  & \colhead{(arcmin$^2$)}  & \colhead{Epochs (MJD)\tablenotemark{b}}
}
\startdata
COSMOS          & 10:00:28 & $+$02:12:04 & 44 &  196.8 $(9\arcmin\times 22\arcmin)$ & [55905], 55953 \\[1mm]
EGS-A           & 14:19:18 & $+$52:49:30 & 25 &  106.2 $(\frac{1}{2}$ of $7\arcmin\times 32\arcmin)$\tablenotemark{c} & [55653], 55703 \\
EGS-B           & 14:19:18 & $+$52:49:30 & 20 &   92.9 $(\frac{1}{2}$ of $7\arcmin\times 32\arcmin)$\tablenotemark{c} & [56387], 56437 \\[1mm]
UDS             & 02:17:38 & $-$05:12:00 & 44 &  207.1 $(9\arcmin\times 22\arcmin)$ & [55512], 55562 \\[1mm]
GOODS-S Wide    & 03:32:42 & $-$27:53:37 & $\sim8$  &  39.4 $(4\arcmin\times 10\arcmin)$ & [55573], 55621 \\
GOODS-S Deep    & 03:32:28 & $-$27:46:01 & $\sim15$ &  66.5 $(7\arcmin\times 10\arcmin)$\tablenotemark{d} & [55480], 55528, 55578, 55624, 55722, 55774, 55821, 55860, 55921, 55974\\[1mm]
GOODS-N Wide NE & 12:37:29 & $+$62:18:40 & $\sim8$ &  38.1 $(4\arcmin\times 10\arcmin)$ & [56183], 56238 \\
GOODS-N Wide SW & 12:36:20 & $+$62:10:25 & $\sim10$ &  49.5 $(5\arcmin\times 10\arcmin)$ & [56020], 56073 \\
GOODS-N Deep    & 12:36:55 & $+$62:14:19 & $\sim15$ &  66.8 $(7\arcmin\times 10\arcmin)$\tablenotemark{d} & [56020], 56073, 56126, 56183, 56238, 56297, 56348, 56402, 56458, 56511\\

\enddata
\tablenotetext{a}{Coordinates give approximate center of each CANDELS IR survey field.}
\tablenotetext{b}{Mean date of observation epoch. First epoch listed [in brackets] provided IR template images.}
\tablenotetext{c}{The CANDELS EGS field was divided into two interlocking halves, observed separately in 2011 and 2013. See \citet{Grogin:2011} for details.}
\tablenotetext{d}{The deep field search areas vary by epoch. The given value reflects the average.}
\ifms \end{deluxetable}
\else \end{deluxetable*} \fi
\end{small}
 
\makeatletter{}\begin{small}
\begin{deluxetable}{llcc} 
\tablecolumns{4}
\tablecaption{Typical Exposures for a Single SN Search Epoch\label{tab:Exposures}}
\tablehead{ 
  \colhead{} & \colhead{} & \colhead{Exposures} & \colhead{Limiting}\\
  \colhead{Camera} & \colhead{Filter} & \colhead{(N$_{exp} \times$ sec)} & \colhead{Magnitude\tablenotemark{a}} }
\startdata
WFC3-IR   & F160W (H)   & 2 $\times$ 600 & 25.4 \\
WFC3-IR   & F125W (J)   & 2 $\times$ 500 & 25.8 \\
WFC3-UVIS & F350LP (W)  & 1 $\times$ 430 & 27.8 \\
ACS-WFC   & F814W (I)   & 2 $\times$ 700 & 27.3 \\
ACS-WFC   & F606W (V)   & 2 $\times$ 350 & 28.1 \\
\enddata
\tablenotetext{a}{Vega magnitude that yields S/N$\sim$5 in the given exposure sequence.}
\end{deluxetable}
\end{small}

In addition to the $\sim$750 \HST\ orbits devoted to survey imaging in
the CANDELS program, an additional 150 orbits were allocated for
target of opportunity (ToO) the \HST\ MCT SN follow-up observations of
newly discovered SN.  Another 52 orbits were provided by the
CLASH program, so the total CANDELS+CLASH SN follow-up allocation was
202 orbits. These follow-up visits provided supplementary imaging and
slitless spectroscopy observations to aid in the classification of SN
candidates, and to measure the light curves of \SNIa, allowing
distance determinations for cosmology.

\subsection{Data Processing Pipeline}\label{sec:Pipeline}

All CANDELS survey images were processed through a
data processing pipeline optimized for the detection of SN by human
searchers. This pipeline is similar in function to the CANDELS and CLASH
pipelines \citep{Koekemoer:2011,Postman:2012}, but includes some
important differences specific to the SN search.  There are four
principal components in the pipeline: calibration, image combination,
template subtraction, and fake SN planting. 

In the calibration stage, {\tt RAW} images from \HST\ are processed into
{\tt FLT} images using the {\tt STSDAS} calibration tools provided by
the Space Telescope Science
Institute.\footnote{\url{http://www.stsci.edu/institute/software\_hardware/pyraf/stsdas}} This includes bias correction, dark subtraction, flat fielding, and
``up-the-ramp'' fitting for cosmic ray rejection, as appropriate for
each camera and detector. 

The image combination step uses the {\tt MultiDrizzle}
software \citep{Koekemoer:2002,Fruchter:2002} to combine multiple
dithered images in the same filter from the same observing epoch,
while also removing the geometric distortion of the \HST\ focal
plane. For each drizzled WFC3-IR image, we then generate a template
image that combines all intersecting images from the prior
epoch(s). These components of the template image are astrometrically
registered using catalog matching to align them with the WFC3-IR image
of the current epoch.  The astrometric registration for the SN search
is done tile-by-tile and the output pixel grid is left in the natural
unrotated frame of the observation. This contrasts with the CANDELS
mosaic imaging pipeline \citep{Koekemoer:2011}, which constructs a
global astrometric solution across the whole field, and rotates every
image to put North up and East to the left. These choices for the SN
pipeline are designed to maximize the precision of the local
inter-epoch registrations and to minimize dilution of the already
undersampled PSF for single-visit drizzled images.

Next, each template image is subtracted from the corresponding search
epoch image, producing the difference images for SN discovery.  Due to
the very stable point spread function (PSF) of \HST, the CANDELS images
do not require any convolution with a PSF kernel to match conditions
across epochs \citep{Alard:1998}, as is commonly done in ground-based
SN surveys.  The CANDELS visits were constructed with small
positioning shifts after each exposure, such that the two H band and
two J band exposures together formed a 4-point ``box'' dither pattern.
This yields better sampling of the PSF and helps in the removal of
detector artifacts from the final combined image.  To take advantage
of the full dither sequence, our SN searching was primarily done on a
combined ``J+H'' image -- simply the sum of the F125W and F160W
difference images for each epoch.

In the final stage of the data processing pipeline, we reprocess all
the search epoch data, this time with fake SN planted into the WFC3-IR
survey images.  These synthetic SN enable a direct measurement of the
detection efficiency of our human searchers (see
Section \ref{sec:DetectionEfficiency}).  Each fake SN consists of a
small image ($\sim50\times50$ pixels) of a simulated point source,
generated using the TinyTim software \citep{Krist:2011}.  The fake SN
images are added to the WFC3-IR images at the FLT stage, 
after image calibration and before drizzling.  These ``faked'' FLT
files are then redrizzled, and the existing template images are
subtracted off, resulting in a parallel set of ``faked'' difference
images.

\subsection{SN Discovery}
\makeatletter{}\begin{figure*}
  \includegraphics[width=0.333\textwidth]{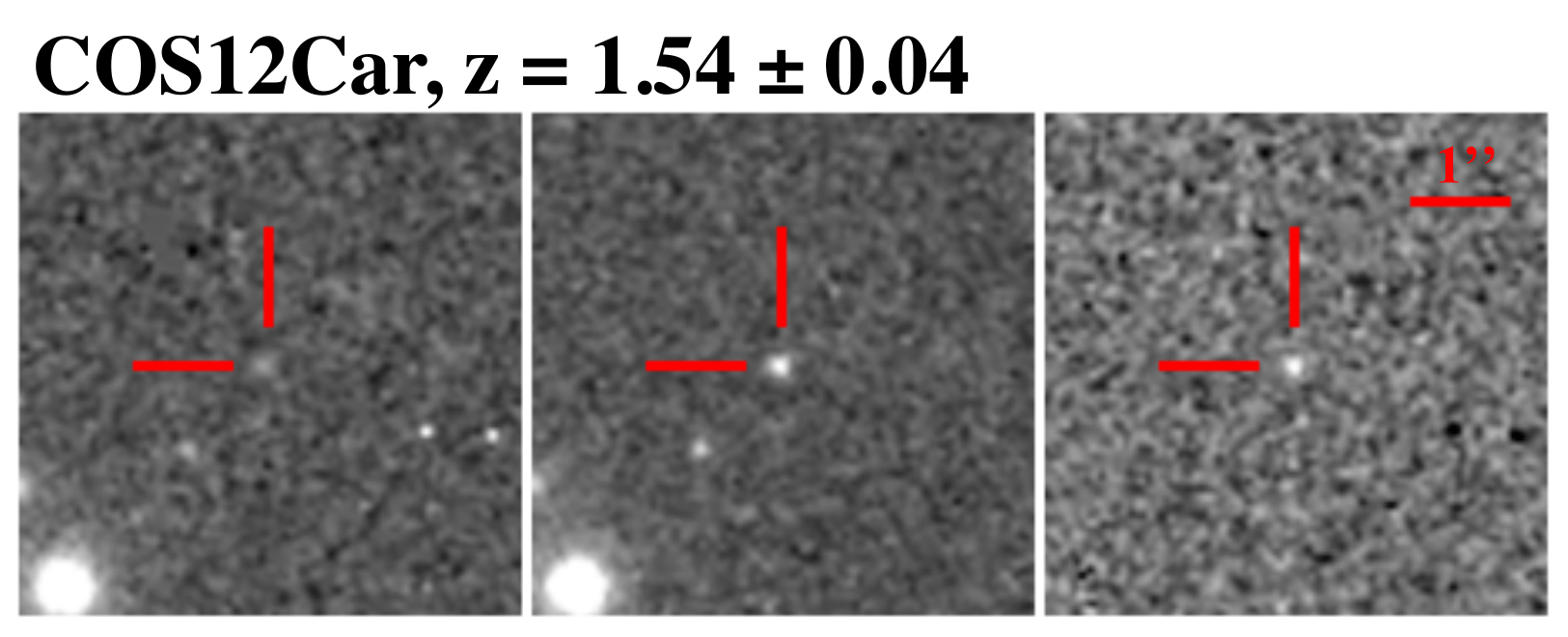}
  \includegraphics[width=0.333\textwidth]{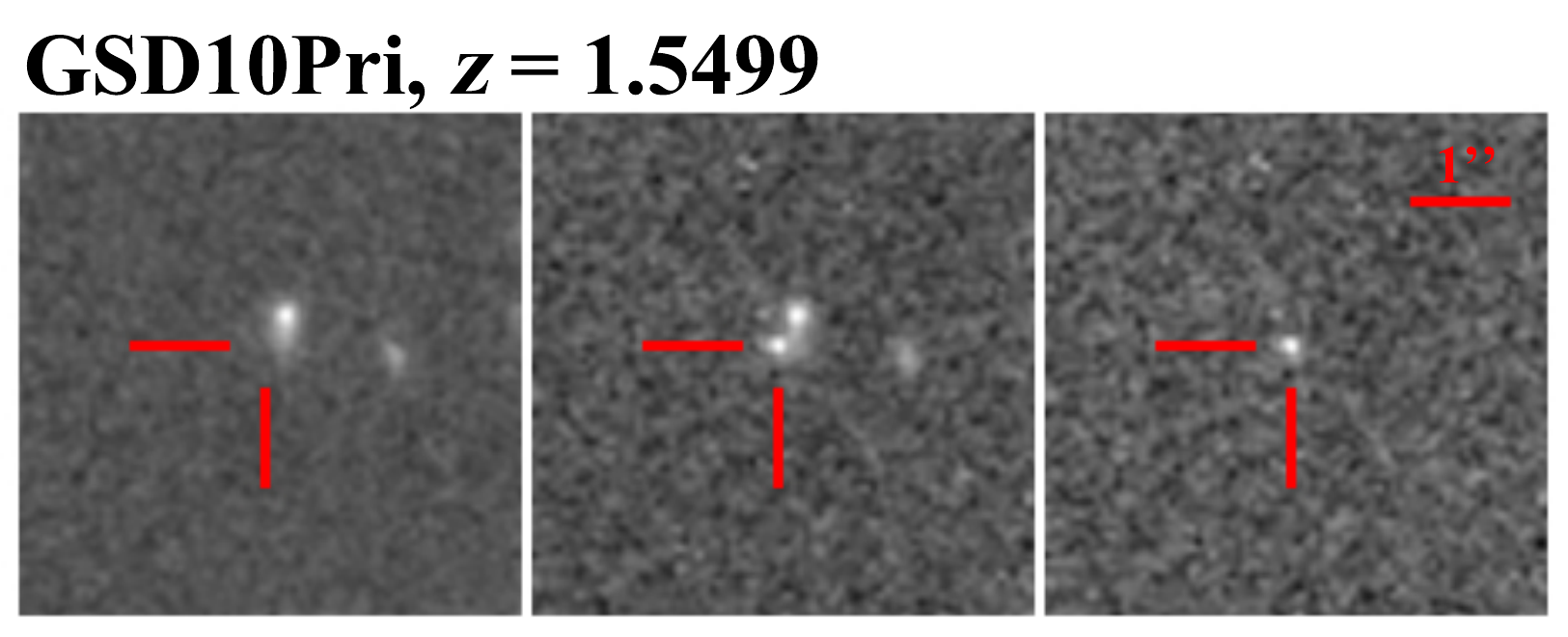}
  \includegraphics[width=0.333\textwidth]{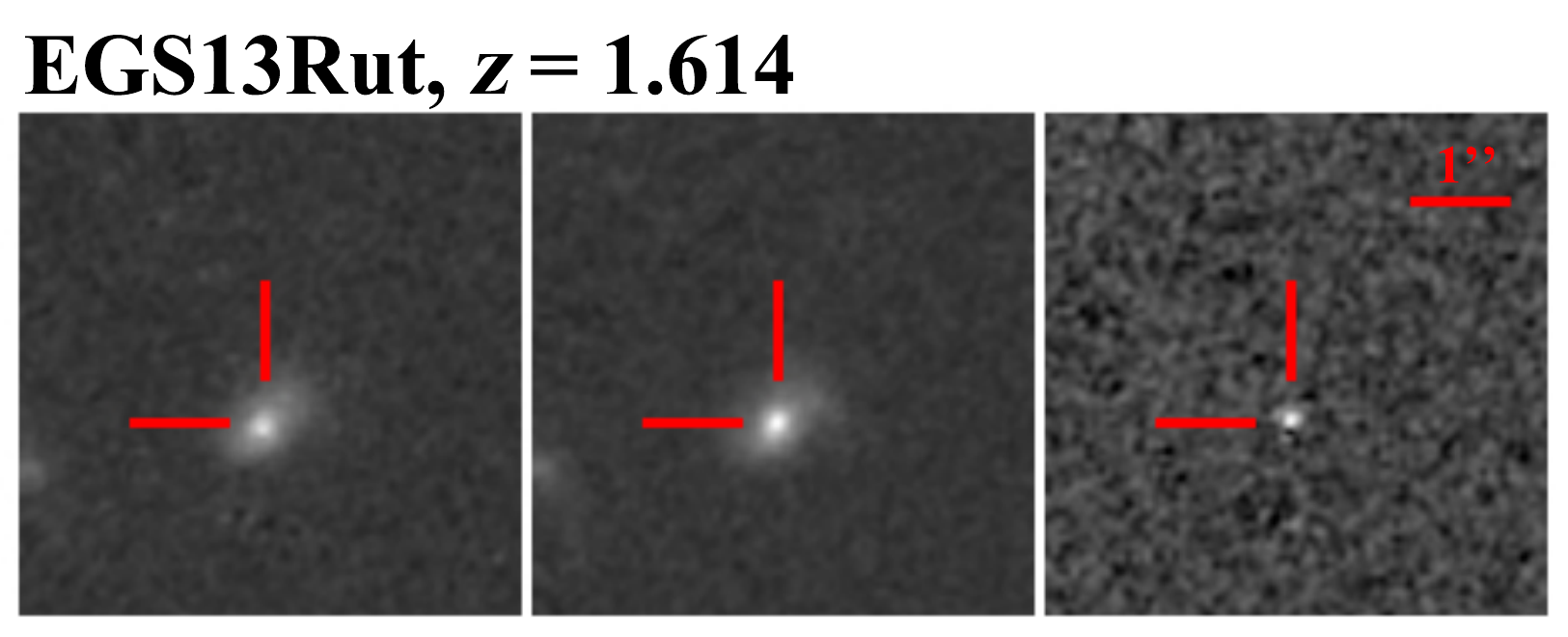}
  \includegraphics[width=0.333\textwidth]{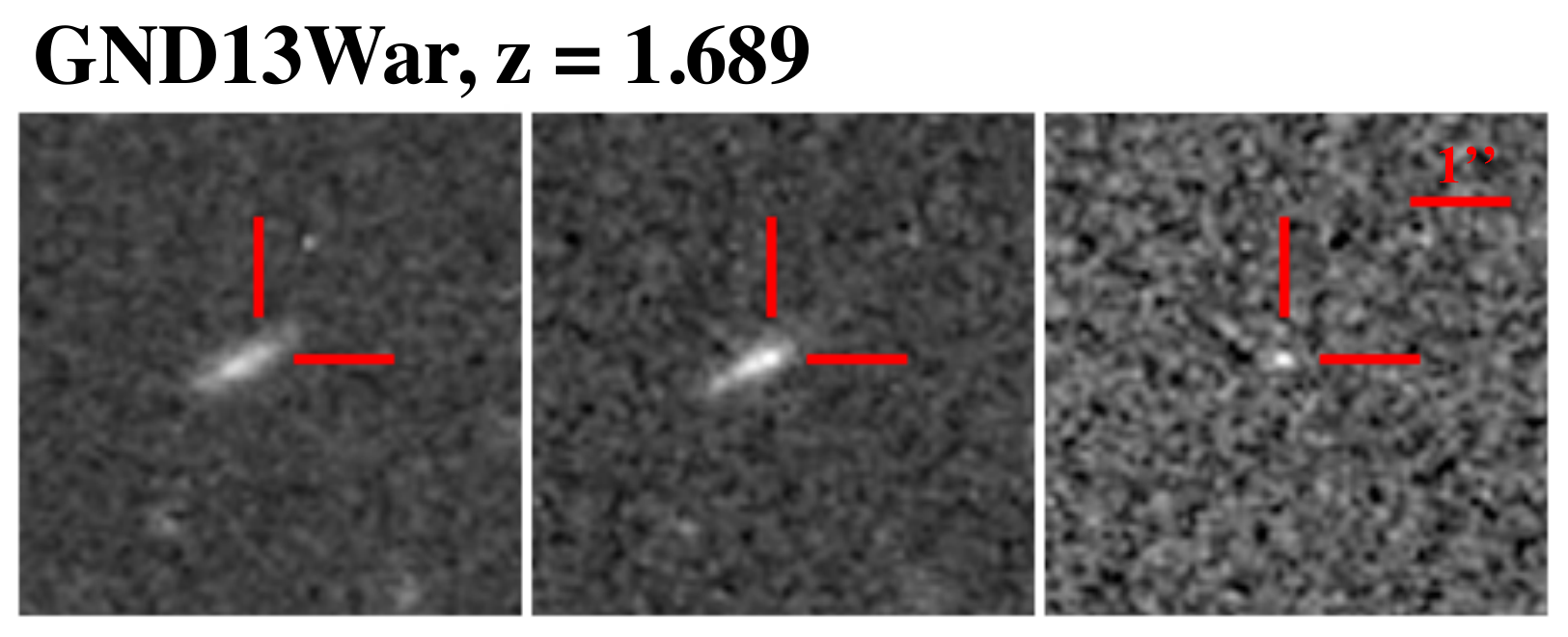}
  \includegraphics[width=0.333\textwidth]{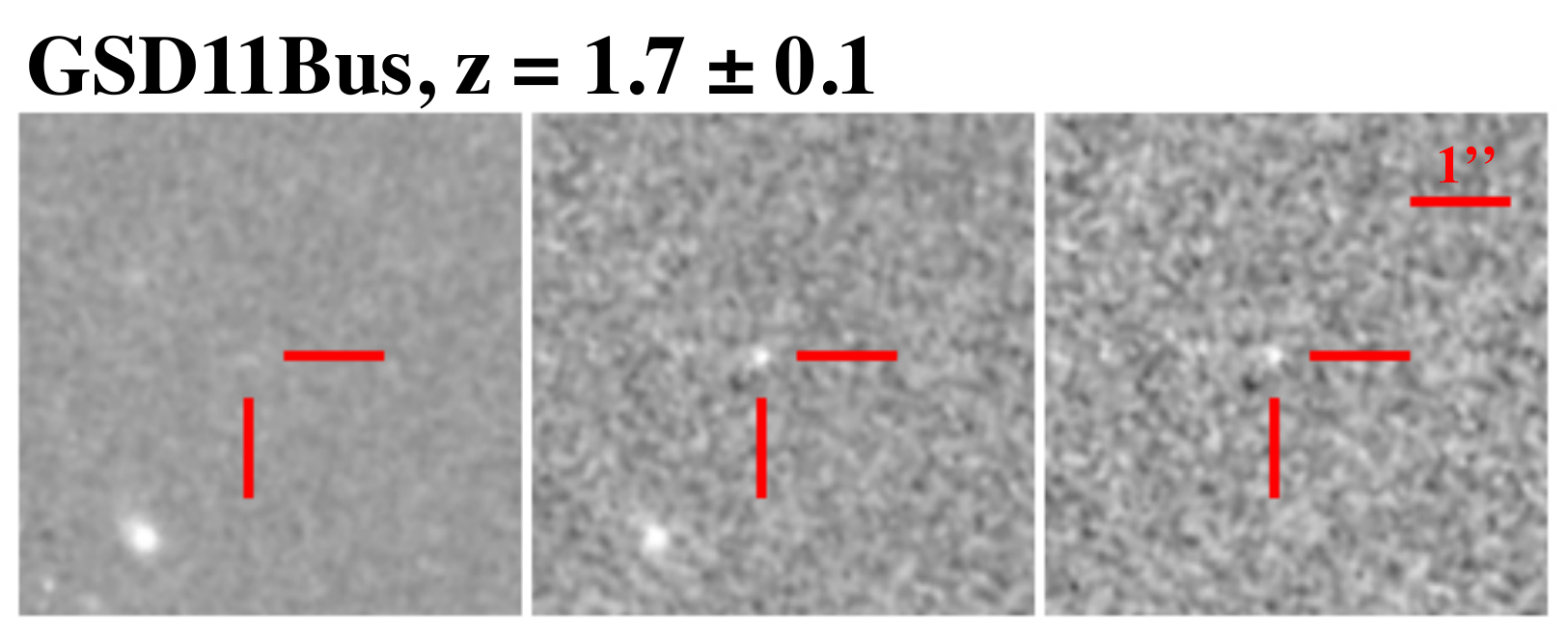}
  \includegraphics[width=0.333\textwidth]{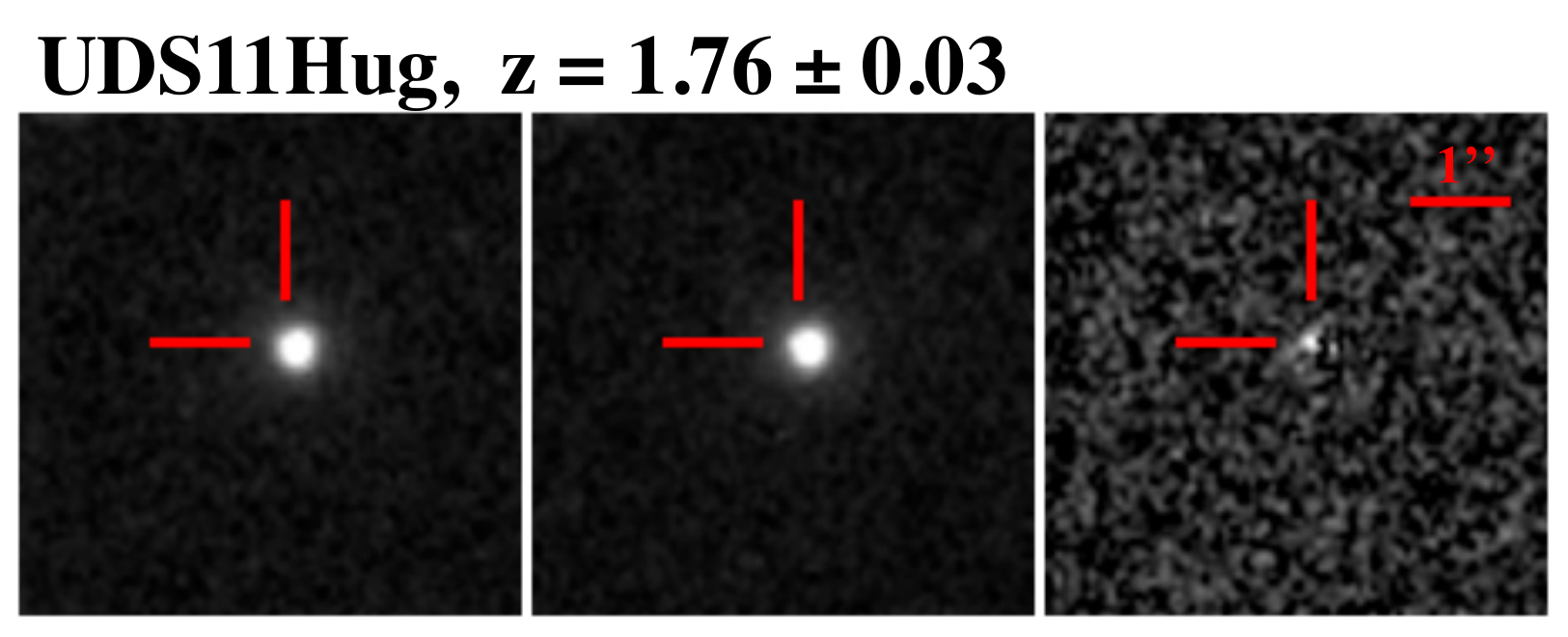}
  \includegraphics[width=0.333\textwidth]{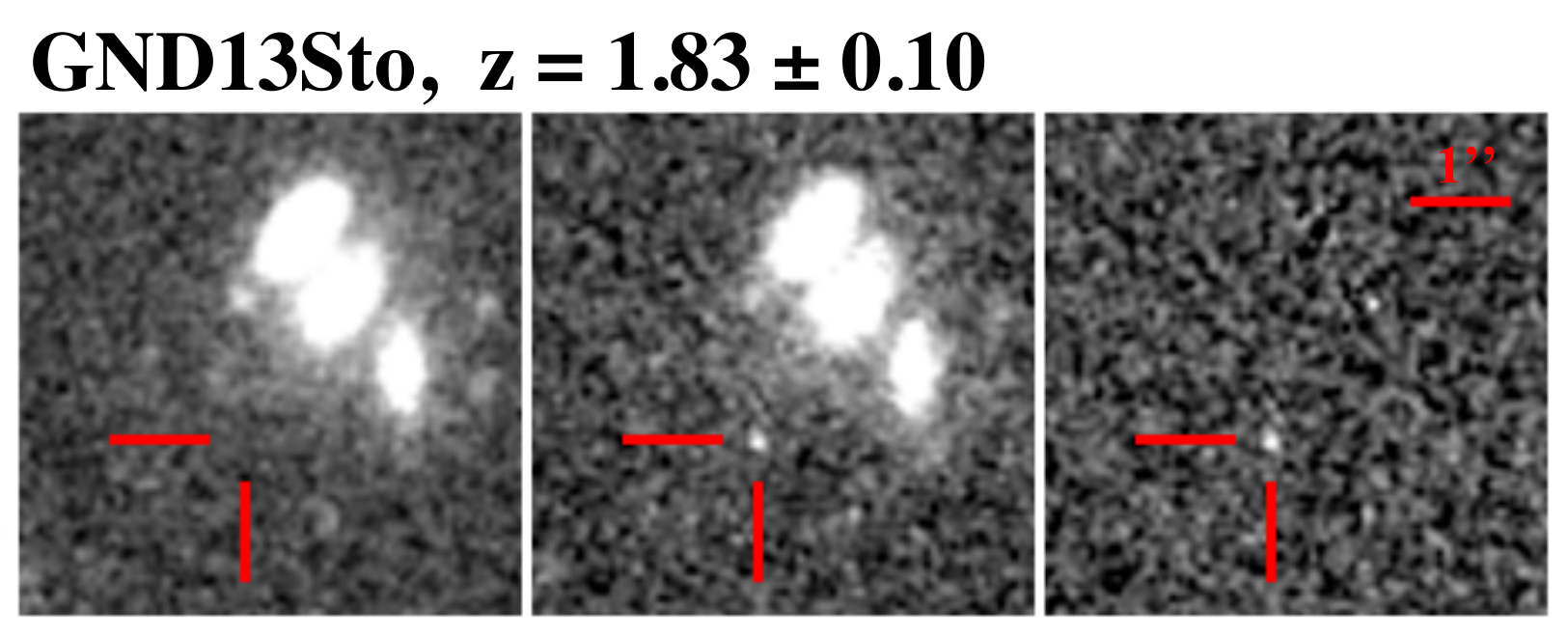}
  \includegraphics[width=0.333\textwidth]{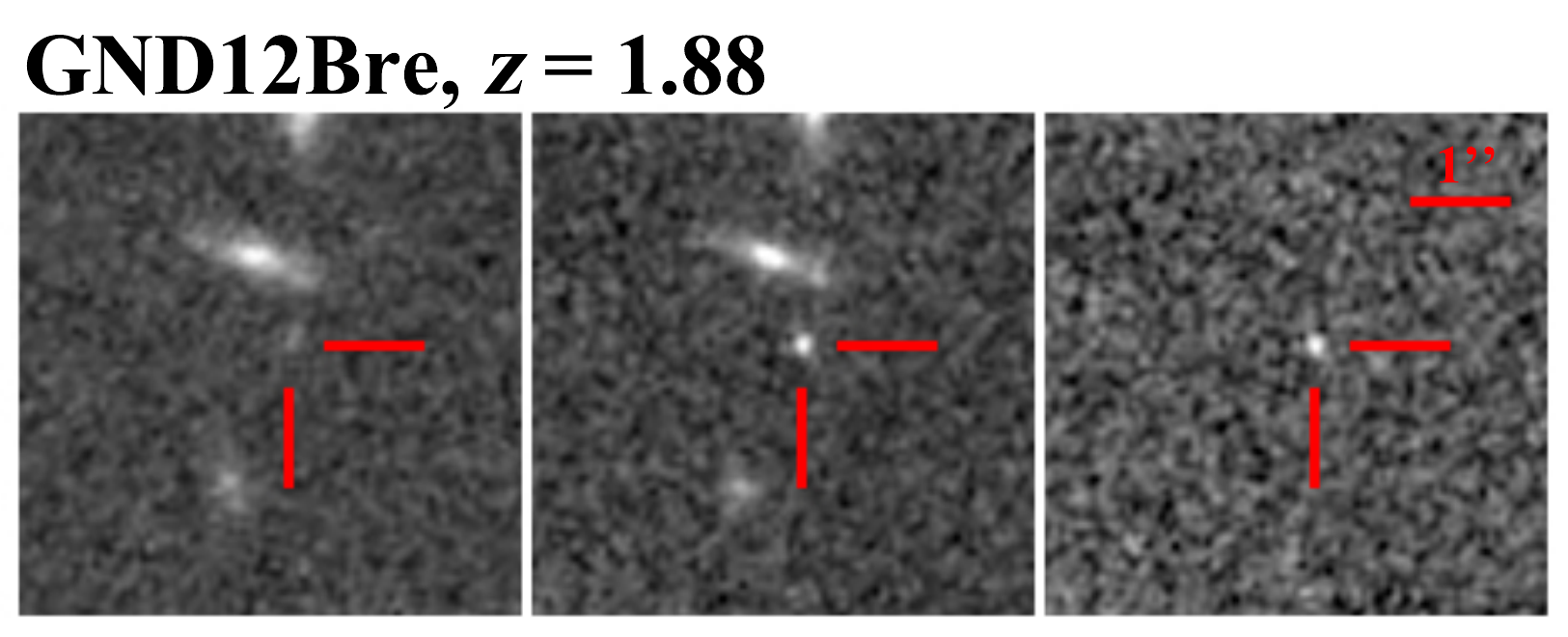}
  \includegraphics[width=0.333\textwidth]{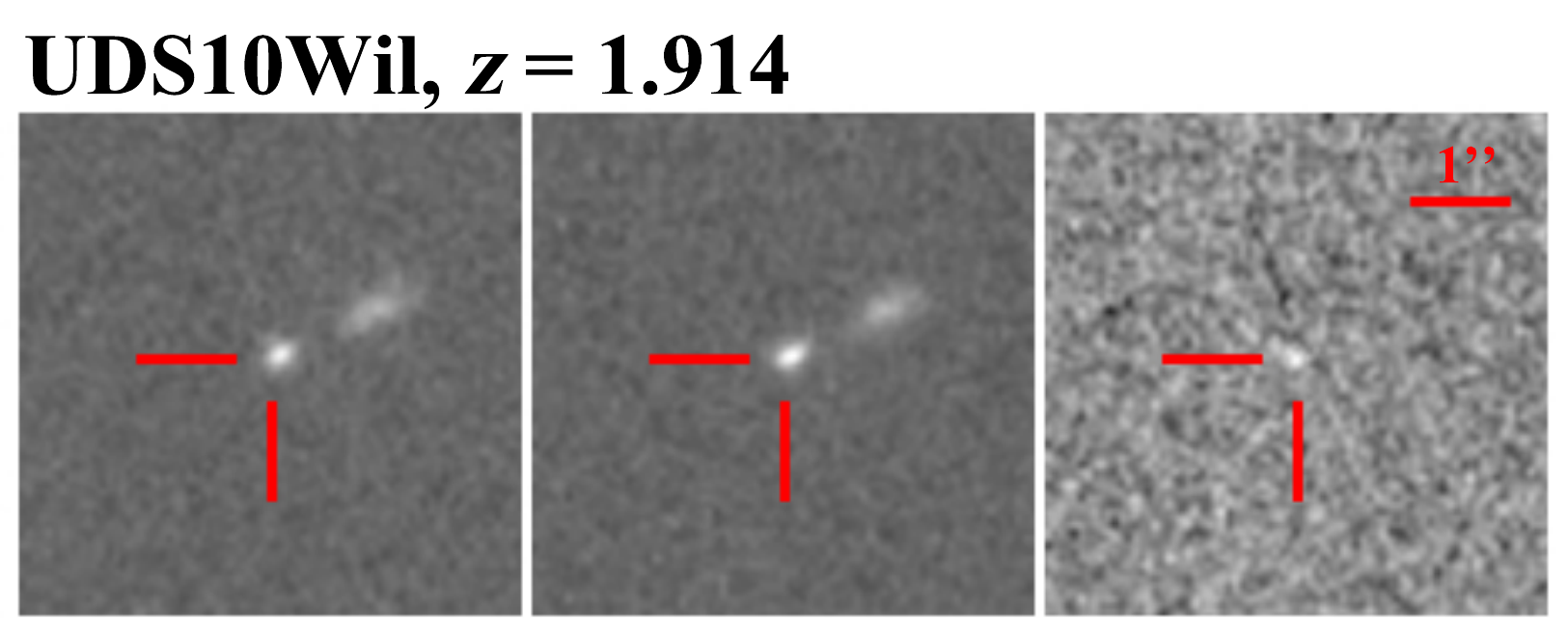}
  \includegraphics[width=0.333\textwidth]{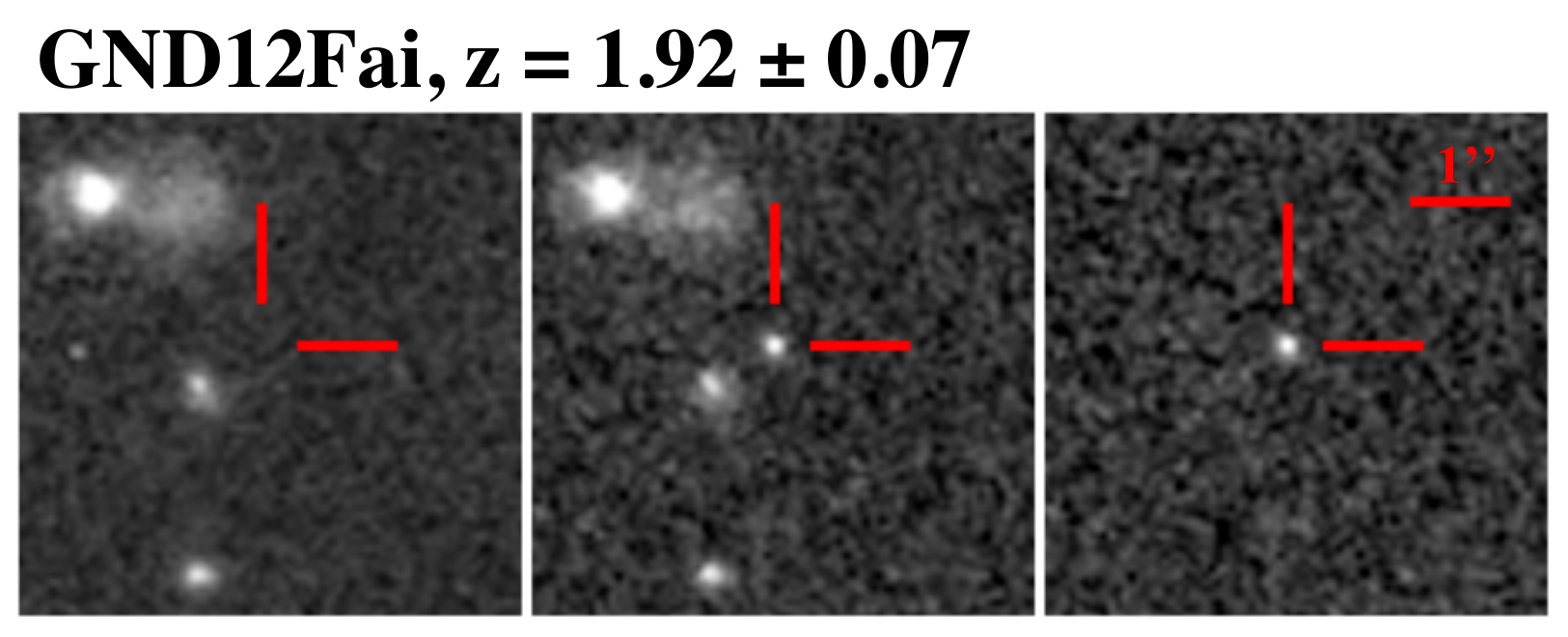}
  \includegraphics[width=0.333\textwidth]{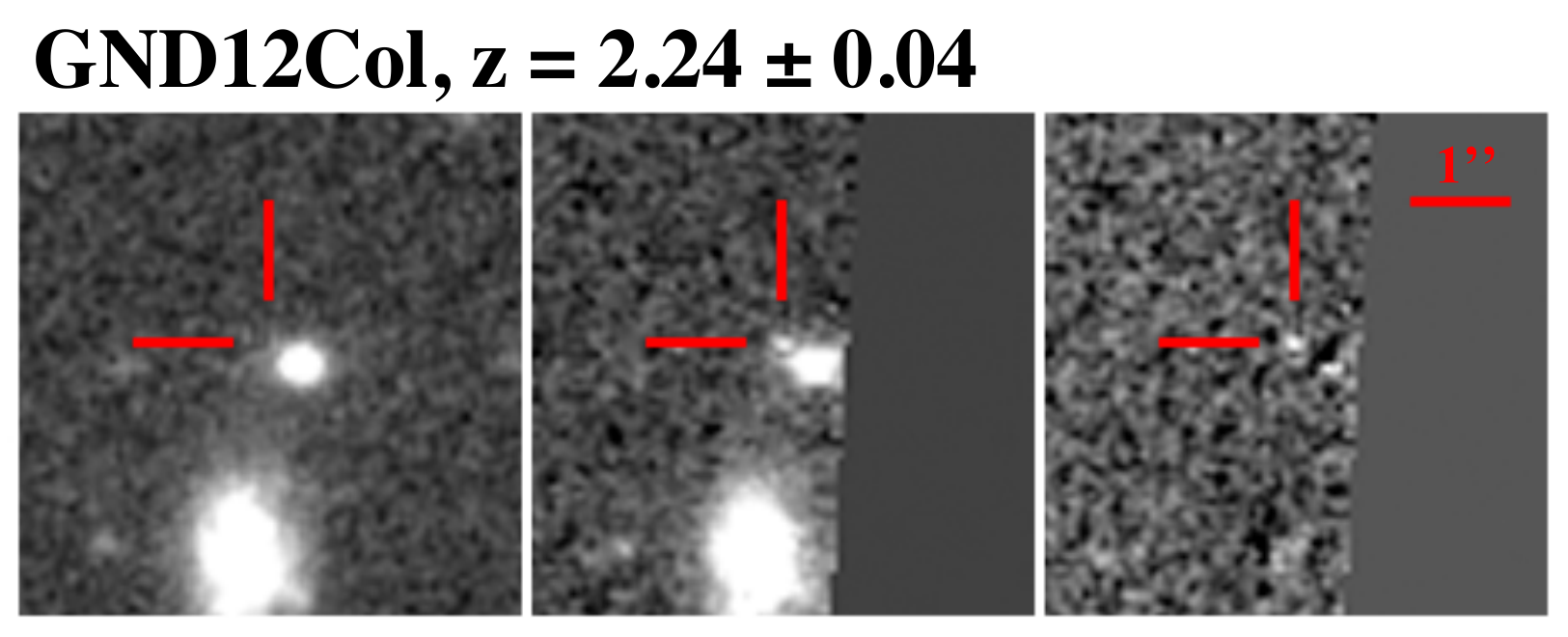}
  \includegraphics[width=0.333\textwidth]{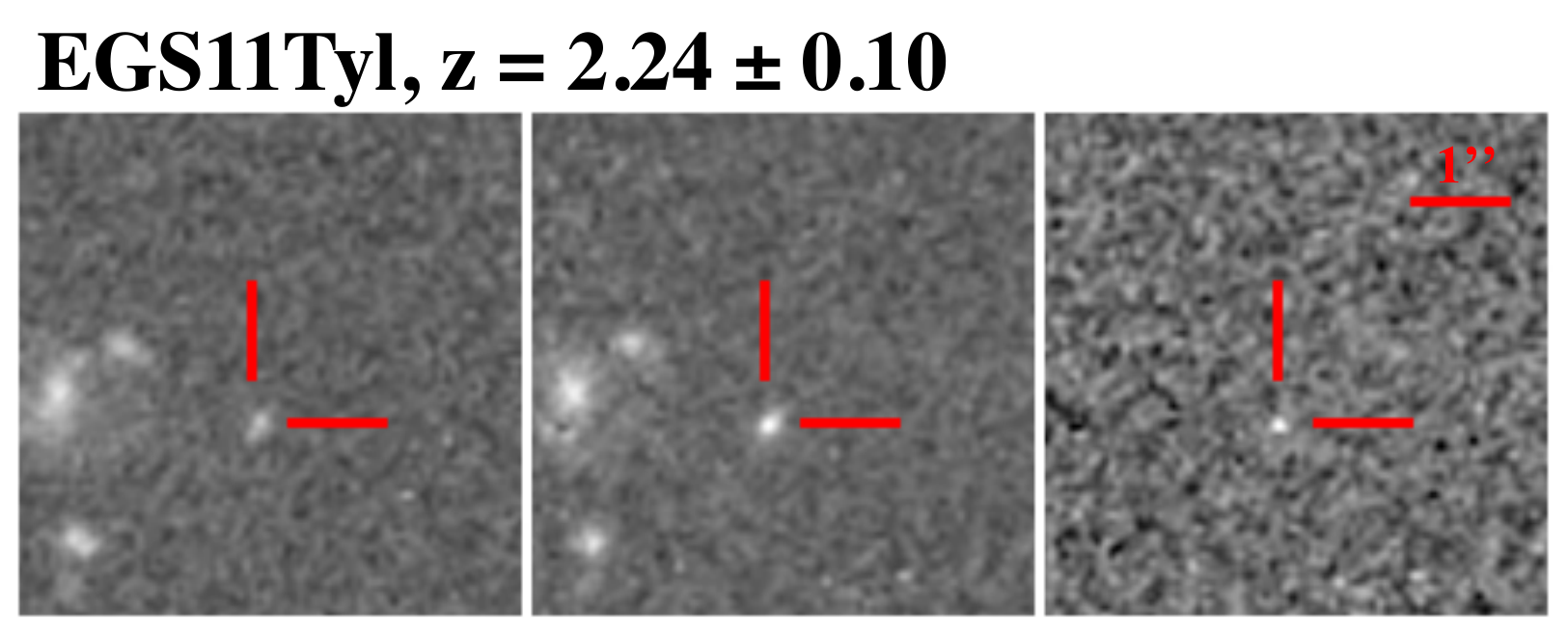}
  \includegraphics[width=0.333\textwidth]{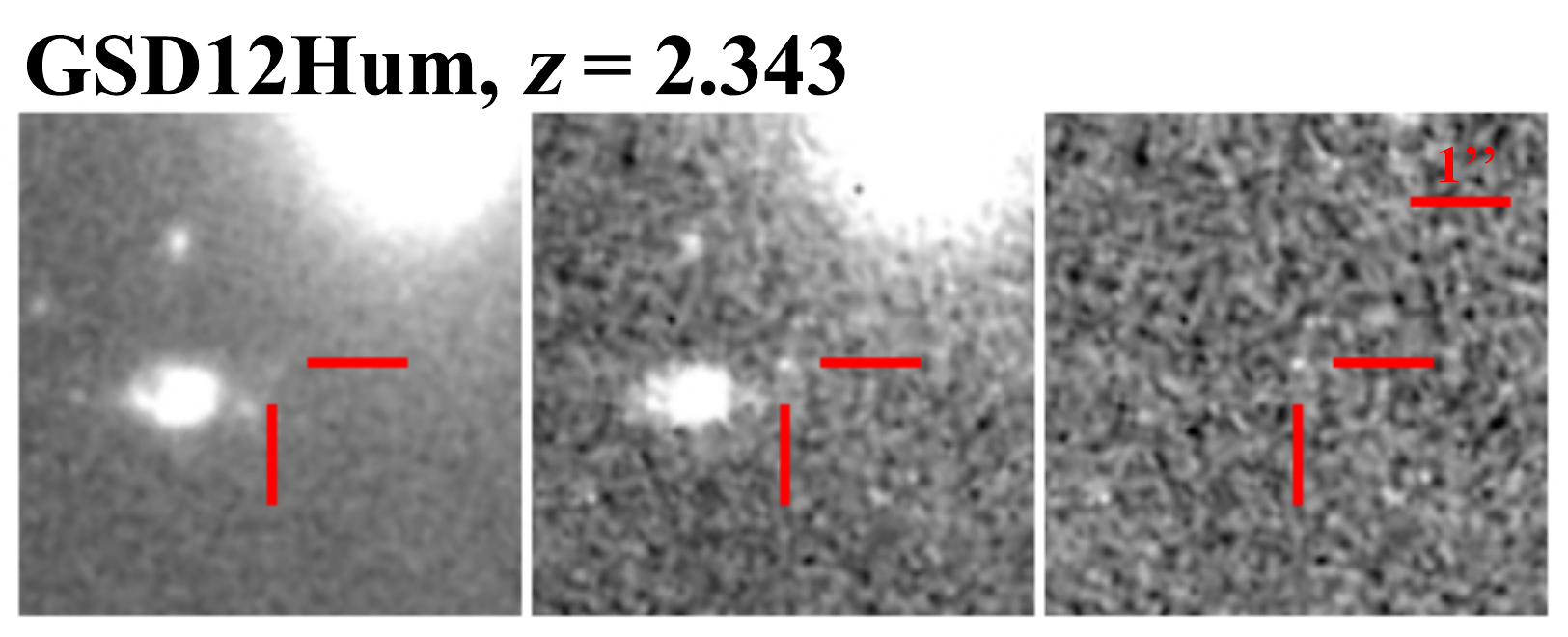}
  \includegraphics[width=0.333\textwidth]{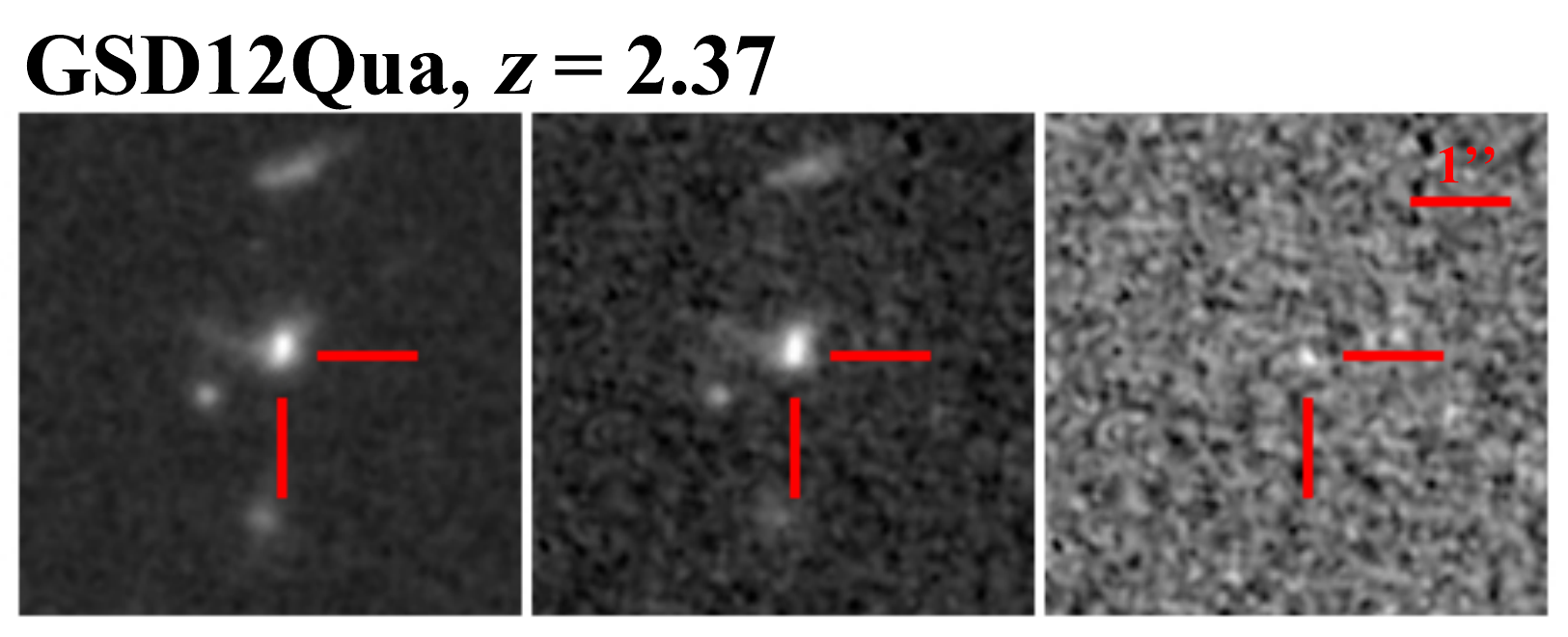}
\caption{ Detection images for 14 SN from the CANDELS fields with
  redshifts $z>1.5$.  Each image triplet shows H band (F160W) images
  with the template image on the left, the discovery epoch image in
  the middle and the difference image on the right. All images have a
  width of about 6 arcseconds, with North up and East to the left. The
  position of the SN is marked by (red) crosshairs in every
  frame. Discovery images for the other 51 SN with $z<1.5$ are
  provided in Appendix~\ref{app:lowz}. \label{fig:highzStamps} }
\end{figure*}

To find SN candidates in the CANDELS WFC3-IR difference images, we
used human searchers, who scanned each image by eye to detect
significant deviations from the noise.  We had $\sim$20 individuals
regularly engaged in searching the CANDELS data, and searching tasks
were assigned so that every WFC3-IR tile was examined by at least two
people.  Searchers recorded the position of all potential transient
object detections, and assigned a quality grade.  All transient
sources receiving a high- or moderate-quality grade were carefully
vetted to pare down to a list of transient sources that are very
likely real SN. The criteria for inclusion in this SN candidate list
are: (1) a profile consistent with a point source; (2) detected in the
J+H difference image and also individually in both the J and H bands,
(3) clean of evidence indicating that it could be a detector artifact
(neighboring bad pixels, on the detector edge, etc).  For the rates
analysis presented here, we also require that the object reached its
peak magnitude in IR bands after the \HST\ template images were
collected (i.e., we reject any SN that has less flux in the search
epoch than in the template epoch).  

Finally, we also discard a total of 6 objects that are positively
classified as AGN.  These 6 were located at the center of a host
galaxy that has observational indicators to classify it as an AGN
(x-ray emission, spectral line broadening, prior optical/IR
variability, etc.).  Our final sample contains 65 SN candidates that
meet these requirements. Table~\ref{tab:highzSN} lists the 14
SN at redshifts $z>1.5$, and Table~\ref{tab:lowzSN} in the Appendix
lists the remaining 51 SN. In keeping with the practice of
past \HST\ SN surveys, we assign each SN a unique 8-digit name that
indicates the field and the year of discovery, with the final 3
letters referencing our team's internal ``nickname'' for each
object.\footnote{The nicknames for the CANDELS SN are mostly derived
from U.S. Presidents and other prominent figures from U.S. history.}
Figure~\ref{fig:highzStamps} shows ``postage stamp images'' with the
detection images for the 14 SN at $z>1.5$, and the remainder
are in the Appendix, in Figures \ref{fig:lowzStamps1}
and \ref{fig:lowzStamps2}.

\subsection{Follow-up Observations} 

Upon discovery, every SN was evaluated for possible follow-up
observations with \HST\ or ground-based telescopes. First, a redshift
probability density function (pdf) was assigned, using pre-existing
spectroscopy of the host galaxy when available and a photometric
redshift (photo-z) when not.  The photo-z estimates were derived from
template fitting to the observed spectral energy distribution (SED) of
each SN host galaxy \citep{Dahlen:2013}.  Then a preliminary SN
classification (Ia or CC) was assigned by comparing the color and
magnitude of the observed SN against a sample of synthetic SN with
redshifts drawn from the best available redshift pdf. These synthetic
SN were generated with the SuperNova ANAlysis
software \citep[SNANA;][]{Kessler:2009a} (see
section~\ref{sec:Classification} for more details).

Any SN with a redshift $z>1$ and a color consistent with a \SNIa\ 
classification was then considered for possible follow-up observations
with \HST. Where necessary and whenever possible, the host galaxies of
these high-priority targets were quickly observed (within $\sim$1 week
of discovery) with ToO spectroscopic observations using ground-based
observatories (primarily Gemini, Keck, and the Very Large Telescope
(VLT)). The host galaxies of other SN candidates (\CCSN\ and those with
$z<1$) were targeted for later spectroscopic observations from the
ground to determine precise redshifts, all reported in
Table~\ref{tab:highzHosts} (and in the Appendix Table~\ref{tab:lowzHosts}). 

\makeatletter{}\renewcommand{\arraystretch}{1.8}
\begin{small}
\ifms
  \begin{deluxetable}{llllllllp{2.1in}}
\else
  \begin{deluxetable*}{llllllllp{2.1in}}
\fi
\tablecolumns{9}
 
\tablecaption{14 Supernovae with $z>1.5$ (see Appendix for the remainder) \label{tab:highzSN}}
\tablehead{\colhead{Name} & \colhead{R.A. (J2000)} & \colhead{Decl. (J2000)} & \colhead{P(Ia$|$D$_{z}$)\tablenotemark{a}} & \colhead{P(Ia$|$D$_{host}$)\tablenotemark{b}} & \colhead{z$_{\mbox{\scriptsize SN}}$\tablenotemark{c}} & \colhead{($\pm$)} & \colhead{$z$ Source\tablenotemark{d}}}
\startdata
COS12Car & 10:00:14.726 & $+$02:11:32.57 & 0.62 $^{+0.09}_{-0.36}$ & 0.80 $^{+0.00}_{-0.07}$ & 1.54  & (0.04)  & SN spec-z + SN phot-z \\
GSD10Pri & 03:32:38.010 & $-$27:46:39.08 & 1.00 $^{+0.00}_{-0.00}$ & 1.00 $^{+0.00}_{-0.00}$ & 1.545 & (0.001) & host+SN spec-z \\
EGS13Rut & 14:20:48.106 & $+$53:04:22.12 & 1.00 $^{+0.00}_{-0.00}$ & 1.00 $^{+0.00}_{-0.00}$ & 1.614 & (0.005) & host spec-z + SN phot-z \\
GND13War & 12:36:54.761 & $+$62:12:16.70 & 0.01 $^{+0.02}_{-0.01}$ & 0.01 $^{+0.00}_{-0.00}$ & 1.689 & (0.005) & host spec-z \\
GSD11Bus & 03:32:42.776 & $-$27:48:07.10 & 0.00 $^{+0.00}_{-0.00}$ & 0.00 $^{+0.00}_{-0.00}$ & 1.7   & (0.1)   & host+SN phot-z \\
UDS11Hug & 02:17:37.427 & $-$05:08:41.43 & 0.82 $^{+0.05}_{-0.21}$ & 1.00 $^{+0.00}_{-0.00}$ & 1.761 & (0.025) & host+SN phot-z \\
GND13Sto & 12:37:16.778 & $+$62:16:41.43 & 1.00 $^{+0.00}_{-0.00}$ & 1.00 $^{+0.00}_{-0.00}$ & 1.83  & (0.10)  & host+SN phot-z \\
GND12Bre & 12:36:55.520 & $+$62:13:58.82 & 0.00 $^{+0.00}_{-0.00}$ & 0.00 $^{+0.00}_{-0.00}$ & 1.880 & (0.001) & host spec-z \\
UDS10Wil & 02:17:46.336 & $-$05:15:24.00 & 1.00 $^{+0.00}_{-0.00}$ & 1.00 $^{+0.00}_{-0.00}$ & 1.914 & (0.001) & host+SN spec-z \\
GND12Fai & 12:36:15.822 & $+$62:15:56.50 & 0.00 $^{+0.00}_{-0.00}$ & 0.00 $^{+0.00}_{-0.00}$ & 1.92  & (0.07) & host+SN phot-z \\
GND12Col & 12:36:37.569 & $+$62:18:32.93 & 1.00 $^{+0.00}_{-0.01}$ & 1.00 $^{+0.00}_{-0.00}$ & 2.24  & (0.04) & host+SN phot-z \\
EGS11Tyl & 14:20:12.944 & $+$52:57:10.60 & 0.24 $^{+0.13}_{-0.15}$ & 0.57 $^{+0.03}_{-0.04}$ & 2.244 & (0.095) & host+SN phot-z \\
GSD12Hum & 03:32:15.500 & $-$27:50:50.02 & 0.00 $^{+0.00}_{-0.00}$ & 0.00 $^{+0.00}_{-0.00}$ & 2.343 & (0.001) & host spec-z \\
GSD12Qua & 03:32:11.723 & $-$27:49:11.72 & 0.00 $^{+0.00}_{-0.00}$ & 0.00 $^{+0.00}_{-0.00}$ & 2.370 & (0.001) & host spec-z \\
\enddata

\tablenotetext{a}{ Type Ia SN classification probability from STARDUST, 
  using the redshift-dependent class prior. Uncertainties reflect systematic
  biases due to the class prior and extinction assumptions
  (Sections~\ref{sec:ClassPrior} and
  \ref{sec:HostAvDistribution}). }
\tablenotetext{b}{ Type Ia SN classification probability from STARDUST, 
  using the {\em galsnid} host galaxy prior. Uncertainties reflect systematic
  biases due to the class prior and extinction assumptions.}
\tablenotetext{c}{ Posterior redshift and uncertainty, as determined 
  by the STARDUST light curve fit.}
\tablenotetext{d}{ The {\it host / SN} values indicate whether the redshift is derived from the host galaxy, the SN itself, or a combination;  {\it spec-z / phot-z} specify a spectroscopic or photometric redshift. A value of {\it host+SN phot-z} means the redshift is derived from a STARDUST light curve fit, with the host galaxy phot-z used as a prior.}

\ifms  \end{deluxetable}
\else  \end{deluxetable*}
\fi
\end{small}
\renewcommand{\arraystretch}{1.0}

\makeatletter{}\renewcommand{\arraystretch}{1.8}
\begin{small}
\ifms
  \begin{deluxetable}{llllrcclll}
\else
  \begin{deluxetable*}{llllrcclll}
\fi
\tablecolumns{10}
 
\tablecaption{Host galaxies of 14 Supernovae with $z>1.5$ (see Appendix \ref{app:lowz} for the remainder)\label{tab:highzHosts}}
\tablehead{\colhead{SN} & \colhead{R.A. (J2000)} & \colhead{Decl. (J2000)} & \colhead{d[\arcsec]} & \colhead{d[kpc]\tablenotemark{a}} & \colhead{Morph.\tablenotemark{b}} & \colhead{SED\tablenotemark{c}} & \colhead{z$_{\mbox{host}}$} & \colhead{($\pm$)} & \colhead{$z$ Reference\tablenotemark{d}}}
\startdata
COS12Car & \nodata & \nodata & \nodata & \nodata & \nodata & \nodata & \nodata & \nodata & \nodata \\
GSD10Pri & 03:32:37.991 & $-$27:46:38.69 & 0.46 & 9.6 & i & SB & 1.545 & 0.001 & \citet{Frederiksen:2012} \\
EGS13Rut & 14:20:48.113 & $+$53:04:22.07 & 0.08 & 1.7 & d & A & 1.614 & 0.005 & HST+WFC3 (A.Riess) \\
GND13War & 12:36:54.787 & $+$62:12:16.60 & 0.21 & 4.3 & di & SB & 1.689 & 0.005 & HST+WFC3 (B.Weiner) \\
GSD11Bus & 03:32:42.776 & $-$27:48:07.10 & 0.00 & 0.0 & u & A & 1.76 & 0.53 & phot-z (T.Dahlen) \\
UDS11Hug & 02:17:37.415 & $-$05:08:41.53 & 0.21 & 4.2 & s & P & 1.82 & 0.13 & phot-z (T.Dahlen) \\
GND13Sto & 12:37:16.823 & $+$62:16:42.65 & 1.26 & 25.5 & u & A & 1.8 & 1.2 & phot-z (T.Dahlen) \\
GND12Bre & 12:36:55.520 & $+$62:13:58.79 & 0.03 & 0.6 & i & SB & 1.880 & 0.005 & Keck+MOSFIRE (J. Trump) \\
UDS10Wil & 02:17:46.332 & $-$05:15:23.90 & 0.12 & 2.4 & s & SB & 1.914 & 0.001 & \citet{Jones:2013} \\
GND12Fai & 12:36:15.934 & $+$62:15:55.91 & 0.98 & 19.9 & sd & SB & 1.77 & 0.25 & phot-z (T.Dahlen) \\
GND12Col & 12:36:37.514 & $+$62:18:32.66 & 0.47 & 9.6 & s & A & 2.1 & 0.2 & phot-z (T.Dahlen) \\
EGS11Tyl & 14:20:12.938 & $+$52:57:10.62 & 0.06 & 1.2 & sd & SB & 1.95 & 0.45 & phot-z (T.Dahlen) \\
GSD12Hum & 03:32:15.585 & $-$27:50:50.43 & 1.20 & 24.3 & di & SB & 2.343 & 0.001 & \citet{Balestra:2010} \\
GSD12Qua & 03:32:11.713 & $-$27:49:11.29 & 0.45 & 9.3 & di & SB & 2.370 & 0.001 & VLT+Xshooter (J.Hjorth) \\
\enddata

\tablenotetext{a}{ Physical separation between the SN and center of the host, 
  computed from the measured angular separation in the preceding column, 
  assuming a flat \LCDM\ cosmology with \Ho=70, \Om=0.3}
\tablenotetext{b}{ Visual classifications for host galaxy morphology: 
   s = spheroid, d = disk, i = irregular}
\tablenotetext{c}{ Template-matching classification of host galaxy SED:
   P = Passive, A = Active, SB = Starburst type}
\tablenotetext{d}{ Unpublished spectroscopic observations are given as {\em
    Observatory}+{\em Instrument} ({\em name of PI}). Host galaxy
  photometric redshifts are marked as {\em phot-z} (Dahlen et al. in
  prep).}

\ifms  \end{deluxetable}
\else  \end{deluxetable*}
\fi
\end{small}
\renewcommand{\arraystretch}{1.0}

Some of the most promising candidates for classification as \SNIa\ at
$z>1.5$ were selected for supplementary imaging and/or grism
spectroscopy with \HST. Two of these, SN \primo\ and
\wilson, have been presented
elsewhere \citep{Rodney:2012,Jones:2013}. Due to the high cost of
grism observations (at least 10 \HST\ orbits are required to reach
sufficient S/N in distant SN), we applied strict criteria for
selecting grism targets: (1) best available redshift $z>1$, preferably
$z>1.5$; (2) observed SN colors consistent with a (possibly
reddened) \SNIa\ at that redshift; (3) observed SN magnitudes within
$\sim$1.5 mag of a \SNIa\ at that redshift (i.e., using a very weak
prior around a standard \LCDM\ cosmology); (4) SN position allows for
a grism observation without severe contamination.

Without a slit to isolate the SN light in WFC3-IR grism spectroscopy,
a high-$z$ \SNIa\ candidate can most productively be observed if the
trace of the SN spectrum can be positioned to avoid contamination from
nearby galaxies. Thus, to satisfy the final criterion (4), the
candidate must be well separated from the core of its host, or located
in a host that is faint relative to the SN.  We also require an
orientation angle that avoids contamination of the SN spectral trace
from the 0th order and 1st order light of other nearby stars and
galaxies.  Of course, this orientation must also be accessible to \HST\
at the time of observation, with suitable guide stars in range.  In
practice, these criteria were satisfied for only 6 SN candidates.  The
results of those observations are described in
Section~\ref{sec:GrismSpectroscopy}.  Another 37 CANDELS SN were
followed with ToO imaging observations.  These imaging targets
included \SNIa\ candidates that satisfied some or all of the first
three criteria, but were not suitable for grism observations, as well
as some likely \CCSN\ that we were able to include in the same field
of view as those primary targets.

\section{Detection Efficiency}\label{sec:DetectionEfficiency}

Translating SN detections into a SN rate measurement requires
characterization of the survey detection efficiency, i.e., the fraction
of SN that are detected by our human searchers.  This recovery
fraction is most strongly influenced by the S/N of the object in the
WFC3-IR difference images. The SN host galaxy is also an important
factor affecting SN detectability, as we discuss further in
Section~\ref{sec:CoreResiduals}.

\ifgrayscale 
  \insertfig{FIG/figDetEff_gray}{\label{fig:DetEff}
  SN detection efficiency measurements as a function of magnitude in the
  ``J+H'' band, taken as an average of the measured F125W and F160W
  magnitudes. Each point represents the fraction of fake SN recovered by
  human searchers, with error bars indicating the standard deviation of
  the efficiency, computed using a Bayesian
  formalism \citep{Paterno:2004}. The best fit model is shown as a solid
  (green) line, with best-fit parameters listed in the lower left. For
  reference, the equivalent best-fit curves for the J and H bands
  individually are shown in dotted and dashed lines,
  respectively. The horizontal and vertical lines mark the 50\%
  efficiency point for J+H detections: $m_{50}=25.4$ mag.  The top axis
  marks the approximate redshift of a normal \SNIa\ with average
  extinction (\Av=0.3) that would reach a peak brightness matching the
  J+H magnitude on the bottom axis.  
  (A color version of this figure appears in the online edition.)
}
\else
  \insertfig{FIG/figDetEff}{\label{fig:DetEff}
  SN detection efficiency measurements as a function of magnitude in the
  ``J+H'' band, taken as an average of the measured F125W and F160W
  magnitudes. Each point represents the fraction of fake SN recovered by
  human searchers, with error bars indicating the standard deviation of
  the efficiency, computed using a Bayesian
  formalism \citep{Paterno:2004}. The best fit model is shown as a solid
  (green) line, with best-fit parameters listed in the lower left. For
  reference, the equivalent best-fit curves for the J and H bands
  individually are shown in blue dotted and red dashed lines,
  respectively. The horizontal and vertical lines mark the 50\%
  efficiency point for J+H detections: $m_{50}=25.4$ mag.  The top axis
  marks the approximate redshift of a normal \SNIa\ with average
  extinction (\Av=0.3) that would reach a peak brightness matching the
  J+H magnitude on the bottom axis.  }
\fi

To measure our SN detection efficiency and explore the associated
systematic biases, we generated a catalog of 2,000 fake SN. The
catalog was drawn from a SNANA Monte Carlo simulation, such that the
F160W magnitudes fill out a uniform distribution covering the range
$21<m_H<28$, and the $J-H$ colors were appropriate for Type Ia and
Core Collapse SN in the redshift range $0.1<z<2.8$.  Each fake SN was
then assigned to a ``host galaxy'' drawn from catalogs of extended
sources in the CANDELS fields. The separation from the host-galaxy
center for each fake SN was then selected randomly from a normal
distribution centered on 0 with a standard deviation of $2\times
R_{50}$, where $R_{50}$ is the radius of an aperture containing 50\%
of the host-galaxy flux. This ensures that the fake SN very roughly
follow the distribution of host light \citep{Kelly:2008}.

With magnitudes, colors and positions defined, we generated synthetic
PSFs for each fake SN using TinyTim, and planted them in the {\tt FLT}
images, as described in section~\ref{sec:Pipeline}.  As the searchers
reviewed each difference image, they were unaware of the number,
brightness, and location of the fake SN, so they recorded fake SN
detections alongside detections of real SN.  After completing each
search, the fake SN detections (and non-detections) were used to
calculate the recovery fraction.

Figure \ref{fig:DetEff} shows the measured detection
efficiency as a function of the ``J+H'' magnitude: the average
of the F125W and F160W magnitudes.  We fit the efficiency
measurements with a functional form similar to that used
by \citet{Sharon:2007}, but we use only a single parameter to
characterize the exponential turnover, and we allow for the peak
efficiency to plateau at a value less than unity: 

\begin{equation}\label{eqn:DetEff}
  \eta_{\rm det}( m ) =  \eta_0 \times \left(1 + \exp\left(\frac{m-m_{50}}{\tau}\right)\right)^{-1},
\end{equation}

\noindent where $m$ is the apparent J+H magnitude, $\eta_0$ is the 
maximum efficiency, $m_{50}$ is the magnitude at which the efficiency
curve passes through the 50\% line, and $\tau$ characterizes the
exponential roll-off.  The best-fit curve shown in
Figure~\ref{fig:DetEff} has $m_{50}$=25.4, $\tau$=0.23, and
$\eta_0$=0.98.

\subsection{Missing SN in Galaxy Cores}\label{sec:CoreResiduals}

One concern for systematic bias entering into these detection
efficiency measurements is the possibility that many SN are obscured
by difference imaging artifacts in the cores of bright galaxies.  The
shot noise in these bright pixels is naturally higher than in the
outskirts, as photon counts are elevated in both the search epoch and
the template.  Additionally, minor cross-epoch registration errors can
result in some residual flux in galaxy cores. In the CANDELS survey
data these effects are both exacerbated by the under-sampled PSF of
our single-epoch WFC3-IR images, as we have only two dithers per
filter.

\insertfig{FIG/figCoreResiduals}{\label{fig:CoreResiduals}
Fraction of CANDELS galaxies showing IR core residuals as a function
of redshift, as determined from visual inspection of difference images
generated by the SN data processing pipeline. 
}

As shown in Figure~\ref{fig:DetEff}, we have measured our maximum
detection efficiency $\eta_0$ to be less than unity even for very
bright SN, due to the fake SN that happen to land in the noisy cores
of bright galaxies. Our SN rate measurements will therefore naturally
account for a small fraction of SN that are missed in this manner.
However, this built-in correction is only valid if the distribution of
positions for the fake SN $-$ relative to their host-galaxy cores $-$
is closely matched to the {\em true} distribution of the \SNIa\
population.  Furthermore, it requires that the galaxies chosen for
``hosting'' our fake SN are themselves representative of the
population of \SNIa\ hosts.  Our fake SN procedures were designed to
meet these requirements at low and intermediate redshifts, but this
does not necessarily carry over into the new high-$z$ regime.

To evaluate whether this effect might be introducing a strong bias at
high-$z$, we visually inspected the CANDELS IR difference images and
identified all galaxies that exhibited strong residuals.  For each
galaxy we tabulated the spectroscopic redshift or the best available
photo-$z$ from CANDELS catalogs.  Comparing this redshift distribution
for core residuals against the count of all galaxies as a function of
redshift gives us a measure of the fraction of (detected) galaxies
that might obscure SN in their bright cores.  As shown in
Figure~\ref{fig:CoreResiduals}, the fraction is less than $\sim10\%$
for all redshifts above 0.01, and less than $\sim2\%$ for $z>1$ --
consistent with the value of $\eta_0$ measured from fake SN.  This
result suggests that any systematic bias from galaxy core residuals is
very minor.  Therefore, in the rates calculation we do not include any
bias correction, and we do not add any contribution to the systematic
uncertainty budget.

\section{Classification}\label{sec:Classification}

To reach the final classification probabilities listed in
Table~\ref{tab:highzSN} (and Table~\ref{tab:lowzSN}), we used a
Bayesian analysis of the observed multi-color light
curves.  This photometric classification approach was used for
our full sample of 65 SN, supplemented by spectroscopic evidence for 6
objects, as described in Section~\ref{sec:GrismSpectroscopy}.  An
early version of this classifier was introduced
in \citet{Jones:2013} with the presentation of SN \wilson.  Here we
will again briefly describe the classification procedure, emphasizing
recent changes.

\subsection{The STARDUST Classifier}

Our photometric classification approach uses SNANA to generate
simulations of \SNIa\ and \CCSN\ light curves.  The \SNIa\
simulations use the SALT2 model \citep{Guy:2010}, 
which has free parameters for the date of peak
(MJD$_{pk}$), redshift ($z$), shape ($x_1$) and color ($\mathcal{C}$).
The simulated \CCSN\ are drawn from the SNANA library of 42 \CCSN\
light curve templates (26 Type II and 16 Type Ib/c). These templates
are derived from the SN samples of the Sloan Digital Sky
Survey \citep{Frieman:2008,Sako:2008,DAndrea:2010}, Supernova
Legacy Survey \citep{Astier:2006}, and Carnegie Supernova
Project \citep{Hamuy:2006,Stritzinger:2009,Morrell:2012}.
Each \CCSN\ template defines the underlying shape and color of the
synthetic light curves, which is then modified with free parameters
for the date of peak, redshift, host extinction (\Av), and luminosity
($\Delta$m, the shift in magnitudes relative to the peak of the
assumed luminosity function).  For this work, 
we fix the SALT2 model
parameters $\alpha=0.135$ and $\beta=4.1$ \citep{Scolnic:2013a}, and 
for all simulated \CCSN we fix the extinction
law to \Rv=3.1.

\makeatletter{}\begin{figure*}
  \includegraphics[width=0.333\textwidth]{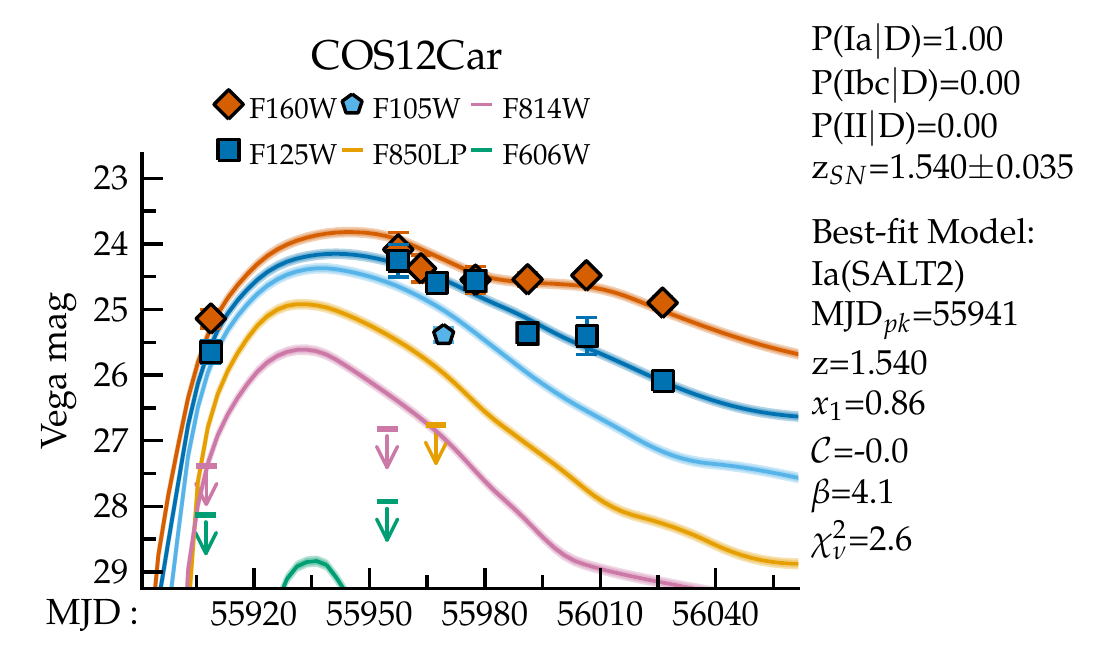}
  \includegraphics[width=0.333\textwidth]{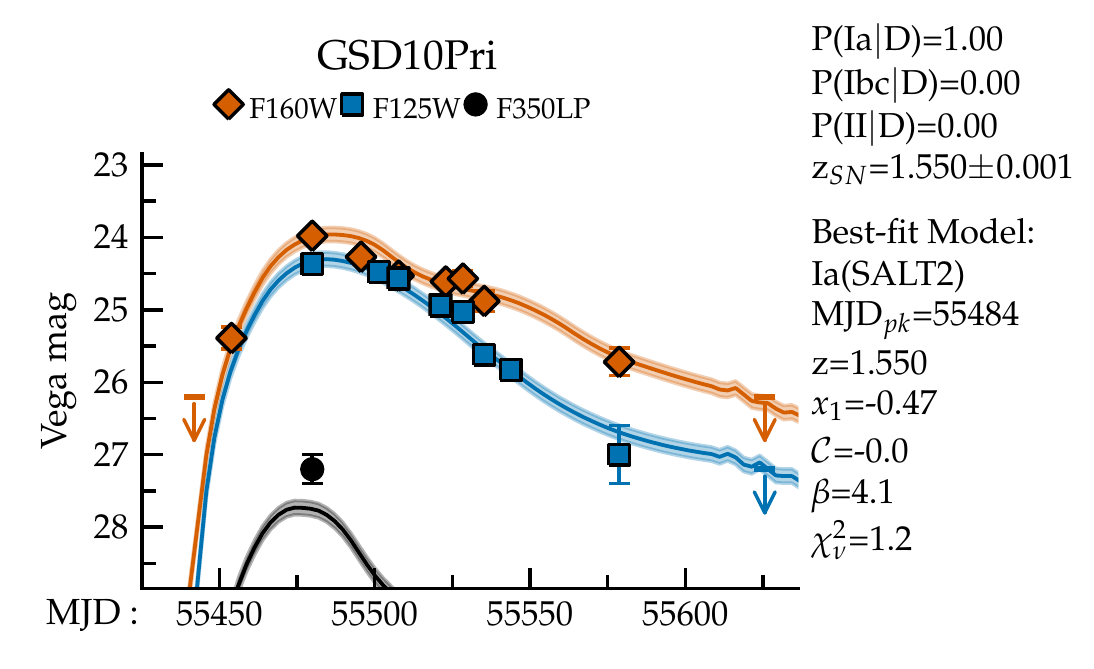}
  \includegraphics[width=0.333\textwidth]{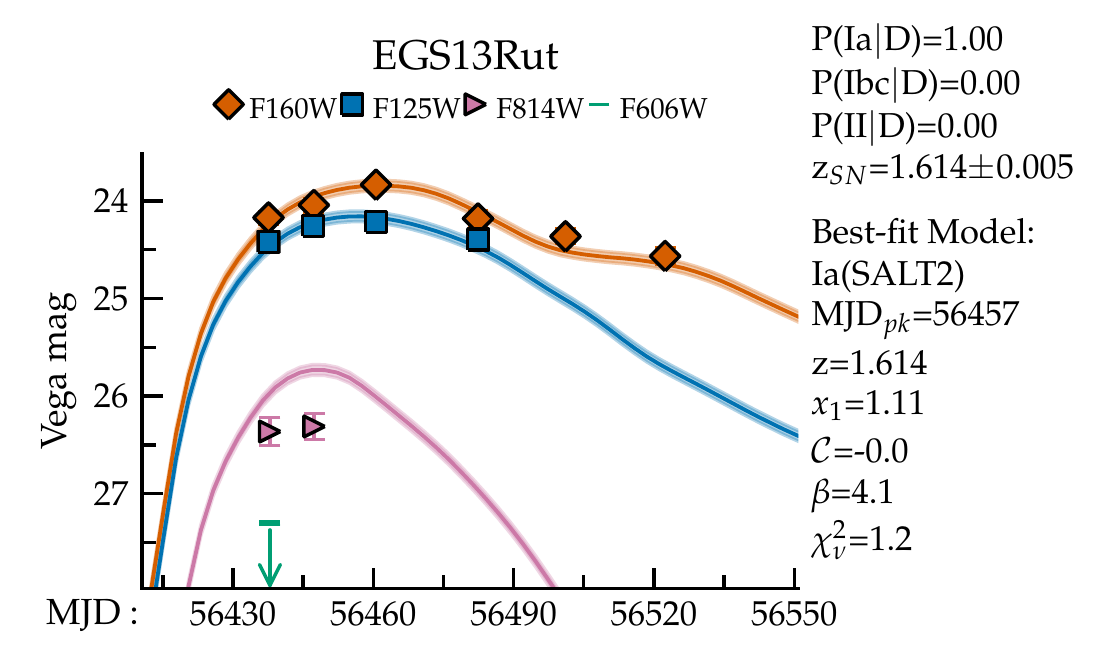}
  \includegraphics[width=0.333\textwidth]{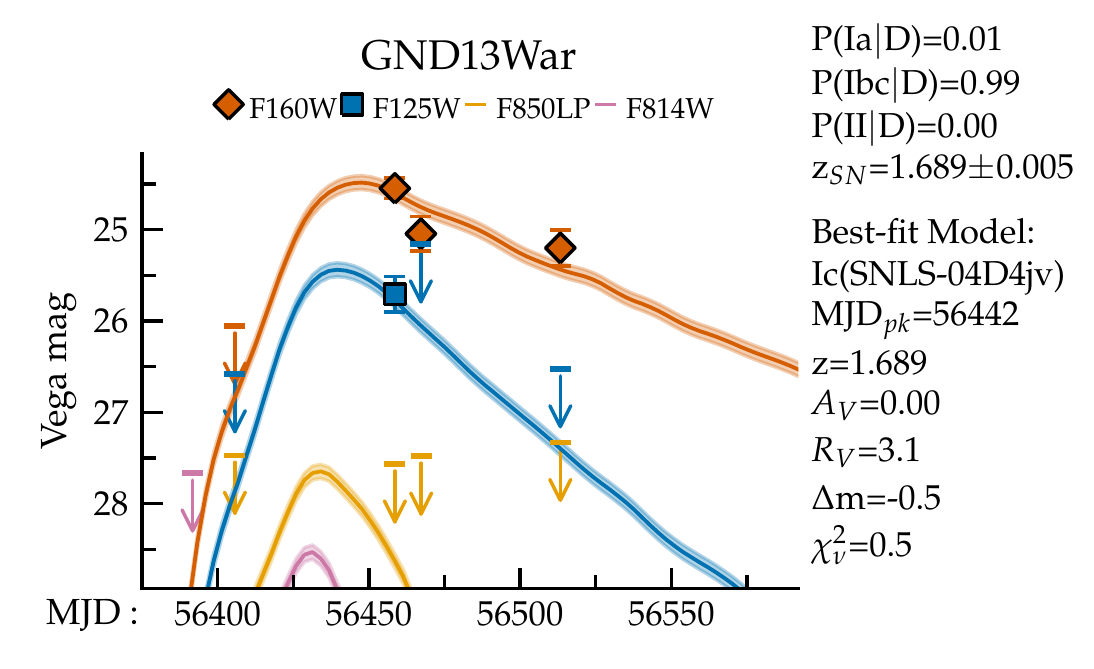}
  \includegraphics[width=0.333\textwidth]{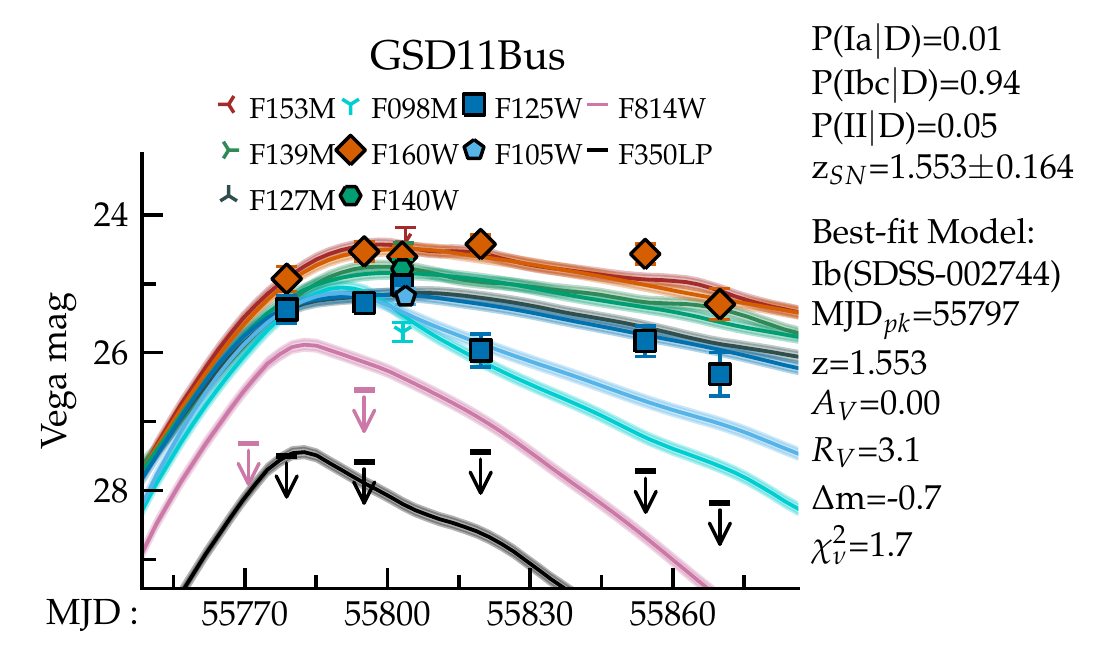}
  \includegraphics[width=0.333\textwidth]{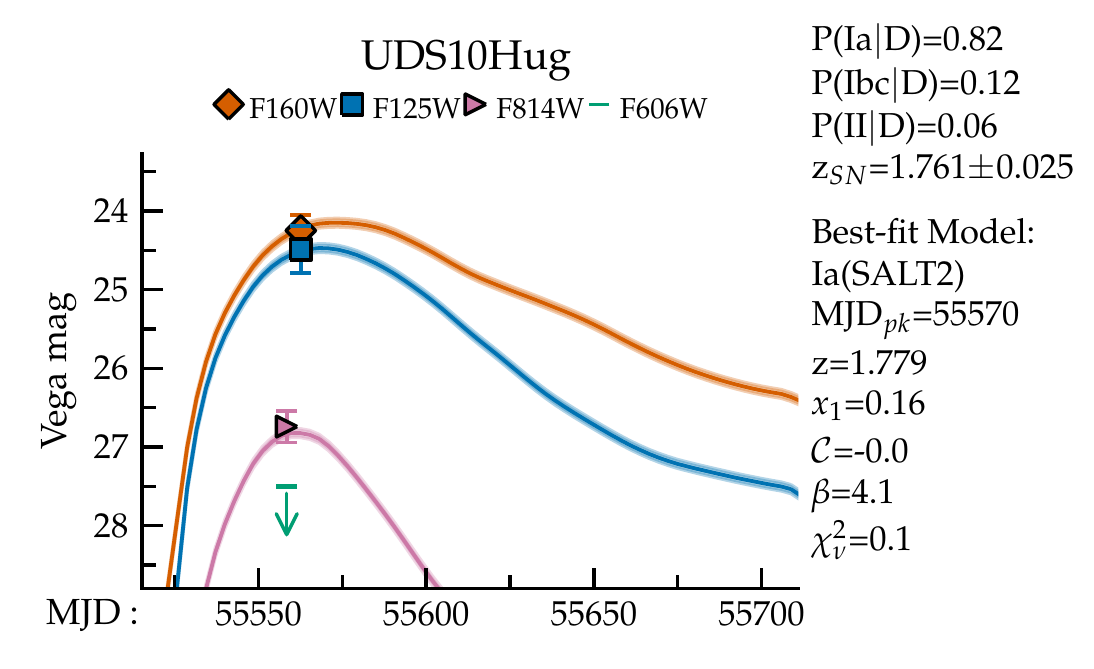}
  \includegraphics[width=0.333\textwidth]{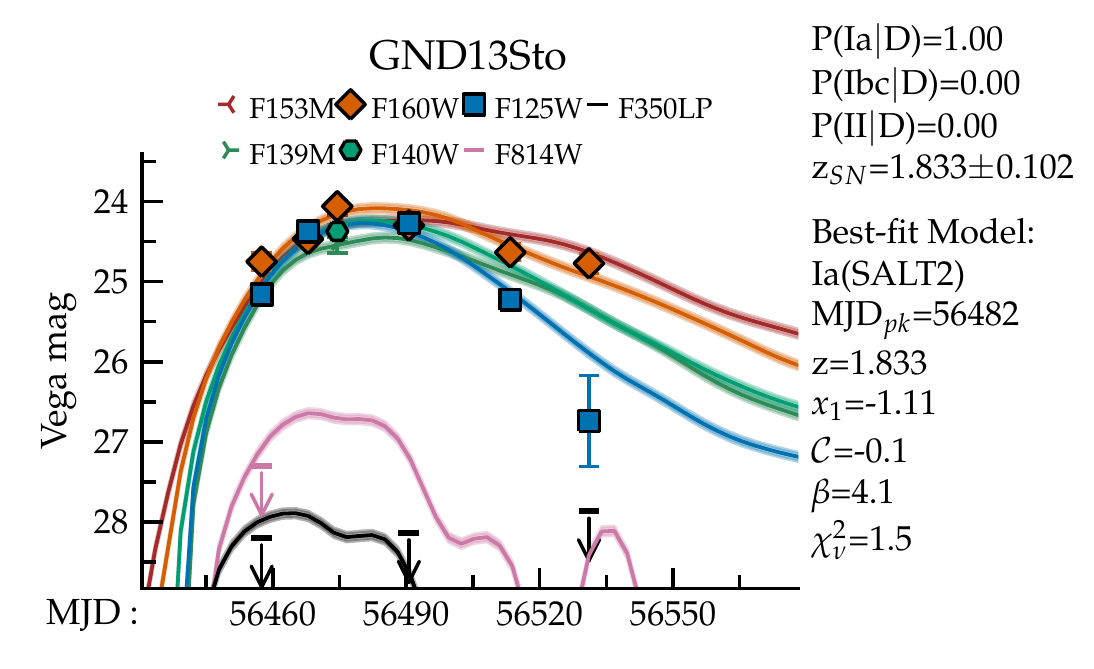}
  \includegraphics[width=0.333\textwidth]{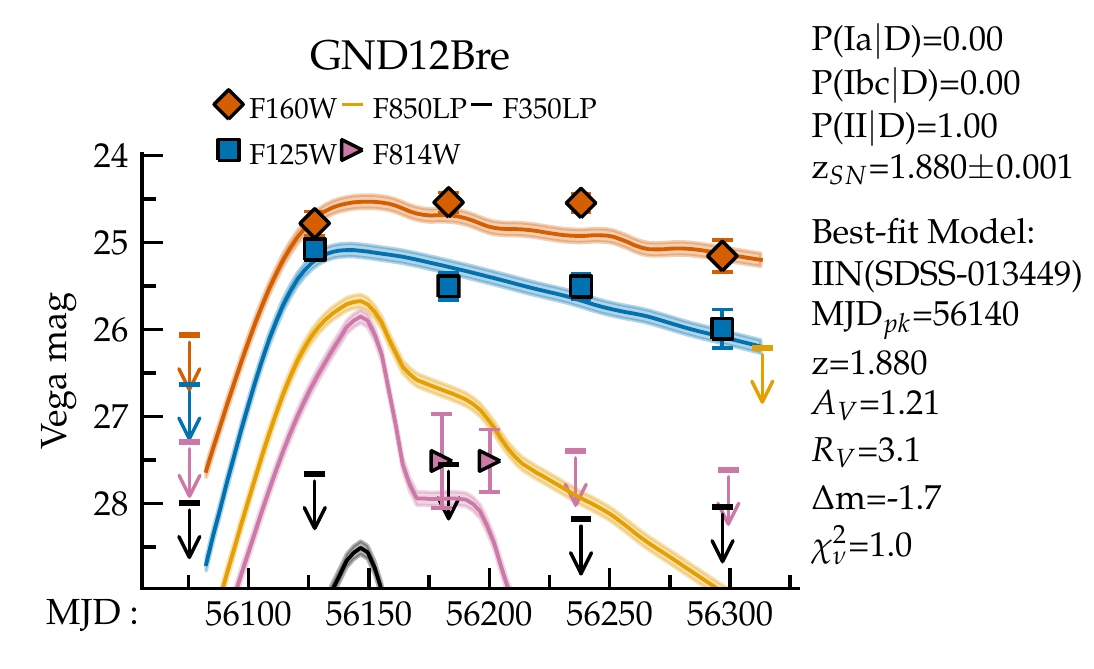}
  \includegraphics[width=0.333\textwidth]{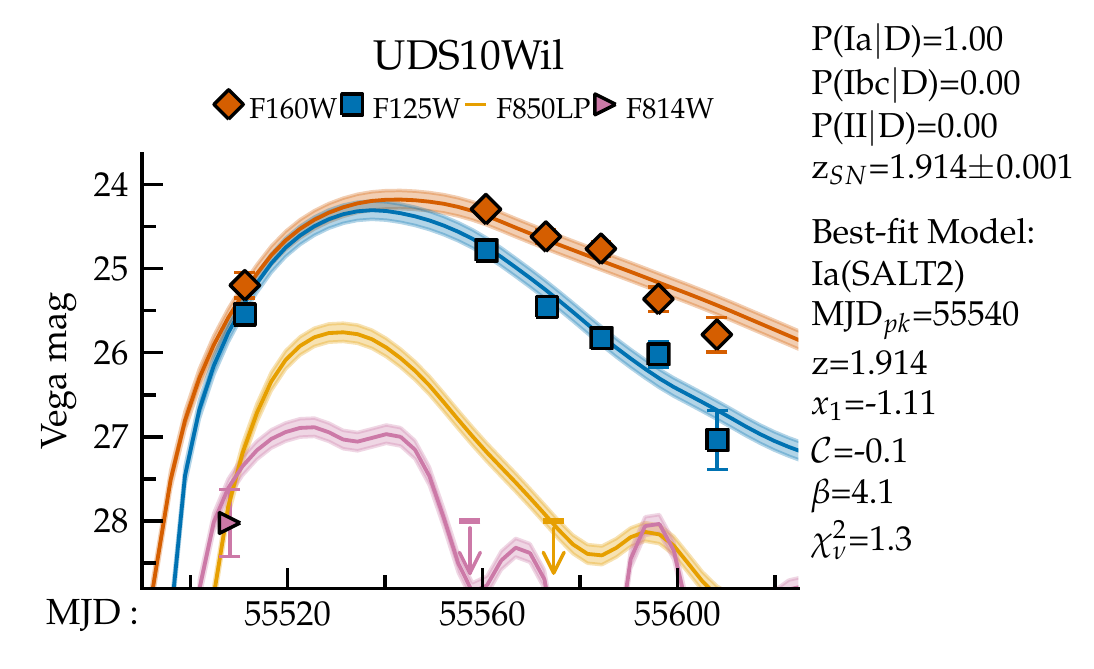}
  \includegraphics[width=0.333\textwidth]{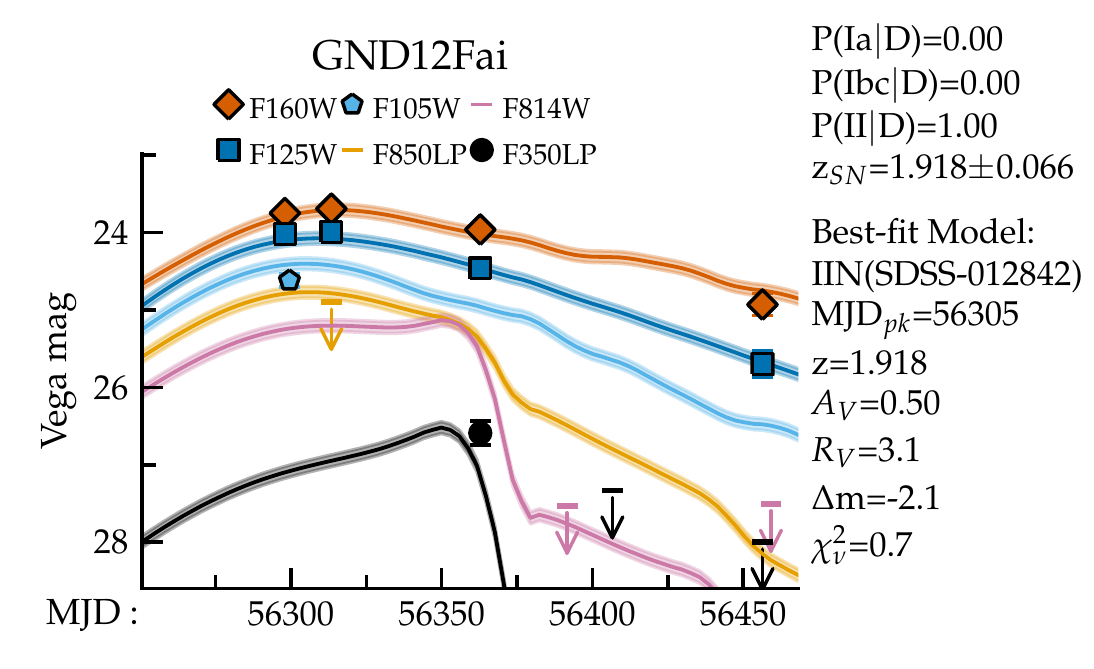}
  \includegraphics[width=0.333\textwidth]{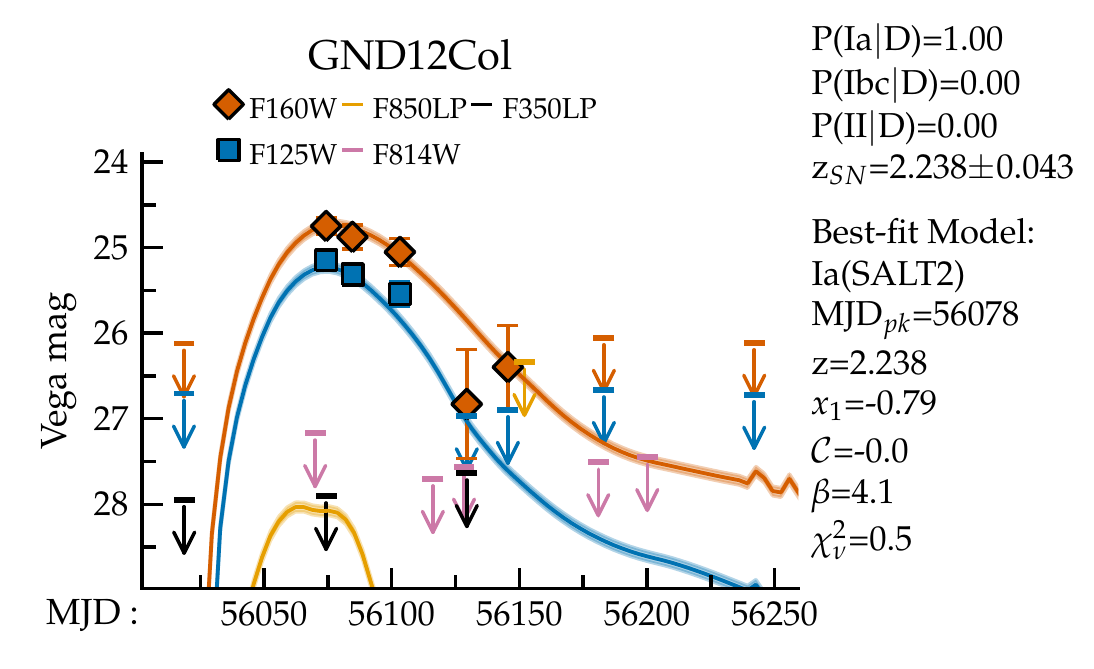}
  \includegraphics[width=0.333\textwidth]{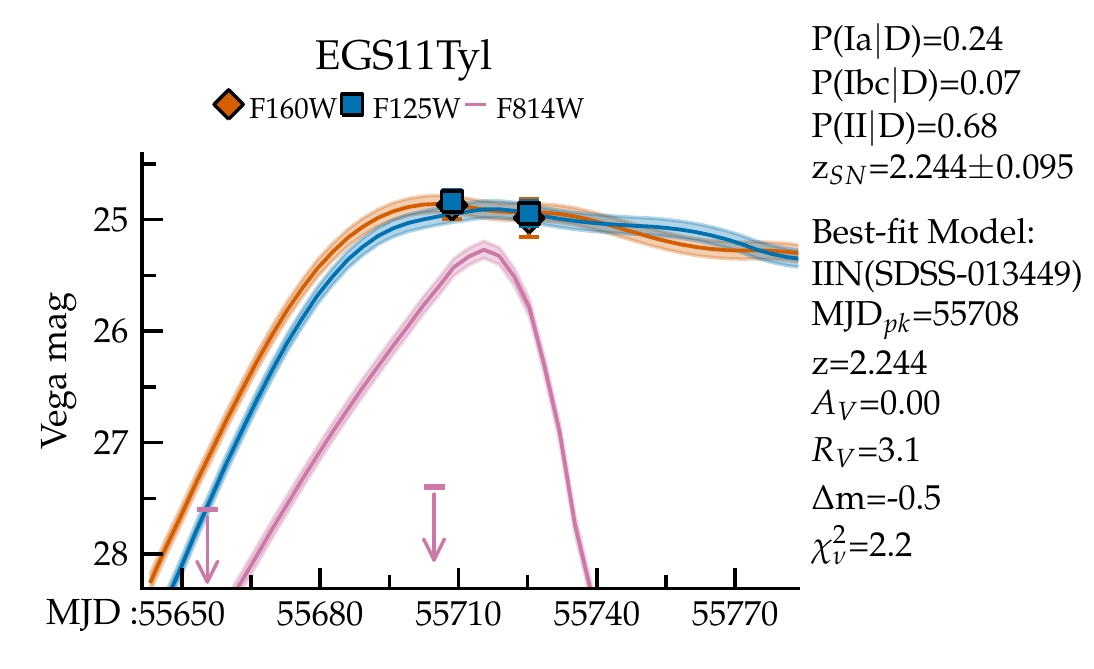}
  \includegraphics[width=0.333\textwidth]{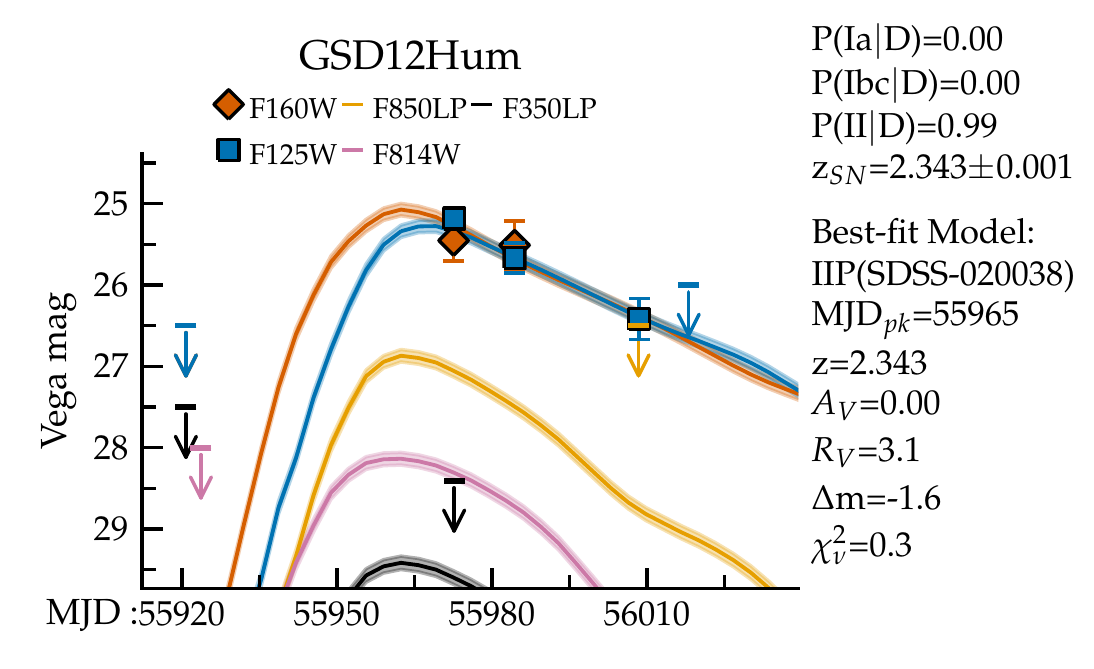}
  \includegraphics[width=0.333\textwidth]{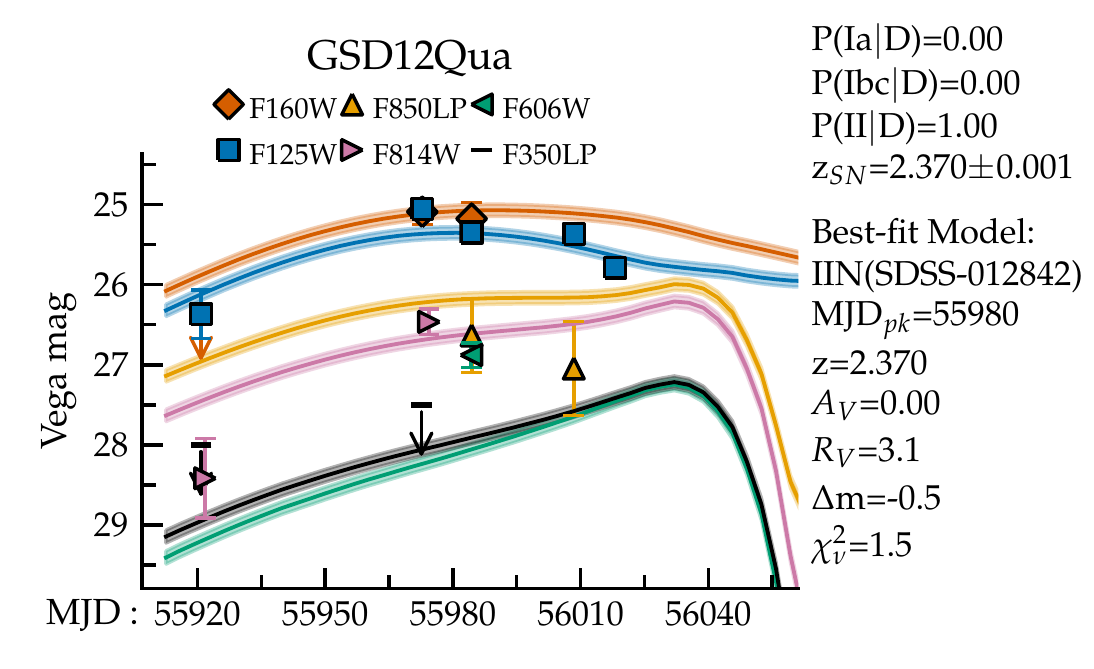}
\caption{ Light curves of 14 SN with $z>1.5$. Each panel
    shows the observed photometry (in Vega mags) from CANDELS imaging
    with filled points and arrows for 3-$\sigma$ upper limits.  Error
    bars are typically less than the size of the points. Curves depict
    the maximum probability light curve fit for the most probable SN
    class, as determined by STARDUST.  Classification probablities and
    redshift as determined by STARDUST are listed on the right side of
    each panel, along with the parameters of the single model that
    delivers the highest posterior probability.  Light curves for SN
  at $z<1.5$ are shown in Appendix~\ref{app:lowz}.
  \label{fig:highzLCfit} }
\end{figure*}

Comparing these synthetic SN to the observed light curves, we compute
a likelihood using the $\chi^2$ statistic: $P({\bf
D_{\rm LC}}|{\bf \theta},Ia) \propto
\exp(-\chi^2/2)$, where the vector ${\bf D_{\rm LC}}$ is the observed SN light
curve and the vector ${\bf \theta}$ gives the parameter values for
each realization of the SNANA models.  We then apply priors for each
model parameter (see \citet{Graur:2014} for a detailed description of
these priors) as well as a redshift-dependent prior for the fraction
of SN that are Type Ia: P(Ia,z) (see Section 3 \ref{sec:ClassPrior}
below). Finally, we derive the total posterior probability that each
object is a \SNIa, $P(Ia|{\bf D_{\rm LC}})$, by marginalizing over the
nuisance parameters, ${\bf \theta}$ and applying Bayes theorem :

\begin{equation}\label{eqn:STARDUST}
P(Ia|{\bf D_{\rm LC}}) = k_{\rm LC}^{-1} \int_{\bf \theta} P(Ia,z) P({\bf
D_{\rm LC}}|{\bf \theta},Ia) d{\bf \theta}.
\end{equation}

\noindent The normalization factor $k_{\rm LC}$ is defined by requiring
that that posterior probabilities for all three primary SN classes
(Ia,Ib/c,II) sum to unity. The custom-built software package that
executes this procedure is named STARDUST: {\it Supernova Taxonomy And
Redshift Determination Using SNANA Templates}.  The STARDUST code will
be presented in full and publicly released in a subsequent paper
(Rodney et al., in prep).  

There are two notable differences between the STARDUST classification
procedure applied here and that described in \citet{Jones:2013}.
First, in this work we do not use a free parameter for flux
scaling,\footnote{The $A$ in Equation 1 of \citealt{Jones:2013}.}  so
the absolute values of the simulated SN fluxes are defined by the SN
luminosity functions and cosmology (\Om=0.3,\OL=0.7,$w$=-1) that are
assumed in the SNANA simulations. To allow for some
uncertainty in this baseline cosmology (or equivalently, introducing
some increased scatter in the assumed SN luminosity functions) we
include a non-zero model uncertainty term in the $\chi^2$
calculation.\footnote{$\sigma_{\rm sim}^2$ in \citealt{Jones:2013}} This
is fixed at 8\% of the simulated flux for \SNIa\ models and 10\% for
all \CCSN\ models. Secondly, when the SN in question does not have a
precise redshift from host-galaxy spectroscopy, we use the host
galaxy's photometric redshift probability distribution (photo-z pdf)
as the redshift prior.

\smallskip

Column 4 of Table~\ref{tab:highzSN} (and Table \ref{tab:lowzSN} in the
Appendix) presents the final SN classification probabilities, which
will be used in Section~\ref{sec:Rates} for the \SNIa\ rate
calculation. Figure~\ref{fig:highzLCfit}
(and \ref{fig:lowzLCfit1}-\ref{fig:lowzLCfit3}) shows the maximum
likelihood light curve fit for each SN, along with the associated
best-fit model parameters. As described below, the systematic
uncertainties associated with each classification probability are
determined by varying two key priors that are not tightly constrained
by observations: the assumed fraction of SN that are of Type Ia and
the distribution of host-galaxy extinction.

\ifgrayscale
  \insertfig{FIG/figRatePriors_gray}{\label{fig:RatePriors}
  Deriving the redshift-dependent prior for SN class fractions. The top
  panel shows \SNIa\ rates and the middle panel shows \CCSN\ rates. Both
  have observed rates plotted as open symbols. In the Ia case these are
  average values from all non-redundant field SN surveys. The \CCSN\
  rate points are the collection from \citet{Dahlen:2012}. In each panel
  the overlaid lines show three versions of a simple empirical
  model for the SN rates, and in the CC panel the thick dashed line traces
  the cosmic star-formation history (see text for details).  The bottom
  panel plots the fraction of all SN explosions that are of Type Ia,
  derived from pairs of curves drawn from the top two panels and
  anchored to $25\pm5\%$ at z=0 \citep{Smartt:2009,Li:2011a}.  These
  relative rate assumptions provide the high-, mid- and low-rate priors
  that are used to derive classification probabilities and associated
  uncertainties for observed SN.  
  (A color version of this figure appears in the online edition.)
}
\else
  \insertfig{FIG/figRatePriors}{\label{fig:RatePriors}
  Deriving the redshift-dependent prior for SN class fractions. The top
  panel shows \SNIa\ rates and the middle panel shows \CCSN\ rates. Both
  have observed rates plotted as open symbols. In the Ia case these are
  average values from all non-redundant field SN surveys. The \CCSN\
  rate points are the collection from \citet{Dahlen:2012}. In each panel
  the overlaid solid lines show three versions of a simple empirical
  model for the SN rates, and in the CC panel the magenta line traces
  the cosmic star-formation history (see text for details).  The bottom
  panel plots the fraction of all SN explosions that are of Type Ia,
  derived from pairs of curves drawn from the top two panels and
  anchored to $25\pm5\%$ at z=0 \citep{Smartt:2009,Li:2011a}.  These
  relative rate assumptions provide the high-, mid- and low-rate priors
  that are used to derive classification probabilities and associated
  uncertainties for observed SN.  }
\fi

\subsection{The Class Prior}\label{sec:ClassPrior}

As with any Bayesian classification approach, the STARDUST classifier
requires an input prior that quantifies the expectation that
any given SN is of Type Ia -- before applying any information from the
SN light curve.  We first assume that our sample is composed entirely
of ``normal'' SN, meaning that we assume no contamination from any
other transient sources.  This is a fairly safe assumption: AGN and
variable stars are excluded by our discovery requirements,
under-luminous SN like the .Ia \citep{Bildsten:2007} or Iax
SN \citep{Foley:2013b} are well below our detection threshold, and
super-luminous SN \citep{Gal-Yam:2012} have an intrinsic rate that is
lower than that of normal SN by a factor of about
10$^4$ \citep{Quimby:2011}.

We then define a redshift-dependent class prior P(Ia,$z$) as the
fraction of all normal SN at any given redshift $z$ that are Type
Ia. Figure~\ref{fig:RatePriors} shows the models used to define this
prior and the associated systematic uncertainty.  The baseline model
(green curve) is anchored at z=0 by the measured Ia
fraction \citep{Smartt:2009,Li:2011a}, and then evolves at higher
redshifts by following simple rate functions that match measured SN
rates and theoretical expectations.

A more complete statistical treatment would define a large number of
plausible models for the fraction of SN that are \SNIa, assigning each
an appropriate weight based on current observations, and then
marginalize over those many discrete priors to get a posterior
probability that is not uniquely guided by the single choice of a
baseline model.  That approach is computationally expensive and will
require further refinement of the STARDUST classifier. For this work,
we have chosen to treat this choice of prior as a component of our
systematic uncertainty budget.  We take the baseline prior described
above as our {\em mid-rate} model and then define two more models,
labeled the {\em high-} and {\em low-}rate priors.  These two
respectively maximize and minimize the fraction of SN that are assumed
to be of Type Ia at any given redshift, and are shown in
Figure~\ref{fig:RatePriors}.  These bounding models offer a
conservative estimate of the systematic uncertainty, because they are
at the extreme limit of plausibility (if either were correct it would
imply that the constraints from past rate \SNIa\ measurements were all
systematically wrong by more than $2\sigma$).

One might be concerned about the apparent circularity of using a
redshift-dependent P(Ia,z) prior based on measured \SNIa\ rates in the
service of a new \SNIa\ rate measurement.  However, the bounding
assumptions for our classification prior should ensure that our
systematic uncertainty estimates account for this.  To test that
assertion, in Appendix~\ref{app:galsnid} we evaluate an alternative
prior that is based on the SN host galaxies and does not evolve with
redshift.  Tables~\ref{tab:highzSN} and \ref{tab:lowzSN} record the
resulting STARDUST classifications using this modified prior
as P(Ia$|$D$_{host}$) in column 5.  Table~\ref{tab:galsnidRates} in the
Appendix reports the final effect of this prior switch on the observed
count of \SNIa\ and the volumetric rates.

\ifgrayscale
  \insertfig{FIG/figDustModels_gray}{\label{fig:DustModels}
  Prior probability distributions for the SN host-galaxy extinction, as
  used in the STARDUST classifier code.  The top panel shows the three
  priors applied to the \SNIa\ models, and the bottom panel shows the
  equivalent priors used for \CCSN\ models. In both cases the {\em
  high-dust} model is shown as a dashed line, the {\em mid-dust}
  model as a solid line, and the {\em low-dust} model as a
  dash-dot line.  Each model is composed of a Gaussian and/or an
  exponential function (see text for details) and the parameters for
  those components are listed in the legends.  Each curve is labeled
  with its expectation value $\langle A_{\rm V} \rangle$, giving the
  ``weighted average'' of the host-galaxy extinction for that model.
  (A color version of this figure appears in the online edition.)
  }
\else 
  \insertfig{FIG/figDustModels}{\label{fig:DustModels}
  Prior probability distributions for the SN host-galaxy extinction, as
  used in the STARDUST classifier code.  The top panel shows the three
  priors applied to the \SNIa\ models, and the bottom panel shows the
  equivalent priors used for \CCSN\ models. In both cases the {\em
  high-dust} model is shown as a red dashed line, the {\em mid-dust}
  model as a green solid line, and the {\em low-dust} model as a
  blue dash-dot line.  Each model is composed of a Gaussian and/or an
  exponential function (see text for details) and the parameters for
  those components are listed in the legends.  Each curve is labeled
  with its expectation value $\langle A_{\rm V} \rangle$, giving the
  ``weighted average'' of the host-galaxy extinction for that model.
  }
\fi

\subsection{Host \Av\ Distribution}\label{sec:HostAvDistribution}

Another prior that can strongly affect the final classification
probabilities is the assumed distribution of
host-galaxy extinctions, $P(\Av)$. 
As with the \SNIa\ fraction prior, we employ a baseline assumption (our
{\em mid-dust} model) and two bounding assumptions ({\em high-dust} and
{\em low-dust}) to constrain the possible systematic bias.  

In keeping with observations, our dust models assume that the \CCSN\
population suffers from significantly more dust extinction than
the \SNIa\ population, at all
redshifts \citep[e.g.,][]{Smartt:2009,Drout:2011,Kiewe:2012,Mattila:2012}.
Our three dust models are generated from the positive half of a
Gaussian distribution centered at $\Av=0$ with dispersion $\sigma$,
plus an exponential distribution of the form $e^{-A_{\rm V}/\tau}$.  The
parameter $\mathcal{R}_0$ gives the ratio of the height of the
Gaussian to the height of the exponential, at $\Av=0$.  The defining
parameters and the expectation values for these three distributions
are summarized in Fig~\ref{fig:DustModels}.

When simulating \SNIa\ with the SALT2 model, the SN color is defined
by the SALT2 $\mathcal{C}$ parameter.  This color term comprises both
the intrinsic SN color as well as reddening from host-galaxy dust.
The distribution of $\mathcal{C}$ values can therefore be described as
a convolution between a narrow Gaussian (the intrinsic dispersion
of \SNIa\ colors) and a function describing the distribution of 
host-galaxy extinctions. Following \citet{Barbary:2012}
and \citet{Scolnic:2013a}, we approximate the
\Av\ distributions of previous \SNIa\ studies by modifying the red side
of the SALT2 $\mathcal{C}$ distribution so that the simulated SN
colors match the output of that convolution. 

Specifically, our high-dust model for \SNIa\ matches the baseline \Av\
distribution used by \citet{Neill:2006}: a Gaussian with
$\sigma=0.62$. The mid-dust model is equivalent to the exponential
distribution of \citet{Kessler:2009b}: $P(\Av)=e^{-A_V/\tau}$, with
$\tau=0.33$.  Our low-dust model for \SNIa\ assumes minimal dust
extinction, using a narrow Gaussian with $\sigma=0.15$ plus a shallow
exponential with $\tau=0.15$.  A more complete treatment of host
galaxy dust would include a prescription for the redshift-dependence
of these extinction distributions, as SN hosts are expected to
be dustier at redshifts approaching $z=2$ \citep{Mannucci:2007}.  As
we will see in Section~\ref{sec:Rates}, this systematic uncertainty is
not a dominant component of the error budget, so redshift dependence
is left for future work.

\smallskip

To determine the combined systematic effects from the \SNIa\ fraction
prior and the dust assumptions, we compute each SN classification
probability 9 times: 3 for each SN rates prior $\times$ 3 for each
dust model.  The {\em mid-rate $+$ mid-dust} combination gives us our
baseline classification probability, which dictates how much each
individual SN contributes to the total count of observed \SNIa,
$N_{\rm obs}$. The extrema from this set of 9 probabilities then provide
the systematic classification errors, which propagate directly into
the systematic uncertainty on the \SNIa\ rate.

\subsection{STARDUST Validation Test}

A full investigation of the accuracy of the STARDUST classification
code is beyond the scope of this paper. It is useful, however, to
examine a simple validation test to demonstrate that this classifier
is not grossly biased or ineffective. To that end, we have applied the
STARDUST classifier to the ``Gold'' sample of 31 SN from the
GOODS and PANS surveys \citep{Strolger:2004,Riess:2007} that
have spectroscopic classifications. These surveys were carried out
using the \HST\ ACS, and share many of the
survey design characteristics of the CANDELS SN
program.  STARDUST correctly classifies 29 of the 31 SN
(93.5\%), using only their redshifts and photometric data.  This
demonstrates that we have a low false negative rate with STARDUST,
i.e., we rarely misclassify a true \SNIa. Unfortunately, this
validation test is not sensitive to false positives -- true \CCSN\
being misclassified as Type Ia -- because we only have a
single spectroscopically confirmed \CCSN\ in this Gold sample.
Preliminary testing of the STARDUST classifier using simulated SN
suggests that the Ia sample purity for photometrically classified SN
could be on the order of 95\% (these validation tests will be
presented in a future paper).

\section{Host Galaxies}\label{sec:HostGalaxies}

Host-galaxy information is recorded in Table~\ref{tab:highzHosts} for
the SN at $z>1.5$, and in the Appendix Table~\ref{tab:lowzHosts} for
the low-redshift SN.  As can be seen in Figure~\ref{fig:highzStamps}
(and Figures~\ref{fig:lowzStamps1}-\ref{fig:lowzStamps2}), most of the
65 CANDELS SN can be unambiguously associated with a host galaxy,
because the host is isolated, or the SN is clearly embedded within the
stellar light of a single galaxy. There are, however, a few exceptions.

For SN COS12Her, there are two host-galaxy candidates: the nearest and
brightest has a photometric redshift of 0.403 $^{+0.04}_{-0.11}$, but
the observed SN colors can not be adequately matched by any normal SN
template in that redshift range. The second COS12Her host candidate
has a photo-z=$1.10^{+0.16}_{-0.19}$.  At this higher redshift, the
STARDUST classifier finds a very good match to the observed light
curve with a Type II-P template.

SN GND13Sto is separated by several arcseconds from all nearby
galaxies.  Of the 6 galaxies within 5 arcseconds of the SN position, 5
have a photo-z distribution that peaks close to z=1.8, including one
with a spectroscopic redshift from the Spitzer Infrared Spectrograph
of $z=1.80\pm0.02$ \citep{Murphy:2009a}.  This is suggestive of a small
cluster or group of galaxies at that redshift, with SN GND13Sto
possibly associated with a low surface brightness group member or
tidal stream.  Indeed, applying STARDUST to the well-sampled SN
GND13Sto light curve (and allowing for a broad redshift range,
z=1.8$\pm$1.2), we find the maximum likelihood match is a \SNIa\
template at z=1.86$\pm0.05$.

There are four SN for which the host galaxy is barely detectable in
the deep IR imaging mosaics from CANDELS.  These are SN GND12Kin,
EGS11Nix, GND13Gar, and GSD11Bus.  All four of these objects lack a
clear spectroscopic redshift from their host, so we are limited to
using photometric redshifts for the STARDUST priors.  In all of these
cases, with STARDUST we find good template matches within the allowed
redshift range.

\subsection{Morphology and SED Type}

The SN host galaxies (along with all CANDELS galaxies) were classified
visually by members of the CANDELS team into three morphological
categories: {\it spheroid}, {\it disk}, and {\it
irregular.}  Visual classifications were done using template
images so that the presence of the SN did not bias the
classification. Each galaxy can be assigned to multiple categories,
so we also include two intermediate categories: {\it spheroid+disk}
and {\it disk+irregular.}  These morphological classes roughly
correspond to broad bins over the Hubble sequence.  This is
appropriate for classifying galaxies at high redshift where
distinguishing between, say, an E and an S0 galaxy is more difficult
and less meaningful.  Full details of the CANDELS morphological
classification procedure will be presented in a forthcoming paper
(Karteltepe, in prep).

We also record the ``SED type'' for each SN host galaxy, determined by
matching the full galaxy SED against a set of templates, using the
GOODZ code \citep{Dahlen:2010}. The GOODZ template library is
segregated into three groups, labeled according to the amount of
ongoing star formation: {\it passive} (early type), {\it active} (late
type), and {\it starburst}.  We use the best-matching SED template for
each CANDELS SN host galaxy to assign it to one of those bins.

For two of the SN (COS12Car and GND12Daw) there is no discernible host
at the location of the SN and no nearby galaxy presents a plausible
host candidate.  For both of these objects we do have spectroscopic
redshift information from the SN themselves, as detailed in
Section~\ref{sec:GrismSpectroscopy}.  For the other 63 SN in our
sample, 10 host galaxies are classified as spheroids, 15 as
spheroid+disk, 17 as disk, 7 as disk+irregular, and 8 as irregular.
For 6 of our SN, the host galaxy is detected, but is too faint for
reliable visual classification, so we report the host morphology as
``unclassifiable''.  For the 63 objects
with detectable host galaxies we have 2 passive, 24 active and
37 starburst-like SEDs.

\section{Grism Spectroscopy}\label{sec:GrismSpectroscopy}

\insertfigquad{FIG/snWashingtonSpec}{FIG/snDawesSpec}{FIG/snGarnerSpec}{FIG/snCarterSpec}{
Observed spectra and spectral template matches for four CANDELS
SN. The spectra for GSD11Was (top left) and GND13Daw (top right) were
collected using the G141 grism on \HST's WFC3-IR detector. The
GND13Gar spectrum (lower left) used the ACS G800L grism, and the
spectrum of COS12Car (lower right) combines observations from both the
G102 and G141 WFC3 grisms. In each panel the observed flux is shown in
gray, with binned points overlaid in blue, and the best-fitting
template spectrum in red. 
\label{fig:SNspec}
}

There are six objects in our sample for which we collected
useful \HST\ grism spectroscopy of the SN themselves.  SN GSD10Pri, a
Type Ia SN at $z=1.55$, was presented in \citet{Rodney:2012} with an
analysis of the host galaxy in \citet{Frederiksen:2012}.  SN UDS11Wil,
a Type Ia SN at $z=1.91$, was described
in \citet{Jones:2013}. Figure~\ref{fig:SNspec} presents grism spectra
for the remaining four: SN GSD11Was, GND12Daw, GND13Gar and
COS12Car. In all of these cases the signal to noise ratios and
rest-frame wavelength coverage are insufficient for a purely
spectroscopic classification.  Rather, as with GSD10Pri and UDS11Wil,
we used the spectroscopic information to supplement the STARDUST
photometric classifier, leading to a more robust classification.

The host galaxy of SN GSD11Was has a photometric redshift of
$z=1.04\pm0.3$.  We obtained a spectrum of SN GSD11Was with the
WFC3-G141 grism, shown in Figure~\ref{fig:SNspec} (top left). Here we
can see hints of an absorption feature at $\sim$14,000 \AA.  At a
redshift of $z\sim 1.3$ this feature can be explained as the
characteristic SiII absorption trough seen at rest-frame
$\sim$6150 \AA\ in \SNIa\ spectra. Photometric classification of this
SN with STARDUST agrees, finding the object is best matched by
a \SNIa\ template at $z=1.3\pm0.05$.  (see the light curve plot in
Appendix~\ref{app:lowz}, Figure~\ref{fig:lowzLCfit3}).

SN GND12Daw, GND12Gar, and COS12Car all have no detectable host galaxy
in any optical or NIR band, and no neighboring galaxies have redshifts
that allow for acceptable light curve template matches in STARDUST.
The most likely explanation is that these SN reside in very low
surface brightness galaxies, too faint for detection even in our deep
\HST\ imaging.   

The spectrum for GND12Daw shows hints of a broad emission feature at
12000 \AA, which could be H$\alpha$ emission, if the object is at
z=0.830.  This could be interpreted as strong Balmer line emission
from an otherwise very faint host galaxy, or it could be showing the
H$\alpha$ emission from the SN itself -- characteristic of Type II-P
spectra.  Given the very low signal to noise ratio in this spectrum
(it was derived from just a single orbit of \HST\ observations) this
alone would be weak support. However, when allowing STARDUST to search
over a redshift range $0.1<z<2.0$, we find that a Type II-P light
curve template consistently provides the strongest match to the broad
light curve shape of this SN, and the solution at $z\approx0.8$ provides
$>90\%$ of the total likelihood.

For GND12Gar (upper right) the absorption trough at $\sim$7700 \AA\
provides a key observable that can anchor the fit and define the age
of the SN.  If this feature corresponds to Ca II absorption, then that
would fix the object's redshift to $z=1.07\pm0.02$.  At this redshift
the light curve is matched very well by \SNIa\ templates, and no other
redshift or SN class can provide a better light curve match.

The strongest spectral constraint for SN classification comes from SN
COS12Car.  For this object we have observations with both the
WFC3-G102 and G141 grisms.  Fitting the composite spectrum with the
SuperNova IDentification (SNID) software \citep{Blondin:2007}, we find
the best template match is a Type Ia SN at z=1.59.  Once
again we find that the STARDUST photometric classification agrees well
with this spectroscopic information: a \SNIa\ light curve at $z\approx1.6$
provides the best available light curve template.

\section{The Volumetric \SNIa\ Rate}\label{sec:Rates}

To convert the observed SN counts into a volumetric rate, we
use an approach similar to \citet{Dahlen:2008}
and \citet{Rodney:2010b}. We first divide up the detected SN into
four redshift bins of width $\Delta z=0.5$. The total contribution to
the \SNIa\ count from each observed SN is equal to the Ia
classification probability for that object. For objects with uncertain
redshifts, this fractional contribution is distributed over multiple
redshift bins according to the integrated area of the redshift pdf.
Adding up all the fractional counts gives us the total observed \SNIa\
count as a function of redshift: $N_{\rm obs}(z)$.  Statistical
uncertainties for each bin are defined by the points encompassing the
central 68\% of the Poisson distribution.

We then use Monte Carlo simulations to compute a ``control count'' for
each bin, $N_{\rm ctrl}(z)$, which is the expected number of \SNIa\ that
would be detected if the cosmic \SNIa\ explosion rate were constant
for all redshifts at 1 SNuVol = \SNuVol. 
By simulating \SNIa\ light curves within the context of the CANDELS
survey, the computation of $N_{\rm ctrl}(z)$ incorporates both
the survey volume and the control time (the time interval over which
any given SN is visible to our survey).

We again use SNANA as our simulation engine, this time generating
100,000 \SNIa\ based on the SALT2 light curve model \citep{Guy:2010}.
The light curve for each synthetic SN is determined by a set of 4
variables: date of peak brightness $mjd_{\rm pk}$, redshift $z$, SALT2 shape parameter $x_1$, and SALT2 color 
parameter $\mathcal{C}$.  The $\mathcal{C}$ parameter in SALT2
includes both intrinsic SN color as well as reddening from host-galaxy
dust.  Each redshift $z$ is
drawn from the range $0<z<2.5$, following the constant volumetric rate
assumption. To translate this redshift into a luminosity distance, we
use our baseline cosmology: \Om=0.3, \OL=0.7, $w$=-1, \Ho=70. Values
for $x_1$ are drawn from a normal distribution with mean and
dispersion from \citet{Kessler:2009b}: $\bar{x}_1=-0.13$,
$\sigma_{x_1}=1.24$.  The color parameters are draw from a bifurcated
Gaussian distribution with $\bar{\mathcal{C}}=0.04$,
$\sigma_\mathcal{C}^-=0.08$, and $\sigma_\mathcal{C}^+=0.25$ $-$
parameters that match the ``mid-dust'' model described in
Section~\ref{sec:HostAvDistribution} and Figure~\ref{fig:DustModels}.

To choose values for $mjd_{\rm pk}$ we first establish the width of the
survey window at any given redshift.  By examining simulated \SNIa\
light curves in F125W and F160W, we find $t_{\rm min}$ and $t_{\rm max}$, the
minimum and maximum dates relative to peak for which each SN would be
detectable to our survey.  Here detectability is defined by measuring
the change in flux relative to a template epoch 52 days prior, and
requiring that the corresponding J+H magnitude is brighter than the
50\% detection threshold seen in Figure~\ref{fig:DetEff}. The allowed
range for the simulated $mjd_{\rm pk}$ values at redshift $z$ is then
[MJD$_{\rm first}$-$t_{\rm max}(z)$, MJD$_{\rm last}$+$t_{\rm min}(z)$], where
MJD$_{\rm first}$ and MJD$_{\rm last}$ are the epochs for the first and last
search epoch, respectively.  For each redshift, random $mjd_{\rm pk}$
values are then drawn from a flat distribution spanning this survey
window.

Each synthetic SN is ``observed'' in the SNANA simulator using survey
parameters that match the actual operations of the CANDELS program, as
given in Tables~\ref{tab:Survey} and \ref{tab:Exposures}.  For the
Wide fields (COSMOS, EGS, UDS and the wings of the GOODS fields) we
only have a single search epoch, so these fields are simulated
together as the ``CANDELS-Wide'' search field.  The 10-epoch GOODS-S
and GOODS-N Deep fields are treated separately, but all observational
parameters are computed in the same way.  Due to the excellent
stability of the \HST\ photometric system, we adopt a single set of
average values for zero points and detector noise.  The total area in
each field reflects the area in which SN searching can be done,
i.e., the area covered by the SN search epoch and at least one prior
epoch.  The cadence between epochs is nominally 52 days, but the
actual separation in time varies from pointing to pointing due to \HST\
scheduling constraints.  For this simulation we use a mean cadence for
each field and each epoch, weighted by the area available for SN
searching.  Finally, we use the detection efficiency curve of
Figure~\ref{fig:DetEff} to define the probability of ``detecting''
each simulated SN in any given epoch.

\makeatletter{}\begin{small}
\ifms 
  \begin{deluxetable}{ c ccc c cc c ccc }
\else 
  \begin{deluxetable*}{c ccc c cc c ccc }
\fi
\tablewidth{0pt}
\tablecolumns{11}
\tablecaption{Observed \SNIa\ Counts and the Volumetric Rate\label{tab:Rates}}
 \tablehead{
    \colhead{} &
    \multicolumn{3}{c}{Observed Count~\tablenotemark{a}} & &
    \multicolumn{2}{c}{Control Count~\tablenotemark{b}} & &
    \multicolumn{3}{c}{SN Rate~\tablenotemark{c}} \\
    \cline{2-4} \cline{6-7} \cline{9-11}\\
    \colhead{Redshift} &
    \colhead{$N_{\rm obs}$} & \colhead{$\delta N_{\rm Poiss}$} & \colhead{$\delta N_{\rm syst}$} & &
    \colhead{$N_{\rm ctrl}$} & \colhead{$\delta N_{\rm syst}$} & &
    \colhead{SNR}  & \colhead{$\delta$SNR$_{stat}$} & \colhead{$\delta$SNR$_{syst}$}
  }
\startdata

$0.25$ &  1.46 &  $^{+2.46}_{-1.06}$ &   $^{+0.48}_{-1.44}$ & &   4.10 & $^{+0.01}_{-0.04}$  & &  0.36  &  $^{+0.60}_{-0.26}$  &  $^{+0.12}_{-0.35}$  \\[0.7em]
$0.75$ &  7.19  &  $^{+3.80}_{-2.62}$ &   $^{+1.94}_{-2.54}$ & &  14.11 & $^{+0.53}_{-1.71}$  & &  0.51  &  $^{+0.27}_{-0.19}$  &  $^{+0.23}_{-0.19}$  \\[0.7em]
$1.25$ &  8.47  &  $^{+4.02}_{-2.85}$ &   $^{+0.45}_{-2.04}$ & &  13.16 & $^{+2.46}_{-4.05}$  & &  0.64  &  $^{+0.31}_{-0.22}$  &  $^{+0.34}_{-0.23}$  \\[0.7em]
$1.75$ &  5.54  &  $^{+3.49}_{-2.28}$ &   $^{+0.17}_{-0.61}$ & &   7.67 & $^{+3.47}_{-2.99}$  & &  0.72  &  $^{+0.45}_{-0.30}$  &  $^{+0.50}_{-0.28}$  \\[0.7em]
$2.25$ &  1.24  &  $^{+2.39}_{-0.96}$ &   $^{+0.13}_{-0.16}$ & &   2.52 & $^{+1.72}_{-1.07}$  & &  0.49  &  $^{+0.95}_{-0.38}$  &  $^{+0.45}_{-0.24}$  \\[0.7em]

\enddata
\tablenotetext{a}{\footnotesize Statistical uncertainties reflect the
  limits that contain 68\% of the Poisson distribution. Systematic
  uncertainties are due to the assumed dust model and rates prior.}
\tablenotetext{b}{\footnotesize Systematic uncertainties are due to
  the assumed dust model.} 
\tablenotetext{c}{\footnotesize The \SNIa\ rate measurements in units
  of SNuVol = \SNuVol.} 
\ifms \end{deluxetable}
\else \end{deluxetable*} \fi
\end{small}

Counting the number of detected synthetic SN in each redshift bin
gives us the control count, which carries units of SNuVol$^{-1}$. The
observed volumetric rate of \SNIa\ explosions is simply the ratio
 
\begin{equation}
 {\rm SNR}_{\rm Ia}(z) = \frac{N_{\rm obs}(z)}{N_{\rm ctrl}(z)} 
  \label{eqn:snr}
\end{equation}

 The measured values for $N_{\rm obs}(z)$, $N_{\rm ctrl}(z)$, and \SNRz\
from the CANDELS survey are given in Table~\ref{tab:Rates} along with
uncertainty estimates due to statistical noise (Poisson errors) and
systematic biases. The total sample size is quite small, with only
$\sim$21 \SNIa\ across all redshifts and fewer than 7 in each bin.
This means that the statistical errors are substantial, roughly equal
to or greater than the systematic uncertainties in every redshift bin.
One cannot infer a clear trend with redshift from these data alone,
but rather we must evaluate them within the context of other rates
measurements and \SNIa\ progenitor models.

\subsection{Systematic Uncertainties}

In preceding sections we have considered three principal sources of
systematic biases: (1) missing SN detections due to subtraction
artifacts in the cores of bright galaxies, (2) the assumed fraction of
SN that are of Type Ia as a function of redshift, and (3) the assumed
distribution of host-galaxy dust extinction values. 
 We have determined that bias from the first source is
negligible. The second is examined in more detail in
Appendix~\ref{app:galsnid}, and is 
reflected in the systematic uncertainty estimates for the count of
observed \SNIa\ (column 4 of Table~\ref{tab:Rates}). The third item
also affects the control count (column 6). 

Other potential sources of systematic bias include: errors in the
luminosity functions for SN sub-classes, biases in the \SNIa\ model or
the \CCSN\ template libraries, and peculiar detection biases from
individual human searchers.  For this work, these contributions to the
systematic error budget are assumed to be insignificant.  Future
analysis with the full CANDELS+CLASH sample will revisit this
assumption and evaluate these systematic error sources.

\section{Testing \SNIa\ Progenitor Models}\label{sec:Discussion}

The measurement of volumetric \SNIa\ rates at high redshift is
principally motivated by two astrophysical investigations. First, it
directly informs our understanding of the cosmic enrichment history, as  
\SNIa\ are a primary source for Fe-group elements in the
universe \citep[e.g.,][]{Wiersma:2011}. Secondly, by measuring the
delay between star formation and SN explosion through the DTD
formalism, one can draw inferences about the nature of \SNIa\
progenitor systems.  In this work we limit our discussion to the
latter, beginning with a comparison to other published \SNIa\ rates,
then evaluating new constraints on progenitor models and finally
making a projection toward future improvements.

\insertfigwide{FIG/figRatesPoints}{\label{fig:Rates}
Measured SNIa rates from CANDELS and other surveys. The CANDELS
volumetric \SNIa\ rates measurements are shown as large orange
squares, spanning 5 redshift bins of width dz=0.5. For these CANDELS
points, the systematic uncertainties are shown as broad orange bars,
while the thin vertical error bars show the combined systematic and
statistical uncertainty. Four other high-redshift SN surveys are
highlighted: gray circles for the Subaru Deep
Field \citep[SDF][]{Graur:2011}, blue downward triangles for
volumetric (not cluster) rates from the Cluster Supernova
Survey \citep[CSS][]{Barbary:2012}, green upward triangles for the
GOODS and PANS surveys \citep{Dahlen:2008}, and magenta diamonds for
the CLASH survey \citep{Graur:2014}.  Assorted ground-based surveys
plotted as gray
circles, as in Figure~\ref{fig:RatesPreCC}.}

\subsection{Comparison to Earlier Rate Measurements}

Figure~\ref{fig:Rates} presents the CANDELS \SNIa\ rates within the
context of other rates measurements from the literature. The CANDELS
rate measurements are shown in 5 bins of width $\Delta z=0.5$ as large
orange squares.  Rate measurements from 13 ground-based surveys are
plotted as small gray circles, reaching out to z=1.1.  Four surveys
that have previously extended the rate measurements to $z\approx1.5$ are
highlighted with larger colored points (see caption for details).

As with past \HST\ surveys, our survey volume is too small to add any
useful new information at $z<1$, but the general agreement with
ground-based surveys is an important validation that our rate
measurements are realistic.  For a more informative comparison, we
turn now to the high-$z$ side of the plot.

Before CANDELS and CLASH, there were just three surveys with
any \SNIa\ rate measurements above z$\approx$1.2.  First were the GOODS
SN surveys, which used the \HST\ ACS camera to measure the \SNIa\ rate
to $z\approx1.8$ \citep[GOODS][]{Strolger:2004,Dahlen:2004}.  These data
were interpreted as showing a flattening or a downturn in the \SNRz\
at $z>1.2$, a trend that garnered support from additional \HST\
observations and independent
analysis \citep{Dahlen:2008,Kuznetsova:2008}, including another
ACS program, the Cluster Supernova
Survey \citep[CSS][]{Barbary:2012}.\footnote{This was a survey of
galaxy clusters, but the work of \citet{Barbary:2012} presented
volumetric \SNIa\ rates from the SN detected outside the clusters.}
Ground-based rate measurements from the Subaru Deep Field (SDF) survey
also reached out to $z\approx1.8$, though these were more susceptible to
systematic biases due to the absence of light curve information and
spectroscopy for SN
classification \citep{Poznanski:2007,Graur:2011}. As the CANDELS and
CLASH surveys began, it was still an open question as to whether we
had now seen the peak in the \SNIa\ rate, or if it was continuing to
rise beyond $z\approx1.2$.

The GOODS, CSS and SDF surveys all used optical bands that correspond
to rest-frame near-UV wavelengths at high redshift.  For a SN at
redshift $z\approx1.5$, observations in the $z$ band ($\sim 9000$~\AA)
are sampling the rest-frame U band, while the observer's $i$ band
($\sim 8000$~\AA) reaches well into the rest-frame near-ultraviolet.
At these wavelengths the available SN light curve templates for use in
photometric classification are poor, because most nearby SN surveys do
not observe SN in the UV.  Both \SNIa\ and \CCSN\ also exhibit more
natural heterogeneity at these blue wavelengths, and this is all
compounded by a greater sensitivity to dust obscuration in the UV.
Thus, optical-wavelength surveys were more susceptible to both of the
components that dominate the systematic uncertainties of
high-$z$ \SNIa\ rate measurements: classification bias and dust
obscuration.  By contrast, the CANDELS survey utilized the $J$ and $H$
IR bands that sample rest-frame optical wavelengths, even out to
redshift $z\approx2.5$.  The CANDELS rates should therefore be less strongly
affected by those systematic biases.

At $z=1.25$ the CANDELS rate is substantially lower than all past
measurements, though still consistent at the 1--2$\sigma$ level.  The
CANDELS rate then climbs slightly in the bin at $z=1.75$, where it is
completely consistent with past measurements.  CANDELS is the only
survey with any detections beyond $z=2$, and there we have only a
single object with a strong probability of being a \SNIa\ (SN GND12Col
in the GOODS-N field, at $z\approx2.24$).  The rate formally shows a
decline to $z=2.25$, although this change is much smaller than the
uncertainties.  The CANDELS rates are fully consistent at all
redshifts with the similarly derived rates from CLASH, which are also
quite low relative to past surveys \citep{Graur:2014}.

Due to the small sample sizes and large uncertainties, none of these
individual high-$z$ surveys has sufficient precision to clearly
delineate the shape of the \SNRz\ curve.  From Figure~\ref{fig:Rates}
we can only say that the \SNIa\ rate rises steadily to $z\approx1.2$, and
then is flat or slowly declining at redshifts $z>1.2$.

Each independent analysis of \SNIa\ rates makes slightly different
assumptions about host-galaxy extinction and each takes a different
approach to SN classification.  These differences become particularly
important at $z>1$ where observed rates are dominated by \HST\ SN
surveys, which have much less complete spectroscopic information.
Here the potential for systematic biases is much greater as a larger
fraction of SN classifications and host-galaxy redshifts rely on
photometric data alone. 

An optimal approach would be to effectively treat the past decade of
\HST\ SN surveys as a single composite survey. One could compute the
rates from all the \HST\ SN surveys together, using the same SN
classifier(s), consistent models for (redshift-dependent) host-galaxy
extinction, and the best available host-galaxy redshift information.
Such an effort is beyond the scope of this work, but will be an
important contribution for future DTD tests.

\subsection{Isolating the Prompt \SNIa\ Fraction}

To examine how the observed \SNIa\ rate can inform the modeling
of \SNIa\ progenitors, we will employ a simple toy model, motivated by
a variety of recent observations and theoretical
predictions.  For a complementary analysis using DTD
predictions from binary population synthesis modeling,
see \citet{Graur:2014}. Multiple lines of evidence now suggest that
the overall shape of the \SNIa\ DTD follows a $t^{-1}$ power law for
times $t>500$ Myr \citep[see][for a recent review]{Maoz:2012}.  At
short delays, $t<500$ Myr, the evidence is much less definitive, and
this is the region where the CANDELS observations may provide unique
new insight. Thus, our primary question is: {\it What fraction
of \SNIa\ explode within 500 Myr of their formation?}.

To isolate this ``prompt \SNIa\ fraction'', we define a bifurcated DTD
model: the long-delay component follows a $t^{-1}$ distribution for
all times $t>500$ Myr, and the prompt component is set to be constant
with time for $t<500$ Myr, down to a lower limit of $t_{\rm min}=40$ Myr (the
shortest possible time to reach explosion, \citealt{Belczynski:2005}):

\begin{equation}
  {\rm SNR}(t) = \begin{cases} 
     0 &   \text{for $t<0.04$ Gyr},\\
     K\,\eta_{\rm Ia}\,\frac{\fp}{1-\fp} & \text{for $0.04<t<0.5$ Gyr},\\
     \eta_{\rm Ia}\,t^{-1} & \text{for $t>0.5$ Gyr},\\
  \end{cases}
  \label{eqn:dtd}
\end{equation}

\noindent  Here $\eta_{\rm Ia}$ indicates the efficiency of generating \SNIa\ 
progenitor systems, in units of \SNIa\,yr$^{-1}$\,\Msun$^{-1}$,
and \fp\ sets the fraction of all \SNIa\ that arise from the prompt
channel. The constant K is defined by the time thresholds that
delineate this model: 

\begin{equation}\label{eqn:DTDK}
 K = \ln(t_{\rm max}/t_{1}) / (t_{1}-t_{\rm min}),
\end{equation}

\noindent where $t_{\rm min}=0.04$ Gyr as defined above, $t_{1}=0.5$ Gyr marks the
abrupt transition from the the constant rate to the power law, and
$t_{\rm max}=13.3$ Gyr is the maximum age of a WD in the current
universe -- using our assumed \LCDM\ cosmology and assuming
star formation began at $z=20$.  For these values, we have $K=7.132$.
With this simple DTD model, we can allow $\eta_{\rm Ia}$ and \fp\ to be
free parameters, and fit to the data to find the observed efficiency
and prompt Ia fraction.

\ifgrayscale
  \insertfig{FIG/figCSFR_gray}{\label{fig:CSFR}
  The cosmic star-formation rate (CSFR) as a function of
  redshift. Points show the compilation of recent CSFR measurements
  from \citep{Behroozi:2013b}, adopting from those authors the
  corrections for dust attenuation and more realistic systematic
  errors. The solid line shows the best-fit double-power law model
  from \citep{Behroozi:2013b}, and the shaded region demarcates the
  1-$\sigma$ systematic uncertainties. 
  (A color version of this figure appears in the online edition.)}
\else
  \insertfig{FIG/figCSFR}{\label{fig:CSFR}
  The cosmic star-formation rate (CSFR) as a function of
  redshift. Points show the compilation of recent CSFR measurements
  from \citep{Behroozi:2013b}, adopting from those authors the
  corrections for dust attenuation and more realistic systematic
  errors. The solid line shows the best-fit double-power law model
  from \citep{Behroozi:2013b}, and the shaded region demarcates the
  1-$\sigma$ systematic uncertainties. }
\fi

To convert from this DTD model into a prediction for \SNIa\ rates, we
convolve this DTD with a parameterized representation of the cosmic
star-formation history, giving us a prediction for the
observable \SNRz.  For this exercise we use the recent compilation of
measurements of the cosmic star-formation rate ($CSFR(z)$)
from \citet{Behroozi:2013b}, shown in Figure~\ref{fig:CSFR}.  The
precise shape of the CSFR curve at $z>2$ is still a matter of debate,
but for our purposes here we take the Behroozi et al. curve and
associated systematic uncertainties to be representative of the
current state of the art \citep[but see][for further evaluation of SFH
variation]{Graur:2014}.

The construction of our bifurcated DTD model is reminiscent of the
two-component ``A+B'' model \citep{Mannucci:2005,Scannapieco:2005},
but it has closer ties to recent theoretical predictions from binary
population synthesis models.
For example, \cite{Ruiter:2013} found that a ``violent merger'' DD
model predicts a $t^{-1}$ power law shape for long delay SN, but also
includes a very prompt component that arises from a distinct subset
of binary systems.  A separate prompt channel for \SNIa\ explosions
could also arise from a single-degenerate pathway with a helium-star
donor \citep{WangB:2009a,Claeys:2014}.

\subsection{DTD model fitting results}

To find the most likely values for our two parameters $\eta_{\rm Ia}$
and \fp, we use three \SNIa\ rate data sets.  First we define the
``All'' data set, utilizing all available (non-redundant) volumetric
rate measurements from the literature \citep[see][for a compilation
table]{Graur:2014}.  Secondly, our ``Ground'' subsample picks out
the 13 independent rate measurements from ground-based surveys.
Finally, our ``CANDELS+CLASH'' sample isolates those 2
companion \HST\ surveys.

The first three columns of Table ~\ref{tab:DTD} summarize the maximum
likelihood values for our DTD parameters $\eta_{\rm Ia}$ and \fp, when
fitting to each of these subsamples. When using all of the
available \SNIa\ survey data, we find $\eta_{\rm Ia}
= \left(1.60 \pmstat ^{0.24}_{0.236} \pmsys ^{0.25}_{0.59}\right) \times
10^{-4}$ \SNIa\ yr$^{-1}$ \Msun$^{-1}$
and \fp=0.53 \pmstat $^{0.09}_{0.10}$ \pmsys $^{0.10}_{0.26}$.

Fitting to the ground-based data alone, we find very similar best-fit
parameters, with the prompt \SNIa\ fraction inching up to \fp=0.59 and
the efficiency remaining at $\eta_{\rm Ia}\approx1.5$.  When we isolate
the \HST\ CANDELS and CLASH surveys, we get much larger uncertainties,
but perhaps also a subtle hint at tension between the ground- and
\HST-based results: from the CANDELS+CLASH sample we
get \fp=0.21 \pmstat$^{0.34}_{0.21}$ \pmsys$^{0.49}_{0.12}$.  The
difference in these best-fit parameters reflects a (very) mild
disagreement between the ground-based, primarily low-$z$ rate
measurements and the high-$z$ constraints from \HST.

\insertfigwide{figRatesDTD}{\label{fig:DTD}
Comparing observed SN rates against DTD models.  Grey markers show the
collection of rates from the literature, with filled points
for the ground-based surveys, and open symbols for past \HST\ surveys.
As in Figure~\ref{fig:Rates}, large purple diamonds show the CLASH
rates from \citet{Graur:2014} and large orange squares show the
CANDELS rates from this work.  In this figure vertical error bars show
only the statistical uncertainties.  Two curves show the \SNIa\ rates
predicted by assuming a DTD that is proportional to $t^{-1}$ for all
times above 500 Myr, but with different assumptions for the fraction
of \SNIa\ that are prompt. The magenta solid line shows the best-fit
to the ground-based data (solid grey points), which has
$\sim$60\% of all \SNIa\ exploding within 500 Myr of birth.  The green
dashed line is the best-fit model when using the CANDELS+CLASH data alone, for
which the prompt \SNIa\ fraction is $\sim$20\%.  }

\makeatletter{}\renewcommand{\arraystretch}{1.2}
\begin{small}
\ifms 
  \begin{deluxetable}{ c lll | l}
\else 
  \begin{deluxetable*}{ c lll | l }
\fi
\tablewidth{0pt}
\tablecolumns{11}
\tablecaption{Observational Constraints on the \SNIa\ Delay Time Distribution\tablenotemark{*}\label{tab:DTD}}
\tablehead{ 
  \colhead{Sample} & 
  \colhead{$\eta_{Ia}$} &
  \colhead{\fP} &
  \colhead{N$_{Ia}$/M$_{*}$ \tablenotemark{a,b}} &
  \colhead{N$_{Ia}$/M$_{*}$ \tablenotemark{b,c}} \\
  & \colhead{$[10^{-4}  \Msun^{-1} yr^{-1}]$} & 
  & \colhead{$[10^{-3}  \Msun^{-1}]$} 
  & \colhead{$[10^{-3}  \Msun^{-1}]$}
}

\startdata
CANDELS \& CLASH    &  
  2.25 $^{+1.36}_{-1.18}$ $^{+0.72}_{-0.15}$  &
  0.21 $^{+0.34}_{-0.21}$ $^{+0.49}_{-0.12}$  & 
  0.79 $^{+1.52}_{-0.50}$ $^{+2.11}_{-0.16}$  & 
  0.60 $^{+0.97}_{-0.37}$ $^{+0.87}_{-0.46}$  \\[0.6em]
Ground  &  
  1.38 $^{+0.24}_{-0.23}$ $^{+0.43}_{-0.21}$  & 
  0.59 $^{+0.09}_{-0.10}$ $^{+0.05}_{-0.04}$  & 
  0.84 $^{+0.39}_{-0.27}$ $^{+0.44}_{-0.54}$  & 
\nodata \\[0.6em]
All     &  
  1.60 $^{+0.24}_{-0.23}$ $^{+0.25}_{-0.59}$  & 
  0.53 $^{+0.09}_{-0.10}$ $^{+0.10}_{-0.26}$  & 
  0.98 $^{+0.43}_{-0.30}$ $^{+0.87}_{-0.46}$  &
  0.79 $^{+0.88}_{-0.72}$ $^{+1.09}_{-0.75}$  \\
\enddata

\tablenotetext{*}{\footnotesize Errors give first statistical then systematic uncertainties.}
\tablenotetext{a}{\footnotesize Assuming a $t^{-1}$ delay time model of the form given in Equation~\ref{eqn:dtd}}
\tablenotetext{b}{\footnotesize Using the \citet{Behroozi:2013b} cosmic star formation history, which assumes a \citet{Chabrier:2003} IMF.}
\tablenotetext{c}{\footnotesize Using the \SNIa\ rate data directly, without any DTD model assumption.}
\ifms \end{deluxetable}
\else \end{deluxetable*} \fi
\end{small}
\renewcommand{\arraystretch}{1.0}

\insertfig{figDTDcontours_HSTvsGround}{\label{fig:DTDcontours}
Constraints on the DTD normalization factor and the fraction of \SNIa\
that are prompt explosions.  Contours show the 68\% and 95\%
confidence regions for the baseline assumptions ({\it mid-dust,
mid-rates}) in the $\eta_{\rm Ia}$ vs. \fp\ parameter space.  The background
color map indicates the time-integrated \SNIa\ efficiency,
$N_{\rm Ia}/M_{*}$, for each point in that parameter space. Dashed
contours show the confidence regions derived from only CANDELS+CLASH data,
reaching to $z\approx 2.5$. Solid contours are from the collection of all
ground-based SN surveys, dominated by measurements at $z<1$.  }

The source of this deviation is easily seen in Figure~\ref{fig:DTD},
where we plot two \SNRz\ curves derived from the bifurcated DTD model.
The (magenta) solid line shows the best fit to the ground based data
alone, with \fp=0.6.  The (green) dashed line sets the prompt
fraction to 20\%, the best fit value for the CANDELS+CLASH
data sample.  These two \HST\ surveys find a relatively
low \SNIa\ rate at all redshifts $z\gtrsim 1$, which pulls the
best-fit curve downward at high redshift, leading to the low
best-fit \fp.  We can also see the slight tension between ground and
recent \HST\ measures in Figure~\ref{fig:DTDcontours}, where we show
confidence regions in the $\eta_{\rm Ia}$ vs. \fp\ parameter space.
The 68\% contours from the ground- and \HST-based surveys fall just
short of overlapping.  This discrepancy is only slightly above
1$\sigma$ in significance, and comes with all the caveats cited above
regarding the method for combining data from disparate
surveys. Nevertheless, these \HST\ data do sample the redshift range
with the greatest leverage for constraining \fp, so the scarcity of
high-$z$ \SNIa\ detections in multiple \HST\ surveys should not be
discounted.

Table ~\ref{tab:DTD} and Figure~\ref{fig:DTDcontours} also present the
total number of \SNIa\ per stellar mass, $N_{\rm Ia}/M_{*}$.  This is
computed by integrating the \SNIa\ rate over a Hubble time, and
dividing by the total mass of formed stars.  For the denominator, we
take the integral of the \citet{Behroozi:2013b} $CSFR(z)$ curve from
Figure~\ref{fig:CSFR} (which assumes a \citet{Chabrier:2003} stellar
initial mass function).  To get the numerator -- the total number
of \SNIa\ explosions in a Hubble time -- we can integrate the best-fit
DTD-based \SNRz\ model for each subsample of rates measurements.
Those values are reported in the fourth column of
Table~\ref{tab:DTD}. In the fifth column we list an alternative
calculation, now directly integrating the \SNRz\ data, without
reference to any DTD model.  This latter approach yields a much less
precise constraint, but it is more appropriate for use as a test of
progenitor models, because it does not presuppose any particular
shape for the DTD.  Note that we do not measure a data-only constraint
from the ground-based subsample because it does not reach a sufficiently 
high redshift.

Figure~\ref{fig:DTDcontours} shows a color map in the background,
reflecting the variation of $N_{\rm Ia}/M_{*}$ within the $\eta_{\rm
Ia}-\fp$ plane (assuming that the DTD follows our two-component toy
model).  The contours derived from both the ground-based and \HST\
surveys are roughly aligned along lines of constant $N_{\rm Ia}/M_{*}$
(a single-color ridge in the color map). Hence the relatively tight
model-dependent constraints on $N_{\rm Ia}/M_{*}$ as reported in
column 4 of Table~\ref{tab:DTD}.

All of the above $N_{\rm Ia}/M_{*}$ measurements are consistent with a
value of roughly $1\times10^{-3} \Msun^{-1}$.  This is fully
consistent with past measures of the volumetric rates, using similar
stellar IMF assumptions \citep[e.g.,][]{Graur:2011}.  Other
observational constraints, such as cluster \SNIa\ rates, have recently
found values closer to $2\times10^{-3} \Msun^{-1}$ \citep{Maoz:2010}
-- still consistent within the large error bars.  However, theoretical
predictions from binary population synthesis models are frequently
lower by factors of 10 or more \citep{Bours:2013}.  This discrepancy
between theory and observation remains one of the key concerns in
the \SNIa\ progenitor problem.

\subsection{Interpretation and Speculation}\label{sec:Interpreting}

As described above, our analysis of all available \SNIa\ rates
measurements suggests that the fraction of \SNIa\ explosions occurring
$<$500 Myr after formation is \fp$\approx$50\%.  This observed value
of \fp\ is broadly consistent with the simplistic assumption of a
t$^{-1}$ DTD that continues without truncation all the way down to 40
Myr, which yields \fp=0.43.  A prompt fraction close to 50\%\ is also
observationally supported by several lines of evidence in the local
universe.  \citet{Mannucci:2006} first proposed that roughly half of
all \SNIa\ explode promptly after formation, based on observations
of \SNIa\ host galaxies at low redshift.  Building on that
work, \citet{Raskin:2009b} used measurements of low-$z$ \SNIa\
environments on sub-galactic scales to infer that most of those
prompt \SNIa\ explode at $200-500$ Myr.  \citet{Mennekens:2013} used
binary population synthesis (BPS) to predict the distribution of
chemical enrichment in our galaxy over time.  Comparing this to
observations of [Fe/H] in nearby G-dwarfs, they infer that prompt
explosions must make up a large fraction of the \SNIa\ population (and
thereby contribute to rapid galactic enrichment). 

Can this measurement of the prompt \SNIa\ fraction be used to
distinguish SD and DD progenitor models?  BPS calculations generally
agree that SD pathways preferentially generate prompt \SNIa\
explosions. In particular, SD progenitor models in which the companion
is a naked He star are found to peak at $t\approx100$ Myr after
formation, while those with a normal main sequence or giant companion
preferentially explode at $200-500$
Myr \citep{WangB:2009a,Mennekens:2010,Greggio:2010,Claeys:2014}.  Some
BPS modeling also finds that DD progenitors could contribute
substantially to the population of \SNIa\ explosions younger than 500
Myr \citep{Ruiter:2009,Greggio:2010,Ruiter:2013}, although recent work
by \citet{Claeys:2014} suggests that the DD pathway does not dominate
the DTD until $t>500$ Myr.  At the moment, we can only say that a
prompt fraction \fp$\approx$50\% is commonly predicted by models that
include both SD and DD progenitors -- but it may be possible in a pure
DD model as well.

When we analyze the \HST\ sample in isolation, we find a hint that \fp\
could be closer to $\sim$20\%, though further analysis and a
larger \SNIa\ sample will be needed to improve this measurement.  Let
us indulge in a bit of speculation and suppose that these future
improvements confirm that \fp\ is close to 20\%\ at high redshifts.
We would then need a theoretical explanation for how the prompt
component of the \SNIa\ population could be suppressed at redshifts
$z>1.5$.  One possible explanation would be that the prompt \SNIa\
component is dominated by a SD progenitor pathway with a strong
metallicity dependence.  Such a metallicity dependence has been
proposed by requiring an optically thick wind from the WD to regulate
the mass transfer rate, allowing the WD to grow to the Chandrasekhar
mass limit \citep{Hachisu:1996,Hachisu:1999}.  This wind would be
absent in low metallicity WDs, resulting in a suppression of
prompt \SNIa\ explosions in the early
universe \citep{Kobayashi:1998,Kobayashi:2009}. No such metallicity
threshold is expected for DD models, so this scenario could provide a
clean way to disentangle the relative contributions of those two
progenitor pathways.

\section{Summary}\label{sec:Summary}

We have presented a sample of 65 SN from the 5 CANDELS fields.  This
sample, collected in concert with the CLASH SN search, is the first to
extend \SNIa\ detections beyond $z=2$, and the first to detect SN at
$z\approx1.5$ in rest-frame optical bands.  These SN have been classified
primarily through the application of STARDUST, a new
Bayesian photometric classifier that is optimized for working
with light curves of high-redshift SN.  We have spectroscopic
redshifts from the SN and/or host galaxy for 82\% of the sample (53 of
65), although we rely on photometric redshifts for 43\% of the
SN at $z>1.5$ (6 of 14).  Our SN classification probabilities are in
general tightly constrained by well-sampled light curves, rest-frame
UV-optical colors, and well defined redshifts.  The primary sources
for potential systematic biases in our classifications are 1) a
redshift-dependent prior describing the relative fraction of SN that
are Type Ia, and 2) the assumed distribution of dust extinction
values.  For the former, a test using host-galaxy information to
replace the class prior indicates that our systematic uncertainty
estimates are appropriate.

From the CANDELS SN sample we have measured the volumetric \SNIa\ rate
in 5 redshift bins reaching to $z=2.5$.  We find that the
CANDELS \SNIa\ rate measurement at $z\approx1.25$ is a factor of 2 lower
($\sim2\sigma$) than past \HST\ rates measurements at the same
redshift \citep{Dahlen:2008,Barbary:2012}, but is consistent with the
concurrently measured \SNIa\ rates from CLASH \citep{Graur:2014}. We
attribute this discrepancy to Poisson noise, due to the very small
sample sizes in all of these \HST\ surveys.  At higher redshifts the
CANDELS rate measurements remain approximately flat with redshift.

Combining these CANDELS results with other surveys from \HST\ and from
the ground, we have examined the constraints that can be placed
on \SNIa\ DTD models (Table~\ref{tab:DTD}, Figure \ref{fig:DTD}).  We
have invoked a simple two-component model with a $t^{-1}$ distribution
for long delay times ($t>500$ Myr), and a constant rate at shorter
times.  We find that the ground-based rates (primarily at $z<1$) and
the full sample of all available \SNIa\ rates both are best matched
with a prompt \SNIa\ fraction of $\fp=0.5$.  When the CANDELS+CLASH
surveys are analyzed in isolation, the best fit for the prompt \SNIa\
fraction falls to $\fp=0.21$$^{+0.34}_{-0.21}$$^{+0.49}_{-0.12}$.  This is
substantially lower, but with very large error bars. 

Collectively, the constraints from all available volumetric SN data
indicate that the prompt \SNIa\ fraction can not be much larger than
about 60\%.  As described above, there is room for substantial 
improvement in our measurement of the prompt \SNIa\ fraction --
without needing to acquire more data.  Systematic uncertainties can be
reduced (or at least better understood) by combining the existing \HST\
surveys in a composite analysis that handles SN classification and
dust obscuration in a consistent manner.  Increasing the
high-$z$ \SNIa\ sample size would also help, of course. The \HST\
Frontier Fields initiative will provide the next opportunity for new
high-$z$ SN discoveries. This program will utilize $\sim$900 orbits
of \HST\ observations for very deep imaging of 6 massive galaxy
clusters over 3 years.  An approved \HST\ program for SN discovery and
follow-up (PI:Rodney, PID:13386) is expected to deliver a \SNIa\
sample that reaches out to $z\approx3$.

\bigskip

We dedicate this work to the memory of our friend and colleague, Tomas
Dahlen. He is dearly missed.

The authors would like to thank the referee, Massimo Della Valle, 
for helpful comments and suggestions.  We are also grateful to the STScI
science support staff for their extraordinary efforts in executing the
MCT programs, especially program coordinators Patricia Royle and Beth
Perriello, as well as the entire STScI scheduling team that made our
\HST\ ToO program possible.  We give thanks to Kelsey Clubb, Ori Fox, 
Patrick Kelly, Isaac Shivvers, Brad Tucker, and WeiKang Zheng for assistance 
with some of the Keck observations; to Ismael Botti, Alice Mortlock, 
and Omar Almaini for redshift and AGN classifications of SN candidates in 
the UDS field; to Stephane Blondin for VLT spectral reductions and 
assistance with SNID modifications; to Jennifer Lotz and Ryan Foley 
for discussions improving the measurement and application of SN host-galaxy
morphologies; and to Thomas Holoien and Mark Ziegler for ``fake SN''
searching assistance.

The CANDELS and SN-MCT programs were supported by NASA through {\it
HST} grants GO-12060 and GO-12099 (respectively) from STScI.  Support
for S.A.R. was provided by NASA through Hubble Fellowship grant
HST-HF-51312.  Support for this research at Rutgers University was
provided in part by NSF CAREER award AST-0847157 to S.W.J.  A.V.F. is
also grateful for the support of National Science Foundation (NSF)
grant AST-1211916, the TABASGO Foundation, and the Christopher
R. Redlich Fund.  The Dark Cosmology Centre is supported by the Danish
National Research Foundation.  R.P.K. thanks the National Science
Foundation for AST-1211196, and the John Simon Guggenheim Foundation
for support.  This work was supported by NASA Keck PI Data Awards (to
Rutgers University, PI: S.W.J.), administered by the NASA Exoplanet
Science Institute. J.M.S. is supported by an NSF Astronomy and
Astrophysics Postdoctoral Fellowship under award AST-1302771.
S.G.W. is supported by the NSF Graduate Research Fellowship under
Grant No. DGE-1232825.

This research was based primarily on observations made with the
NASA/ESA {\it Hubble Space Telescope}, delivered by the data archive team 
at the Space Telescope Science Institute (STScI), which is operated by
the association of Universities for Research in Astronomy, Inc. under
NASA contract NAS 5-26555. We used the {\it HST} Science
Archive hosted by the Canadian Astronomy Data Centre (CADC/NRC/CSA),
as well as the Mikulski Archive for Space Telescopes (MAST).

Based in part on observations made with ESO telescopes at the La Silla
Paranal Observatory under program IDs 086.A-0660 and 088.A-0708.

Some of the data presented herein were obtained at the W. M. Keck
Observatory, partly from telescope time allocated to NASA
through the agency's scientific partnership
with the California Institute of Technology and the University of
California. The Observatory was made possible by the generous
financial support of the W. M. Keck Foundation.  The authors wish to
recognize and acknowledge the very significant cultural role and
reverence that the summit of Mauna Kea has always had within the
indigenous Hawaiian community.  We are most fortunate to have the
opportunity to conduct observations from this mountain.

Based in part on observations obtained at the Gemini Observatory,
which is operated by the Association of Universities for Research in
Astronomy, Inc., under a cooperative agreement with the NSF on behalf
of the Gemini partnership: the National Science Foundation (United
States), the National Research Council (Canada), CONICYT (Chile), the
Australian Research Council (Australia), Minist\'{e}rio da
Ci\^{e}ncia, Tecnologia e Inova\c{c}\~{a}o (Brazil) and Ministerio de
Ciencia, Tecnolog\'{i}a e Innovaci\'{o}n Productiva (Argentina).
Contributing data came from Gemini programs GN-2011A-Q-14,
GN-2011B-Q-18, GN-2012A-Q-32, GN-2013A-Q-25, GS-2011A-Q-16,
GS-2011B-Q-18, GS-2012A-Q-17, and GS-2013A-Q-19.

This research has made use of the VO Datascope, developed with the
support of the NSF under Cooperative Agreement
AST--0122449 with the Johns Hopkins University, and hosted by the
Astrophysics Science Division and the High Energy Astrophysics Science
Archive Research Center (HEASARC), a service of Goddard Space Flight
Center and the Smithsonian Astrophysical Observatory. We also used 
the SIMBAD database, operated at CDS, Strasbourg,
France; NASA's Astrophysics Data System
Bibliographic Services; and Astropy, a
community-developed core Python package for Astronomy
\citep{Astropy:2013}.

%\medskip
%{\it Facilities:} 
%\facility{HST (WFC3)} \facility{Keck (DEIMOS)} 
%\facility{Keck (LRIS)} \facility{VLT (X-shooter)} 
%\facility{VLT (FORS2)} \facility{Gemini (GMOS-N)}
%\facility{Gemini (GMOS-S)}

\bibliographystyle{apj}

\begin{appendix}

\section{The Host-Based Class Prior}\label{app:galsnid}

In this Appendix we perform a test to see if our systematic uncertainty
estimates are accurately reflecting the bias that could
arise from using an incorrect redshift-dependent prior P(Ia,z) in the
STARDUST classifier.  For this test we measure the change in
the volumetric \SNIa\ rate that occurs when we adopt a
redshift-independent prior based on host-galaxy data. 

It is well established that \CCSN\ are almost never observed in
galaxies dominated by older stellar populations (passive, red
ellipticals).  The rate of \SNIa\ per unit mass increases in galaxies
with young stellar populations, but not as sharply as the
specific \CCSN\ rate.  The \galsnid\ SN classifier \citep{Foley:2013a}
exploits these relationships between SN and their host environments to
define a posterior classification probability that is completely
independent of the SN photometry.

The \galsnid\ approach relies on a database of observed SN host-galaxy 
properties (morphology, color, luminosity, etc.) from the Lick
Observatory Supernova Survey \citep[LOSS][]{Leaman:2011}.  These data
are used to define a set of likelihoods P({\bf D}$_{h}|$Type)
giving the probability of observing a set of host-galaxy
properties, {\bf D}$_{h}$, if the SN hosted in that galaxy is of the
given Type. The \galsnid\ posterior classification
probability for SN~Ia is

\begin{equation}\label{eqn:galsnid}
P(Ia|{\bf D}_{h}) = k_{h}^{-1}P(Ia)\prod_{i=1}^n P(D_{i}|Ia), 
\end{equation}

\noindent where P(Ia) is the SN~Ia classification prior,  $D_{h,i}$ are 
the $n$ observed host-galaxy properties, and $k_h$ is a normalization
term, which ensures that the sum of posterior classification
probabilities for all \SNIa\ and \CCSN\ types is unity.  For our
implementation of \galsnid, the prior P(Ia) -- and the corresponding
priors for \CCSN\ -- are fixed to match the observed fraction of SN in
a volume-limited sample that are of each type: P(Ia)=0.25,
P(Ib/c)=0.19, P(II)=0.57 \citep{Smartt:2009,Li:2011a}.  For the 
host-galaxy observables, {\bf D$_{h}$}, we employ the two quantities that
provide the strongest discriminatory power according
to \citet{Foley:2013a}: morphology and color.  

\makeatletter{}\begin{small}
  \begin{deluxetable}{lcc}
    \tablecolumns{3}
    \tablecaption{Host galaxy likelihood distributions for defining host-based classification priors \tablenotemark{a}\label{tab:HostPriors}}
    \tablehead{ 
      \colhead{Category} & 
      \colhead{$P(D|Ia)$} & 
      \colhead{$P(D|CC)$} 
    }
    \startdata
   \cutinhead{Morphology} 
   spheroid (E,S0)         &  0.25 $\pm$ 0.03 & 0.01 $\pm$ 0.01 \\
   spheroid+disk (S0,Sa)   &  0.26 $\pm$ 0.03 & 0.15 $\pm$ 0.02 \\
   disk (Sb,Sbc,Sc)        &  0.35 $\pm$ 0.03 & 0.53 $\pm$ 0.03 \\
   disk+irregular (Sc,Scd) &  0.10 $\pm$ 0.02 & 0.20 $\pm$ 0.02 \\
   irregular (Scd,Irr)     &  0.04 $\pm$ 0.01 & 0.11 $\pm$ 0.02 \\

   \cutinhead{SED Type} 
   Passive  (B-K$>$3.75)      &  0.35 $\pm$ 0.06 &  0.08 $\pm$ 0.06 \\
   Active (2.75$<$B-K$<$3.75) &  0.52 $\pm$ 0.08 &  0.62 $\pm$ 0.08 \\ 
   Starburst (B-K$<$2.75)     &  0.13 $\pm$ 0.04 &  0.30 $\pm$ 0.04 \\ 
    \enddata
    \tablenotetext{a}{Likelihood estimates follow the {\it galsnid} 
derivation \citep{Foley:2013a}, using LOSS host galaxy data compiled in \citet{Leaman:2011}.}
\tablenotetext{b}{Approximate rest-frame B-K colors are given in Vega mags. In AB mags the thresholds are 0.82 and 1.82. }
  \end{deluxetable}
\end{small}
 
\makeatletter{}
\begin{small}
  \begin{deluxetable}{ cccc }
\tablewidth{0pt}
\tablecolumns{3}
\tablecaption{Change in Observed \SNIa\ Counts and the Volumetric Rate when adopting the Host Galaxy Prior\label{tab:galsnidRates}}
 \tablehead{
    \colhead{Redshift} & \colhead{$\Delta$N$_{obs}$} & \colhead{$\Delta$SNR} & \% of Sys. Err.
 }
  \startdata
 0.25  &  +0.13  &  +0.03  &  26  \\
 0.75  &  +1.99  &  +0.14  &  62  \\
 1.25  &  +0.93  &  +0.07  &  21  \\
 1.75  &  +0.41  &  +0.05  &  10  \\
 2.25  &  +0.33  &  +0.13  &  28  \\
  \enddata
\end{deluxetable}
\end{small}

Table~\ref{tab:HostPriors} translates from the CANDELS categories for
host-galaxy morphology (spheroid,disk,irregular) into their
approximate counterparts on the  Hubble sequence (E,S0,etc).  That
Table also describes the translation from the CANDELS SED types
(passive, active, or starburst) into the rest-frame (B-K) color. 

For SN in faint hosts where the host-galaxy morphology is undefined,
the $\galsnid$ posterior reflects only the constraint from
color/SED-type.  In the two cases where the SN host galaxy is totally
undetected, the photometric redshifts are unconstrained and \galsnid\
simply reflects back the input prior, P(Ia) in
Equation~\ref{eqn:galsnid}.
The likelihood values in columns 2 and 3 of Table~\ref{tab:HostPriors} mimic
Table 1 of \citet{Foley:2013a} and provide all the information needed
to determine P(Ia,host) for any CANDELS SN host galaxy.

\subsection{Combining \galsnid\ and STARDUST}

To incorporate the \galsnid\ information into STARDUST, we adopt the
\galsnid\ posterior, P(Ia$|${\bf D$_h$}), as a {\em redshift independent}
prior P(Ia,host), replacing P(Ia,z) in Equation~\ref{eqn:STARDUST}. Note that this assumption is assuredly
incorrect: at redshift $z\approx2$ the fraction of \SNIa\ appearing in
red early-type galaxies must be much lower than it is locally, simply
because there are far fewer of those passive old galaxies at high
redshift.  However, for the purpose of this systematic test, we will
make the brazen assumption that the relationships between SN types and
their host-galaxy properties do not evolve with redshift.

To define our baseline rate measurement, we have relied on the
redshift-dependent {\it mid-rate} P(Ia,z) prior (the green solid curve
in Figure~\ref{fig:RatePriors}).  Because the \galsnid-based
prior P(Ia,host) prior is redshift independent, we can use it to
check for strong redshift biases in the the rates-based
P(Ia,z) prior. In principle, these two priors could be combined into a
{\it redshift-dependent, host-based} prior -- but that is beyond the
scope of this paper.

In column 5 of Tables~\ref{tab:highzSN} and \ref{tab:lowzSN} we have
reported the STARDUST probabilities derived using this host-galaxy-based 
prior. As one should expect, SN for which we have abundant
spectroscopic and photometric information are barely affected.  Thus,
classification probabilities that were close to 0 or 1 using the
redshift-dependent prior typically do not shift.  For objects with
intermediate probabilities, the Ia classification probability is
almost uniformly increased, but the change is mostly within the range
allowed by the classification uncertainties.  The total change in the
count of observed \SNIa\ and the resulting change in the volumetric SN
rate are presented in Table~\ref{tab:galsnidRates}.  Again, the shift
is uniformly positive, but relatively small when compared to
systematic uncertainties.  The final column of this table reports the
change in the \SNIa\ rate measurement as a fraction of the systematic
uncertainty estimate, and we see that it is less than unity in every
redshift bin.  While not definitive, this result suggests that our
baseline rates are not heavily biased by the redshift-dependent class
prior, and any existing bias has been sufficiently accounted for in
our systematic uncertainty estimates.

\section{Supplementary Data for SN at $\ensuremath{z}<1.5$}\label{app:lowz}

\begin{figure*}
  \includegraphics[width=0.333\textwidth]{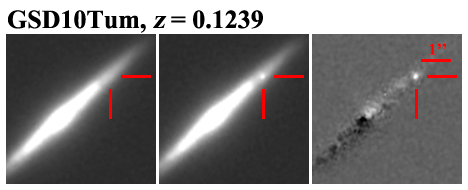}
  \includegraphics[width=0.333\textwidth]{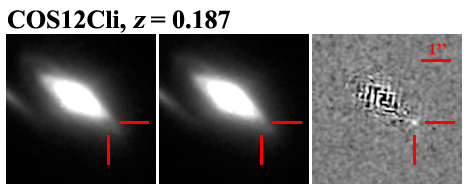}
  \includegraphics[width=0.333\textwidth]{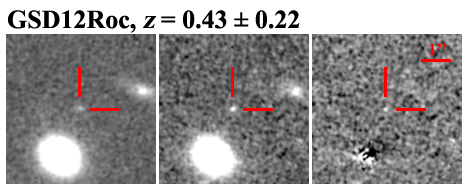}
  \includegraphics[width=0.333\textwidth]{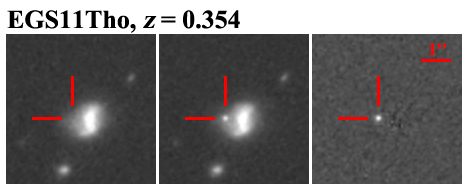}
  \includegraphics[width=0.333\textwidth]{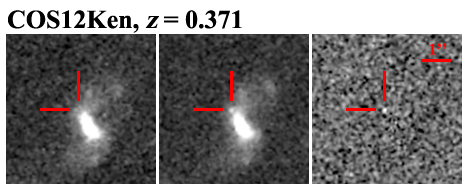}
  \includegraphics[width=0.333\textwidth]{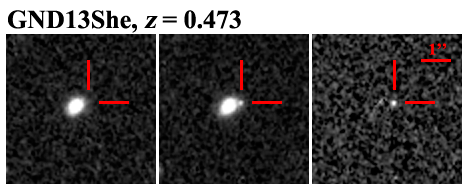}
  \includegraphics[width=0.333\textwidth]{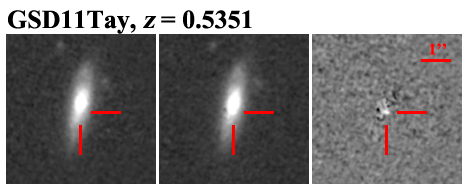}
  \includegraphics[width=0.333\textwidth]{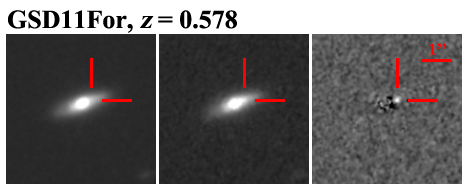}
  \includegraphics[width=0.333\textwidth]{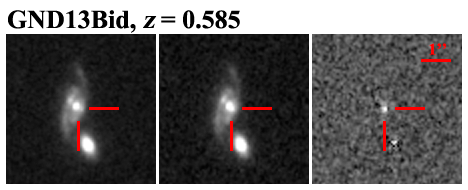}
  \includegraphics[width=0.333\textwidth]{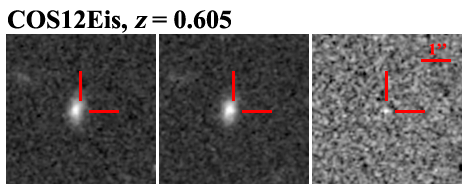}
  \includegraphics[width=0.333\textwidth]{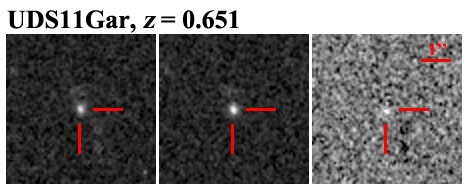}
  \includegraphics[width=0.333\textwidth]{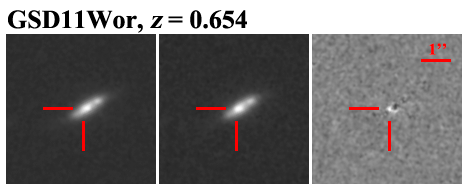}
  \includegraphics[width=0.333\textwidth]{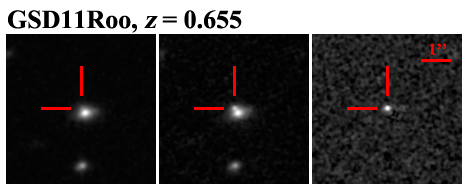}
  \includegraphics[width=0.333\textwidth]{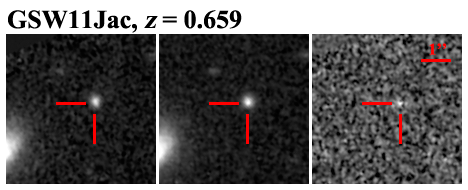}
  \includegraphics[width=0.333\textwidth]{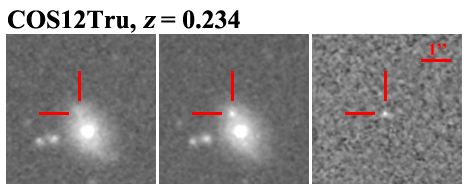}
  \includegraphics[width=0.333\textwidth]{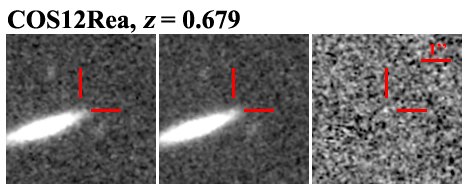}
  \includegraphics[width=0.333\textwidth]{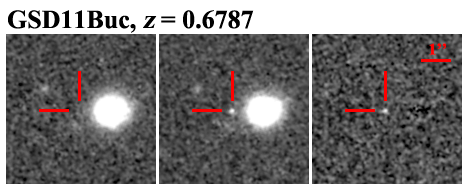}
  \includegraphics[width=0.333\textwidth]{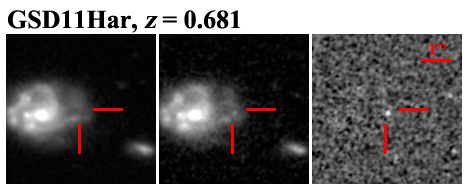}
  \includegraphics[width=0.333\textwidth]{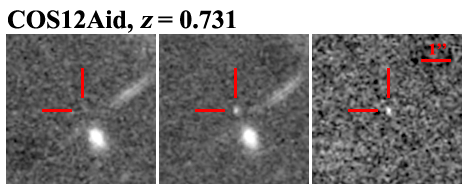}
  \includegraphics[width=0.333\textwidth]{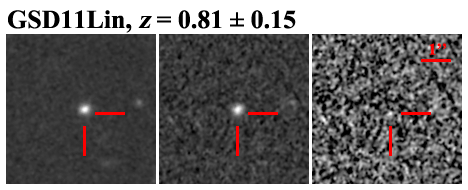}
  \includegraphics[width=0.333\textwidth]{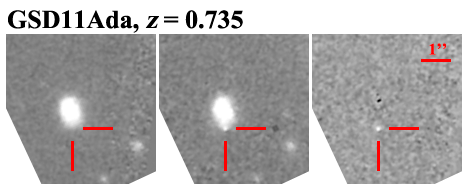}
  \includegraphics[width=0.333\textwidth]{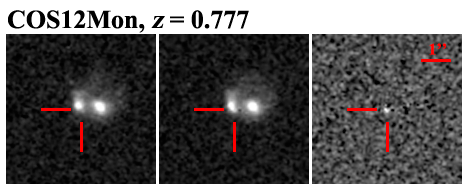}
  \includegraphics[width=0.333\textwidth]{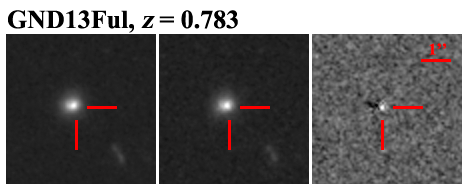}
  \includegraphics[width=0.333\textwidth]{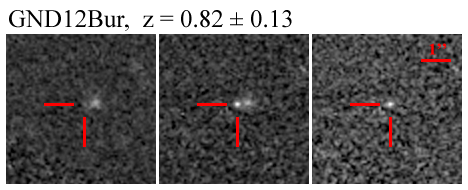}
  \includegraphics[width=0.333\textwidth]{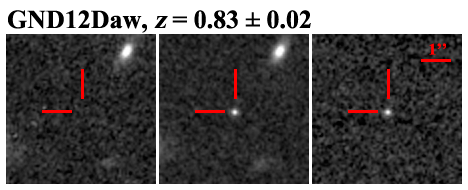}
  \includegraphics[width=0.333\textwidth]{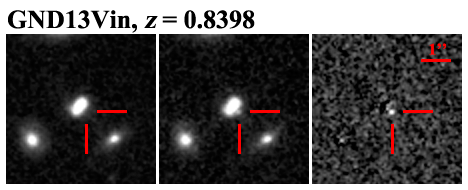}
  \includegraphics[width=0.333\textwidth]{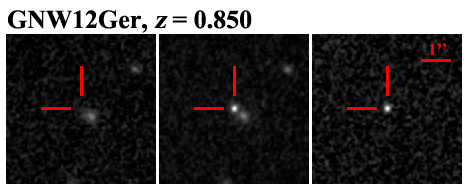}
\caption{ Detection images for 27 SN from the CANDELS fields with
  redshifts $z\leq 0.85$.  Each image triplet shows H band (F160W) images
  with the template image on the left, the discovery epoch image in
  the middle and the difference image on the right. All images have a
  width of about 6 arcseconds, with North up and East to the left. The
  position of the SN is marked by (red) crosshairs in every
  frame. \label{fig:lowzStamps1} }
\end{figure*}

\begin{figure*}
  \includegraphics[width=0.333\textwidth]{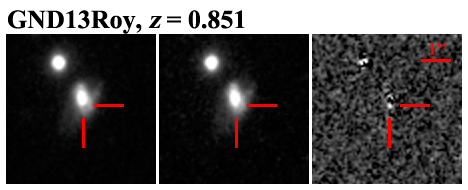}
  \includegraphics[width=0.333\textwidth]{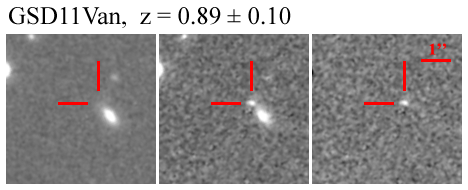}
  \includegraphics[width=0.333\textwidth]{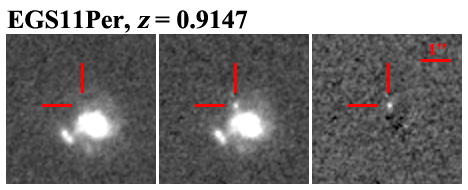}
  \includegraphics[width=0.333\textwidth]{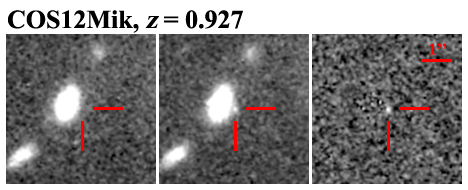}
  \includegraphics[width=0.333\textwidth]{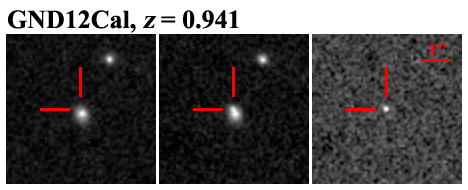}
  \includegraphics[width=0.333\textwidth]{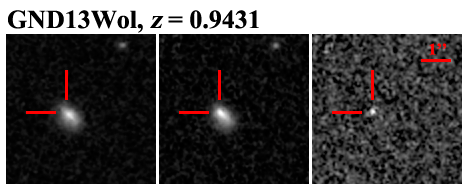}
  \includegraphics[width=0.333\textwidth]{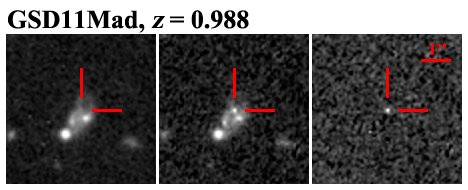}
  \includegraphics[width=0.333\textwidth]{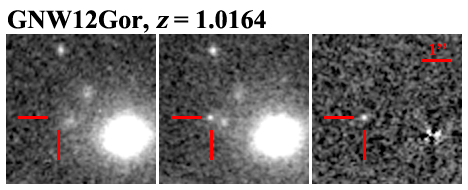}
  \includegraphics[width=0.333\textwidth]{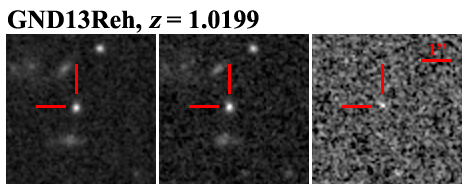}
  \includegraphics[width=0.333\textwidth]{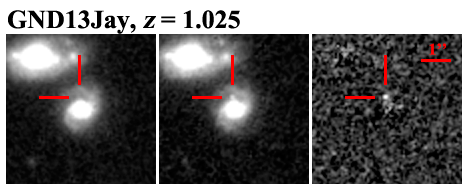}
  \includegraphics[width=0.333\textwidth]{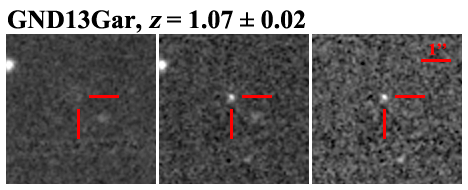}
  \includegraphics[width=0.333\textwidth]{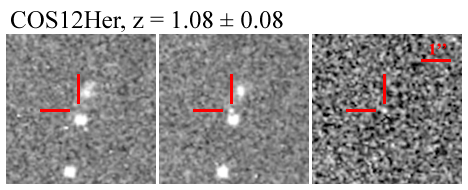}
  \includegraphics[width=0.333\textwidth]{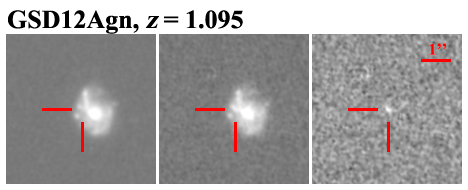}
  \includegraphics[width=0.333\textwidth]{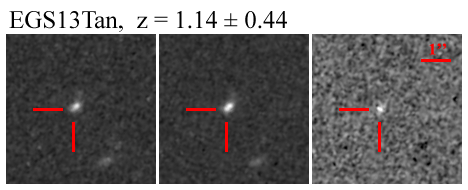}
  \includegraphics[width=0.333\textwidth]{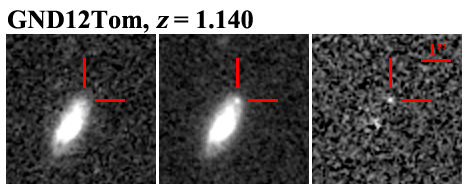}
  \includegraphics[width=0.333\textwidth]{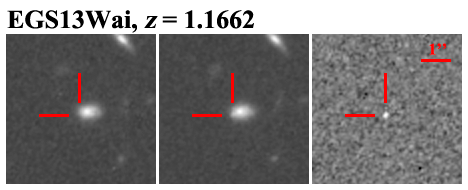}
  \includegraphics[width=0.333\textwidth]{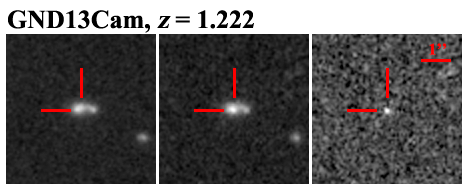}
  \includegraphics[width=0.333\textwidth]{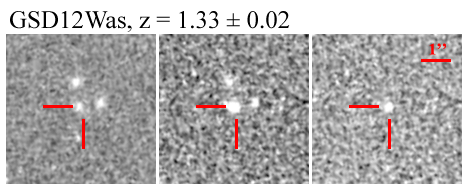}
  \includegraphics[width=0.333\textwidth]{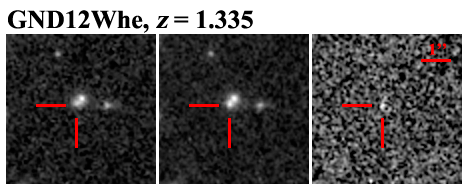}
  \includegraphics[width=0.333\textwidth]{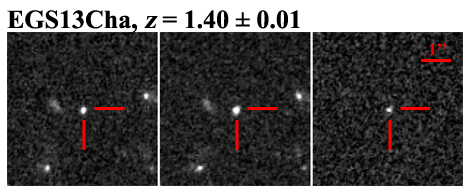}
  \includegraphics[width=0.333\textwidth]{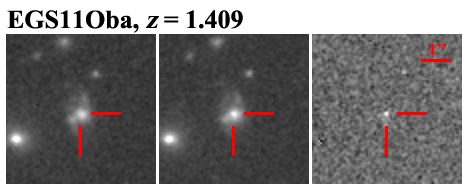}
  \includegraphics[width=0.333\textwidth]{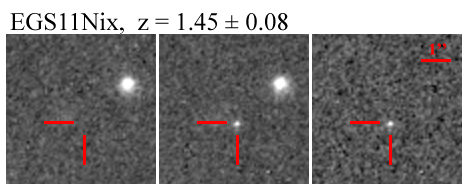}
  \includegraphics[width=0.333\textwidth]{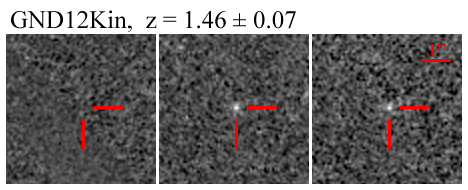}
  \includegraphics[width=0.333\textwidth]{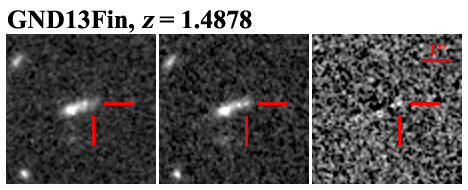}
\caption{ Detection images for 24 SN from the CANDELS fields with
  redshifts $0.85<z<1.5$.  Each image triplet shows H band (F160W) images
  with the template image on the left, the discovery epoch image in
  the middle and the difference image on the right. All images have a
  width of about 6 arcseconds, with North up and East to the left. The
  position of the SN is marked by (red) crosshairs in every
  frame. \label{fig:lowzStamps2} }
\end{figure*}

\begin{figure*}[p]
  \includegraphics[width=0.333\textwidth]{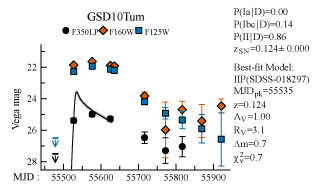}
  \includegraphics[width=0.333\textwidth]{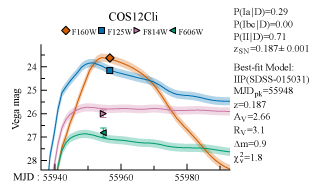}
  \includegraphics[width=0.333\textwidth]{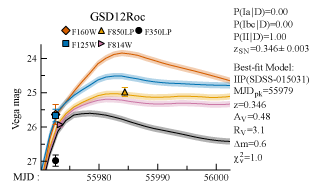}
  \includegraphics[width=0.333\textwidth]{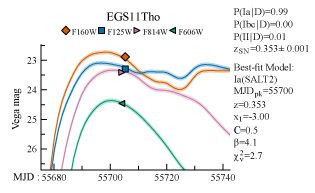}
  \includegraphics[width=0.333\textwidth]{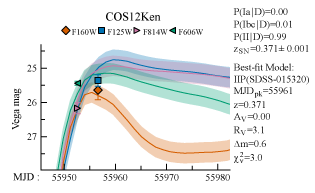}
  \includegraphics[width=0.333\textwidth]{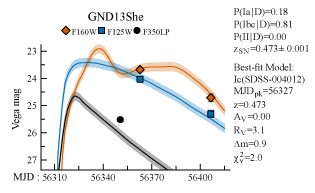}
  \includegraphics[width=0.333\textwidth]{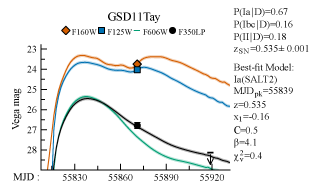}
  \includegraphics[width=0.333\textwidth]{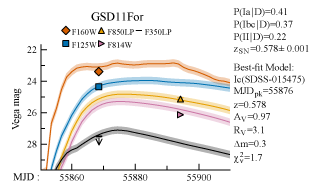}
  \includegraphics[width=0.333\textwidth]{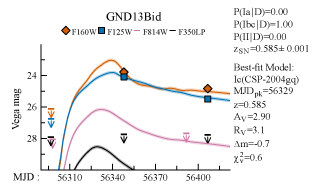}
  \includegraphics[width=0.333\textwidth]{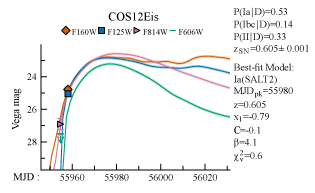}
  \includegraphics[width=0.333\textwidth]{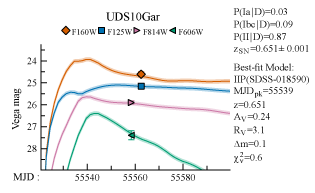}
  \includegraphics[width=0.333\textwidth]{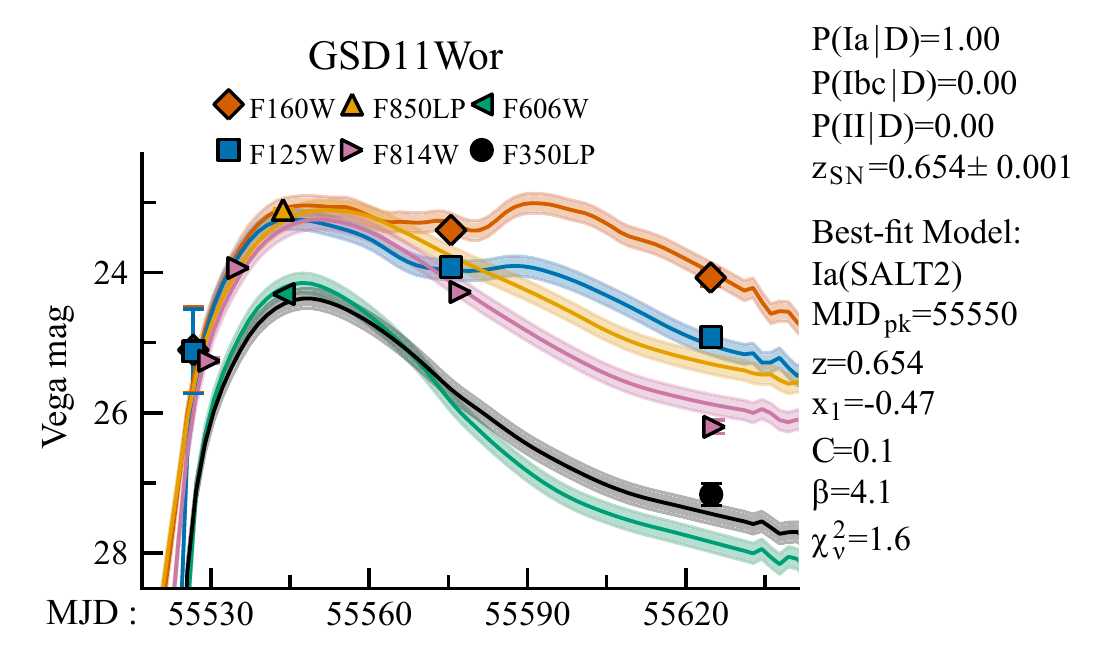}
  \includegraphics[width=0.333\textwidth]{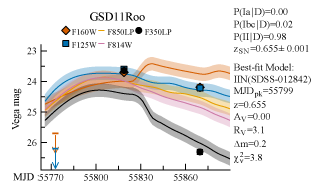}
  \includegraphics[width=0.333\textwidth]{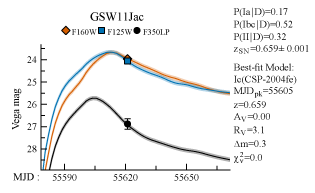}
  \includegraphics[width=0.333\textwidth]{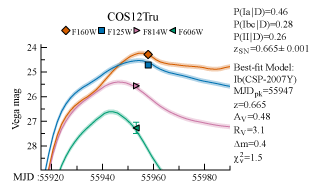}
  \includegraphics[width=0.333\textwidth]{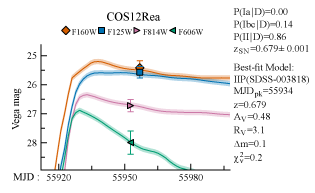}
  \includegraphics[width=0.333\textwidth]{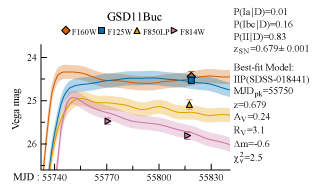}
  \includegraphics[width=0.333\textwidth]{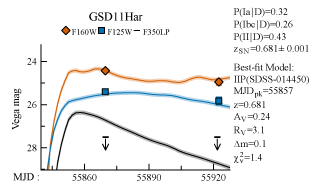}
\caption{ STARDUST light curve matches for the first 18 SN from the CANDELS fields in redshift order, with redshifts $z<0.7$, as in
    Figure~\ref{fig:highzLCfit}. \label{fig:lowzLCfit1} }
\end{figure*}

\begin{figure*}[p]
  \includegraphics[width=0.333\textwidth]{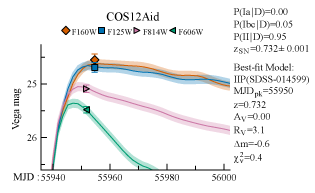}
  \includegraphics[width=0.333\textwidth]{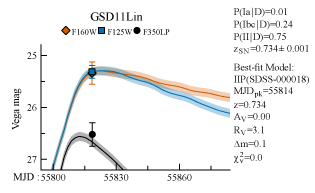}
  \includegraphics[width=0.333\textwidth]{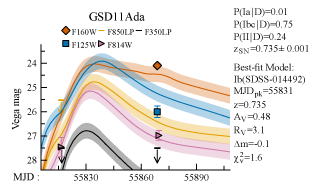}
  \includegraphics[width=0.333\textwidth]{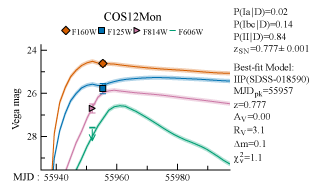}
  \includegraphics[width=0.333\textwidth]{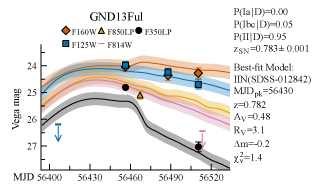}
  \includegraphics[width=0.333\textwidth]{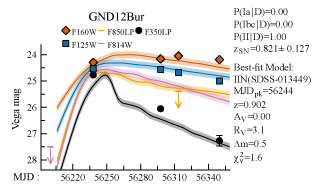}
  \includegraphics[width=0.333\textwidth]{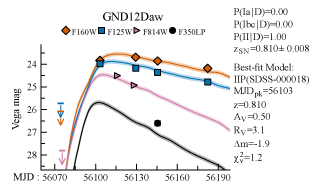}
  \includegraphics[width=0.333\textwidth]{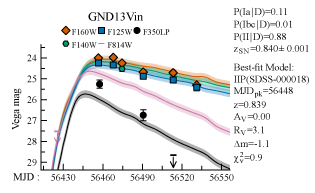}
  \includegraphics[width=0.333\textwidth]{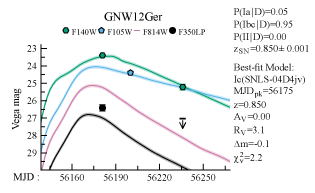}
  \includegraphics[width=0.333\textwidth]{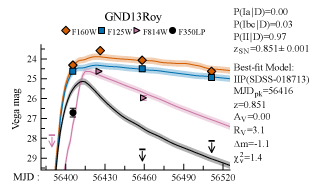}
  \includegraphics[width=0.333\textwidth]{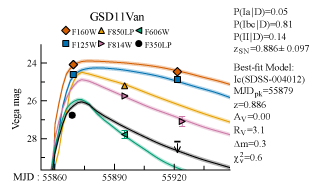}
  \includegraphics[width=0.333\textwidth]{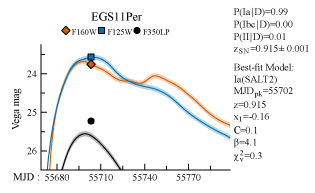}
  \includegraphics[width=0.333\textwidth]{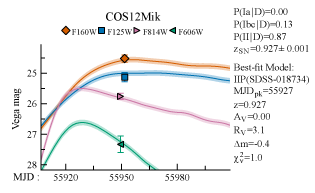}
  \includegraphics[width=0.333\textwidth]{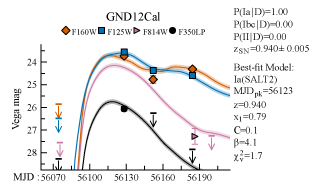}
  \includegraphics[width=0.333\textwidth]{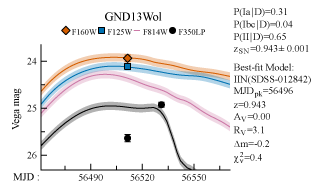}
  \includegraphics[width=0.333\textwidth]{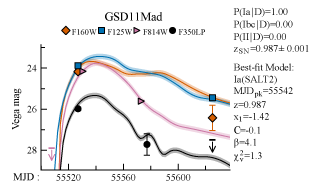}
\caption{ STARDUST light curve matches for the 16 SN from the CANDELS fields with redshifts $0.7<z<1.0$, as in
    Figure~\ref{fig:highzLCfit}. \label{fig:lowzLCfit2}}
\end{figure*}

\begin{figure*}[p]
  \includegraphics[width=0.333\textwidth]{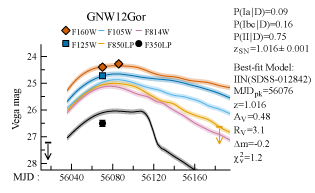}
  \includegraphics[width=0.333\textwidth]{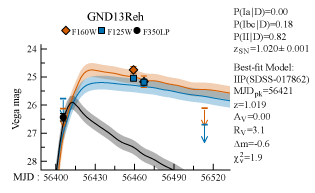}
  \includegraphics[width=0.333\textwidth]{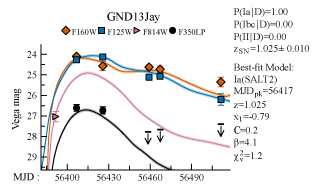}
  \includegraphics[width=0.333\textwidth]{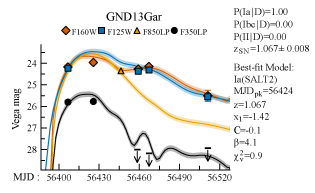}
  \includegraphics[width=0.333\textwidth]{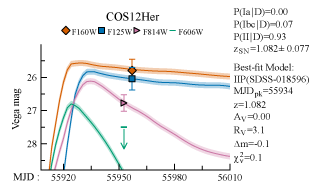}
  \includegraphics[width=0.333\textwidth]{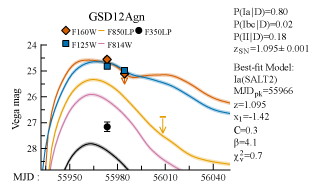}
  \includegraphics[width=0.333\textwidth]{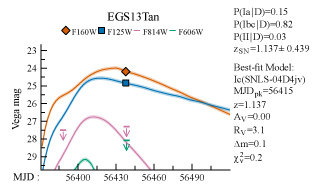}
  \includegraphics[width=0.333\textwidth]{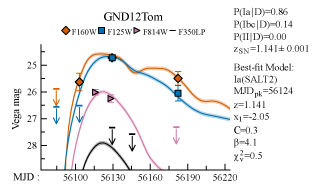}
  \includegraphics[width=0.333\textwidth]{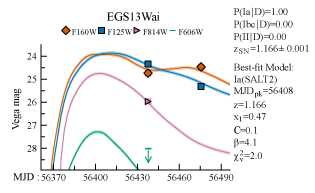}
  \includegraphics[width=0.333\textwidth]{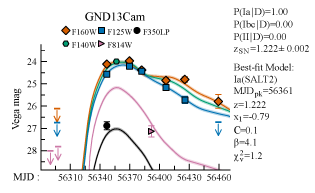}
  \includegraphics[width=0.333\textwidth]{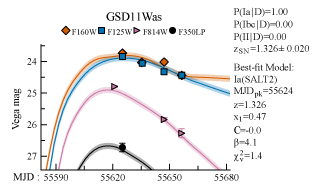}
  \includegraphics[width=0.333\textwidth]{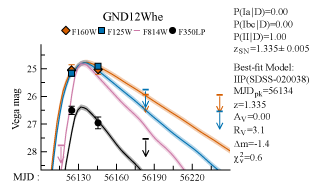}
  \includegraphics[width=0.333\textwidth]{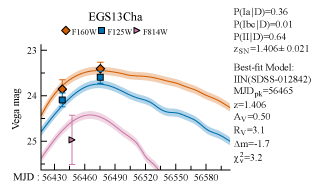}
  \includegraphics[width=0.333\textwidth]{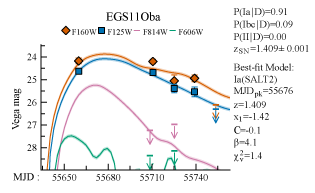}
  \includegraphics[width=0.333\textwidth]{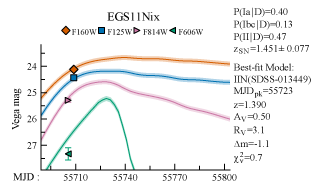}
  \includegraphics[width=0.333\textwidth]{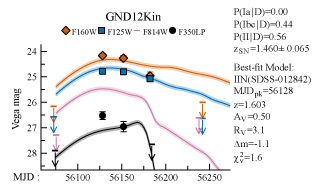}
  \includegraphics[width=0.333\textwidth]{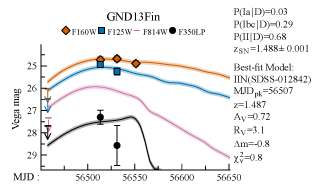}
\caption{ STARDUST light curve matches for the 17 SN from the
  CANDELS fields with redshifts $1.0<z<1.5$, as in
    Figure~\ref{fig:highzLCfit}.\label{fig:lowzLCfit3}}
\end{figure*}

Discovery images for the 51 SN with redshift $z<1.5$ are shown in
Figures~\ref{fig:lowzStamps1} and \ref{fig:lowzStamps2}.  The names,
positions, classification probabilities and redshifts of those 51
low-$z$ SN are given in Table~\ref{tab:lowzSN} (as in
Table~\ref{tab:highzSN}). Host-galaxy information is provided in
Table~\ref{tab:lowzHosts} (as in Table~\ref{tab:highzHosts}). Light
curves and best-fitting template matches from STARDUST are shown
in Figures~\ref{fig:lowzLCfit1}, \ref{fig:lowzLCfit2},
and \ref{fig:lowzLCfit3}.

  To reduce file size, Figures~\ref{fig:lowzStamps1}, \ref{fig:lowzStamps2}, 
  \ref{fig:lowzLCfit1}, \ref{fig:lowzLCfit2}, and \ref{fig:lowzLCfit3} have been
  included at reduced resolution for the arXiv version of this paper.  The full 
  paper with all figures included at full resolution is 10 MB.  It will be available 
  from the AJ site upon publication, and is currently available here: 
  \url{pha.jhu.edu/~srodney/papers/candelsrates.pdf}

\makeatletter{}
 
\makeatletter{}
 
\makeatletter{}\renewcommand{\arraystretch}{1.0}
\begin{footnotesize}
\begin{deluxetable*}{llllllllp{2.1in}}
\tablecolumns{9}
 
\tablecaption{51 Supernovae with $z<1.5$ \label{tab:lowzSN}}
\tablehead{\colhead{Name} & \colhead{R.A. (J2000)} & \colhead{Decl. (J2000)} & \colhead{P(Ia$|$D$_{z}$)\tablenotemark{a}} & \colhead{P(Ia$|$D$_{host}$)\tablenotemark{b}} & \colhead{z$_{\mbox{\scriptsize SN}}$\tablenotemark{c}} & \colhead{($\pm$)} & \colhead{$z$ Source\tablenotemark{d}}}
\startdata
GSD10Tum & 03:32:17.705 & $-$27:50:57.50 & 0.00 $^{+0.00}_{-0.00}$ & 0.00 $^{+0.00}_{-0.00}$ & 0.124 & (0.001) & host spec-z \\
COS12Cli & 10:00:15.958 & $+$02:12:36.27 & 0.29 $^{+0.14}_{-0.28}$ & 0.29 $^{+0.03}_{-0.29}$ & 0.187 & (0.001) & host spec-z \\
GSD12Roc & 03:32:06.368 & $-$27:47:26.63 & 0.00 $^{+0.00}_{-0.00}$ & 0.00 $^{+0.00}_{-0.00}$ & 0.346 & (0.003) & host spec-z \\
EGS11Tho & 14:19:31.775 & $+$52:51:56.16 & 0.99 $^{+0.01}_{-0.98}$ & 0.99 $^{+0.00}_{-0.98}$ & 0.354 & (0.001) & host spec-z \\
COS12Ken & 10:00:36.010 & $+$02:15:26.53 & 0.00 $^{+0.00}_{-0.00}$ & 0.00 $^{+0.00}_{-0.00}$ & 0.373 & (0.001) & host spec-z \\
GND13She & 12:36:09.877 & $+$62:14:06.00 & 0.18 $^{+0.33}_{-0.18}$ & 0.31 $^{+0.25}_{-0.31}$ & 0.473 & (0.001) & host spec-z \\
GSD11Tay & 03:32:54.505 & $-$27:47:04.23 & 0.67 $^{+0.21}_{-0.64}$ & 0.72 $^{+0.12}_{-0.64}$ & 0.535 & (0.001) & host spec-z \\
GSD11For & 03:32:14.300 & $-$27:47:13.30 & 0.41 $^{+0.26}_{-0.41}$ & 0.43 $^{+0.13}_{-0.42}$ & 0.578 & (0.001) & host spec-z \\
GND13Bid & 12:36:41.325 & $+$62:11:42.21 & 0.00 $^{+0.00}_{-0.00}$ & 0.00 $^{+0.00}_{-0.00}$ & 0.585 & (0.001) & host spec-z \\
COS12Eis & 10:00:47.233 & $+$02:11:50.50 & 0.53 $^{+0.14}_{-0.18}$ & 0.69 $^{+0.04}_{-0.01}$ & 0.605 & (0.001) & host spec-z \\
UDS11Gar & 02:17:39.633 & $-$05:11:37.14 & 0.03 $^{+0.09}_{-0.03}$ & 0.08 $^{+0.10}_{-0.08}$ & 0.651 & (0.001) & host spec-z \\
GSD11Wor & 03:32:10.739 & $-$27:48:07.17 & 1.00 $^{+0.00}_{-0.00}$ & 1.00 $^{+0.00}_{-0.00}$ & 0.654 & (0.001) & host spec-z \\
GSD11Roo & 03:32:31.587 & $-$27:46:12.53 & 0.00 $^{+0.00}_{-0.00}$ & 0.01 $^{+0.00}_{-0.00}$ & 0.655 & (0.001) & host spec-z \\
GSW11Jac & 03:32:32.300 & $-$27:54:20.55 & 0.17 $^{+0.13}_{-0.11}$ & 0.18 $^{+0.04}_{-0.06}$ & 0.659 & (0.001) & host spec-z \\
COS12Tru & 10:00:38.322 & $+$02:11:36.41 & 0.46 $^{+0.32}_{-0.46}$ & 0.52 $^{+0.22}_{-0.52}$ & 0.665 & (0.001) & host spec-z \\
COS12Rea & 10:00:31.917 & $+$02:14:16.15 & 0.00 $^{+0.00}_{-0.00}$ & 0.00 $^{+0.00}_{-0.00}$ & 0.679 & (0.001) & host spec-z \\
GSD11Buc & 03:32:28.800 & $-$27:52:32.00 & 0.01 $^{+0.01}_{-0.01}$ & 0.10 $^{+0.06}_{-0.10}$ & 0.679 & (0.001) & host spec-z \\
GSD11Har & 03:32:30.441 & $-$27:45:18.70 & 0.32 $^{+0.28}_{-0.32}$ & 0.25 $^{+0.12}_{-0.25}$ & 0.681 & (0.001) & host spec-z \\
COS12Aid & 10:00:15.252 & $+$02:17:32.12 & 0.00 $^{+0.00}_{-0.00}$ & 0.00 $^{+0.00}_{-0.00}$ & 0.731 & (0.001) & host spec-z \\
GSD11Lin & 03:32:29.812 & $-$27:49:19.71 & 0.00 $^{+0.00}_{-0.00}$ & 0.01 $^{+0.00}_{-0.01}$ & 0.734 & (0.001) & host spec-z \\
GSD11Ada & 03:32:19.805 & $-$27:54:10.04 & 0.01 $^{+0.02}_{-0.01}$ & 0.01 $^{+0.01}_{-0.01}$ & 0.735 & (0.001) & host spec-z \\
COS12Mon & 10:00:26.737 & $+$02:15:13.74 & 0.02 $^{+0.06}_{-0.02}$ & 0.49 $^{+0.19}_{-0.49}$ & 0.777 & (0.001) & host spec-z \\
GND13Ful & 12:36:19.201 & $+$62:15:12.58 & 0.00 $^{+0.00}_{-0.00}$ & 0.00 $^{+0.00}_{-0.00}$ & 0.783 & (0.001) & host spec-z \\
GND12Bur & 12:36:32.536 & $+$62:15:32.62 & 0.00 $^{+0.00}_{-0.00}$ & 0.00 $^{+0.00}_{-0.00}$ & 0.82  & (0.13) & host+SN phot-z \\
GND12Daw & 12:36:41.340 & $+$62:18:52.50 & 0.00 $^{+0.00}_{-0.00}$ & 0.00 $^{+0.00}_{-0.00}$ & 0.83  & (0.02) & SN spec-z + phot-z \\
GND13Vin & 12:37:06.354 & $+$62:15:17.79 & 0.11 $^{+0.18}_{-0.11}$ & 0.13 $^{+0.09}_{-0.13}$ & 0.840 & (0.001) & host spec-z \\
GNW12Ger & 12:37:17.023 & $+$62:20:38.67 & 0.05 $^{+0.02}_{-0.03}$ & 0.66 $^{+0.00}_{-0.10}$ & 0.850 & (0.001) & host spec-z \\
GND13Roy & 12:36:46.223 & $+$62:15:27.13 & 0.00 $^{+0.00}_{-0.00}$ & 0.00 $^{+0.00}_{-0.00}$ & 0.851 & (0.001) & host spec-z \\
GSD11Van & 03:32:19.037 & $-$27:47:17.90 & 0.05 $^{+0.12}_{-0.05}$ & 0.19 $^{+0.18}_{-0.19}$ & 0.886 & (0.097) & host+SN phot-z \\
EGS11Per & 14:18:28.318 & $+$52:42:45.88 & 0.99 $^{+0.00}_{-0.01}$ & 1.00 $^{+0.00}_{-0.00}$ & 0.915 & (0.001) & host spec-z \\
COS12Mik & 10:00:31.687 & $+$02:26:13.73 & 0.00 $^{+0.00}_{-0.00}$ & 0.01 $^{+0.00}_{-0.00}$ & 0.927 & (0.001) & host spec-z \\
GND12Cal & 12:37:10.487 & $+$62:15:47.67 & 1.00 $^{+0.00}_{-0.00}$ & 1.00 $^{+0.00}_{-0.00}$ & 0.941 & (0.005) & host spec-z \\
GND13Wol & 12:36:58.946 & $+$62:18:10.15 & 0.31 $^{+0.08}_{-0.12}$ & 0.58 $^{+0.02}_{-0.05}$ & 0.943 & (0.001) & host spec-z \\
GSD11Mad & 03:32:18.781 & $-$27:52:42.05 & 1.00 $^{+0.00}_{-0.00}$ & 1.00 $^{+0.00}_{-0.00}$ & 0.988 & (0.001) & host spec-z \\
GNW12Gor & 12:36:20.704 & $+$62:08:45.08 & 0.09 $^{+0.10}_{-0.08}$ & 0.23 $^{+0.07}_{-0.15}$ & 1.016 & (0.001) & host spec-z \\
GND13Reh & 12:36:54.452 & $+$62:11:52.47 & 0.00 $^{+0.00}_{-0.00}$ & 0.09 $^{+0.03}_{-0.06}$ & 1.019 & (0.001) & host spec-z \\
GND13Jay & 12:36:41.380 & $+$62:11:30.02 & 1.00 $^{+0.00}_{-0.01}$ & 1.00 $^{+0.00}_{-0.00}$ & 1.03  & (0.01) & host spec-z \\
GND13Gar & 12:36:40.806 & $+$62:11:14.16 & 1.00 $^{+0.00}_{-0.00}$ & 1.00 $^{+0.00}_{-0.00}$ & 1.067 & (0.008) & SN spec-z + phot-z \\
COS12Her & 10:00:47.446 & $+$02:15:17.74 & 0.01 $^{+0.00}_{-0.00}$ & 0.02 $^{+0.03}_{-0.01}$ & 1.082 & (0.077) & host+SN phot-z \\
GSD12Agn & 03:32:25.902 & $-$27:50:19.62 & 0.80 $^{+0.13}_{-0.71}$ & 0.82 $^{+0.07}_{-0.63}$ & 1.095 & (0.001) & host spec-z \\
EGS13Tan & 14:20:02.098 & $+$53:00:16.36 & 0.15 $^{+0.05}_{-0.08}$ & 0.35 $^{+0.07}_{-0.07}$ & 1.137 & (0.439) & host+SN phot-z \\
GND12Tom & 12:36:42.543 & $+$62:18:21.93 & 0.86 $^{+0.09}_{-0.75}$ & 0.85 $^{+0.06}_{-0.63}$ & 1.14  & (0.001) & host spec-z \\
EGS13Wai & 14:20:28.534 & $+$53:04:58.61 & 1.00 $^{+0.00}_{-0.00}$ & 1.00 $^{+0.00}_{-0.00}$ & 1.166 & (0.001) & host spec-z \\
GND13Cam & 12:37:07.354 & $+$62:10:26.90 & 1.00 $^{+0.00}_{-0.01}$ & 1.00 $^{+0.00}_{-0.00}$ & 1.222 & (0.002) & host spec-z \\
GSD11Was & 03:32:20.856 & $-$27:49:41.48 & 1.00 $^{+0.00}_{-0.00}$ & 1.00 $^{+0.00}_{-0.00}$ & 1.33  & (0.02) & SN spec-z + phot-z \\
GND12Whe & 12:36:24.220 & $+$62:17:39.70 & 0.00 $^{+0.00}_{-0.00}$ & 0.00 $^{+0.00}_{-0.00}$ & 1.335 & (0.005) & host spec-z \\
EGS13Cha & 14:20:24.200 & $+$53:03:37.50 & 0.36 $^{+0.01}_{-0.14}$ & 0.80 $^{+0.01}_{-0.11}$ & 1.40  & (0.01) & host spec-z \\
EGS11Oba & 14:20:32.663 & $+$53:02:48.18 & 0.91 $^{+0.02}_{-0.17}$ & 0.90 $^{+0.00}_{-0.05}$ & 1.409 & (0.001) & host spec-z \\
EGS11Nix & 14:20:48.603 & $+$53:00:26.47 & 0.40 $^{+0.01}_{-0.11}$ & 0.58 $^{+0.04}_{-0.10}$ & 1.451 & (0.077) & host+SN phot-z \\
GND12Kin & 12:37:13.005 & $+$62:16:30.83 & 0.00 $^{+0.00}_{-0.00}$ & 0.00 $^{+0.00}_{-0.00}$ & 1.46  & (0.07) & host+SN phot-z \\
GND13Fin & 12:37:18.173 & $+$62:17:09.65 & 0.03 $^{+0.06}_{-0.03}$ & 0.02 $^{+0.01}_{-0.01}$ & 1.488 & (0.001) & host spec-z \\
\enddata
\tablenotetext{a}{ Type Ia SN classification probability from STARDUST, 
  using the redshift-dependent class prior. Uncertainties reflect systematic
  biases due to the class prior and extinction assumptions
  (Sections~\ref{sec:ClassPrior} and \ref{sec:HostAvDistribution}). }
\tablenotetext{b}{ Type Ia SN classification probability from STARDUST, 
  using the {\em galsnid} host galaxy prior. Uncertainties reflect systematic
  biases due to the class prior and extinction assumptions.}
\tablenotetext{c}{ Posterior redshift and uncertainty, as determined 
  by the STARDUST light curve fit.}
\tablenotetext{d}{ The {\it host / SN} values indicate whether the redshift is derived from the host galaxy, the SN itself, or a combination;  {\it spec-z / phot-z} specify a spectroscopic or photometric redshift. A value of {\it host+SN phot-z} means the redshift is derived from a STARDUST light curve fit, with the host galaxy phot-z used as a prior.}

\ifms  \end{deluxetable}
\else  \end{deluxetable*}
\fi
\end{footnotesize}
\renewcommand{\arraystretch}{1.0}

\makeatletter{}\renewcommand{\arraystretch}{1.0}
\begin{small}
\begin{deluxetable*}{llllrccllp{1.5in}}
\tablecolumns{10}
 
\tablecaption{Host galaxies of 51 Supernovae with $z<1.5$\label{tab:lowzHosts}}
\tablehead{\colhead{SN} & \colhead{R.A. (J2000)} & \colhead{Decl. (J2000)} & \colhead{d[\arcsec]} & \colhead{d[kpc]\tablenotemark{a}} & \colhead{Morph.\tablenotemark{b}} & \colhead{SED\tablenotemark{c}} & \colhead{z$_{\mbox{host}}$} & \colhead{($\pm$)} & \colhead{$z$ Reference\tablenotemark{d}}}
\startdata
GSD10Tum & 03:32:17.871	& $-$27:50:59.48 & 2.96 & 57.9 & d & A  & 0.124 & 0.001 & \citet{LeFevre:2004} \\
COS12Cli & 10:00:16.060 & $+$02:12:37.38 & 1.89 & 37.8 & d & SB & 0.187 & 0.001 & \citet{Trump:2009} \\
GSD12Roc & 03:32:06.368 & $-$27:47:26.63 & 0.00 & 0.0 & i & SB & 0.346 & 0.003 & Keck+DEIMOS (B.Mobasher, S.Jha) \\
EGS11Tho & 14:19:31.685 & $+$52:51:56.05 & 0.82 & 16.8 & di & A & 0.354 & 0.001 & Keck+LRIS (A. Filippenko), Gemini+GMOS (S. Jha) \\
COS12Ken & 10:00:35.978 & $+$02:15:25.81 & 0.87 & 17.6 & di & SB & 0.373 & 0.001 & \citet{Lilly:2009} \\
GND13She & 12:36:09.907 & $+$62:14:05.79 & 0.30 & 6.1 & sd & A & 0.473 & 0.001 & \citet{Wirth:2004} \\
GSD11Tay & 03:32:54.502 & $-$27:47:03.67 & 0.56 & 11.5 & d & A & 0.535 & 0.001 & \citet{LeFevre:2004} \\
GSD11For & 03:32:14.320 & $-$27:47:13.15 & 0.30 & 6.3 & d & A & 0.578 & 0.001 & \citet{Mignoli:2005} \\
GND13Bid & 12:36:41.417 & $+$62:11:42.53 & 0.72 & 14.7 & i & A & 0.585 & 0.001 & \citet{Cohen:2000} \\
COS12Eis & 10:00:47.275 & $+$02:11:50.04 & 0.78 & 15.9 & sd & SB & 0.605 & 0.001 & Keck+DEIMOS (B.Mobasher) \\
UDS11Gar & 02:17:39.631 & $-$05:11:37.00 & 0.14 & 3.0 & sd & SB & 0.651 & 0.001 & Keck+DEIMOS (B.Mobasher) \\
GSD11Wor & 03:32:10.730 & $-$27:48:07.14 & 0.12 & 2.6 & d & A & 0.654 & 0.001 & \citet{Mignoli:2005} \\
GSD11Roo & 03:32:31.581 & $-$27:46:12.71 & 0.20 & 4.1 & s & SB & 0.655 & 0.001 & VLT+FORS2 (B.Leibundgut) \\
GSW11Jac & 03:32:32.310 & $-$27:54:20.46 & 0.16 & 3.3 & d & A & 0.659 & 0.001 & VLT+Xshooter (J.Hjorth) \\
COS12Tru & 10:00:38.293 & $+$02:11:35.60 & 0.92 & 18.7 & d & A & 0.665 & 0.001 & Keck+DEIMOS (B.Mobasher) \\
COS12Rea & 10:00:32.021 & $+$02:14:15.43 & 1.72 & 34.4 & d & SB & 0.679 & 0.001 & \citet{Lilly:2009} \\
GSD11Buc & 03:32:28.714 & $-$27:52:32.00 & 1.14 & 23.1 & s & A & 0.679 & 0.001 & \citet{LeFevre:2004} \\
GSD11Har & 03:32:30.570 & $-$27:45:18.35 & 1.75 & 35.0 & d & SB & 0.681 & 0.001 & \citet{LeFevre:2004}, \citet{Mignoli:2005} \\
COS12Aid & 10:00:15.212 & $+$02:17:30.84 & 1.41 & 28.5 & sd & SB & 0.731 & 0.001 & Keck+DEIMOS (B.Mobasher) \\
GSD11Lin & 03:32:29.799 & $-$27:49:19.26 & 0.48 & 9.9 & sd & SB & 0.734 & 0.001 & Keck+DEIMOS (A.Filippenko, S.Jha) \\
GSD11Ada & 03:32:19.785 & $-$27:54:09.15 & 0.93 & 18.9 & sd & SB & 0.735 & 0.001 & \citet{Balestra:2010}, \citet{Popesso:2009} \\
COS12Mon & 10:00:26.746 & $+$02:15:14.02 & 0.31 & 6.4 & s & SB & 0.777 & 0.001 & \citet{Lilly:2009} \\
GND13Ful & 12:36:19.227 & $+$62:15:12.76 & 0.26 & 5.3 & sd & SB & 0.783 & 0.001 & \citet{Wirth:2004} \\
GND12Bur & 12:36:32.464 & $+$62:15:32.75 & 0.52 & 10.7 & i & SB & 0.69 & 0.4 & phot-z (T.Dahlen) \\
GND12Daw & \nodata & \nodata & \nodata & \nodata & \nodata & \nodata & \nodata & \nodata & \nodata \\
GND13Vin & 12:37:06.308 & $+$62:15:18.08 & 0.43 & 8.9 & i & A & 0.840 & 0.001 & \citet{Cohen:2000} \\
GNW12Ger & 12:37:16.965 & $+$62:20:38.34 & 0.52 & 10.7 & s & A & 0.850 & 0.005 & Keck+DEIMOS (A.Filippenko) \\
GND13Roy & 12:36:46.232 & $+$62:15:27.46 & 0.34 & 6.9 & i & SB & 0.851 & 0.001 & \citet{Cooper:2011} \\
GSD11Van & 03:32:19.004 & $-$27:47:18.33 & 0.61 & 12.6 & d & P & 0.74 & 0.15 & phot-z (T.Dahlen) \\
EGS11Per & 14:18:28.248 & $+$52:42:45.07 & 1.03 & 20.9 & d & A & 0.915 & 0.001 & \citet{Barro:2011} \\
COS12Mik & 10:00:31.733 & $+$02:26:13.92 & 0.72 & 14.6 & sd & SB & 0.927 & 0.001 & Keck+DEIMOS (B.Mobasher) \\
GND12Cal & 12:37:10.476 & $+$62:15:47.40 & 0.28 & 5.8 & s & SB & 0.941 & 0.005 & HST+WFC3 (B.Weiner) \\
GND13Wol & 12:36:58.933 & $+$62:18:10.04 & 0.14 & 3.0 & sd & SB & 0.9431 & 0.0002 & \citet{Barger:2008} \\
GSD11Mad & 03:32:18.767 & $-$27:52:42.45 & 0.44 & 9.1 & d & SB & 0.988 & 0.001 & Keck+LRIS (A.Filippenko) \\
GNW12Gor & 12:36:20.619 & $+$62:08:44.91 & 0.62 & 12.7 & u & A & 1.0164 & 0.0005 & \citet{Wirth:2004}, \citet{Treu:2005} \\
GND13Reh & 12:36:54.445 & $+$62:11:52.43 & 0.06 & 1.3 & s & SB & 1.019 & 0.001 & \citet{Barger:2008} \\
GND13Jay & 12:36:41.370 & $+$62:11:29.57 & 0.46 & 9.4 & d & A & 1.03 & 0.01 & HST+WFC3 (B.Weiner) \\
GND13Gar & 12:36:40.813 & $+$62:11:14.34 & 0.19 & 3.9 & u & SB & 1.07 & 0.02 & phot-z (T.Dahlen) \\
COS12Her & 10:00:47.438 & $+$02:15:17.35 & 0.41 & 8.4 & s & SB & 1.1 & 0.4 & Keck+DEIMOS (B.Mobasher) \\
GSD12Agn & 03:32:25.861 & $-$27:50:19.81 & 0.58 & 11.8 & d & SB & 1.095 & 0.001 & \citet{Vanzella:2008} \\
EGS13Tan & 14:20:02.010 & $+$53:00:17.04 & 1.05 & 21.3 & sd & A & 1.39 & 0.8 & phot-z (T.Dahlen) \\
GND12Tom & 12:36:42.636 & $+$62:18:20.74 & 1.36 & 27.4 & d & A & 1.140 & 0.001 & \citet{Barger:2008} \\
EGS13Wai & 14:20:28.495 & $+$53:04:58.72 & 0.37 & 7.6 & di & SB & 1.1662 & 0.0004 & \citet{Barro:2011} \\
GND13Cam & 12:37:07.357 & $+$62:10:26.94 & 0.05 & 0.9 & sd & SB & 1.222 & 0.002 & HST+WFC3 (B.Weiner) \\
GSD11Was & 03:32:20.856 & $-$27:49:41.48 & 0.00 & 0.0 & d & SB & 1.30 & 0.05 & HST+WFC3 (A.Riess) \\
GND12Whe & 12:36:24.206 & $+$62:17:39.92 & 0.24 & 5.0 & sd & SB & 1.335 & 0.001 & Keck+DEIMOS (C.Papovich) \\
EGS13Cha & 14:20:24.200 & $+$53:03:37.50 & 0.00 & 0.0 & sd & A & 1.40 & 0.05 & Keck+MOSFIRE (M.Cooper) \\
EGS11Oba & 14:20:32.666 & $+$53:02:48.10 & 0.08 & 1.8 & di & A & 1.409 & 0.001 & Keck+LRIS (A.Filippenko) \\
EGS11Nix & 14:20:48.607 & $+$53:00:26.56 & 0.10 & 2.0 & u & A & 1.73 & 0.51 & phot-z (T.Dahlen) \\
GND12Kin & 12:37:13.013 & $+$62:16:30.86 & 0.06 & 1.3 & u & SB & 1.9 & 0.4 & phot-z (T.Dahlen) \\
GND13Fin & 12:37:18.322 & $+$62:17:09.34 & 1.08 & 22.0 & i & SB & 1.4878 & 0.0005 & \citet{Wirth:2004}, \citet{Reddy:2006} \\
\enddata

\tablenotetext{a}{ Physical separation between the SN and center of the host, 
  computed from the measured angular separation in the preceding column, 
  assuming a flat \LCDM\ cosmology with \Ho=70, \Om=0.3}
\tablenotetext{b}{ Visual classifications for host galaxy morphology: 
   s = spheroid, d = disk, i = irregular}
\tablenotetext{c}{ Template-matching classification of host galaxy SED:
   P = Passive, A = Active, SB = Starburst type}
\tablenotetext{d}{ Unpublished spectroscopic observations are given as {\em
    Observatory}+{\em Instrument} ({\em name of PI}). Host galaxy
  photometric redshifts are marked as {\em phot-z} (Dahlen et al. in
  prep).}

\ifms  \end{deluxetable}
\else  \end{deluxetable*}
\fi
\end{small}
\renewcommand{\arraystretch}{1.0}

\makeatletter{}
 
\makeatletter{}

\makeatletter{}

\end{appendix}

\end{document}